\pdfoutput=1
\documentclass{aa}  
\usepackage{graphicx}
\usepackage{setspace}
\usepackage{amssymb}
\usepackage{longtable,lscape}
\usepackage{rotating}
\usepackage{natbib}
\usepackage{color}
\usepackage{natbib}
\bibpunct{(}{)}{;}{a}{}{,}

\usepackage{graphicx}
\usepackage{txfonts}
\begin{document} 

\hyphenation{a-na-ly-sis mo-le-cu-lar pre-vious e-vi-den-ce dif-fe-rent pa-ra-me-ters ex-ten-ding a-vai-la-ble ca-li-bra-tion ca-li-bra-te con-ti-nu-um}

\title{A near-IR spectroscopic survey of massive jets towards EGOs}
\subtitle{
\thanks{Based on observations collected at the European Southern Observatory
 La Silla, Chile, 080.C-0573(A), 083.C-0846(A)}
}
\author{A. Caratti o Garatti \inst{1}, B. Stecklum \inst{2}, H. Linz \inst{3}, R. Garcia Lopez \inst{1}, \and A. Sanna \inst{1}}

\offprints{A. Caratti o Garatti, \email{acaratti@mpifr-bonn.mpg.de}}

\institute{
Max-Planck-Institut f\"{u}r Radioastronomie, Auf dem H\"{u}gel 69, D-53121 Bonn, Germany\\
\email{acaratti;rgarcia;asanna@mpifr-bonn.mpg.de}\\
\and
Th\"uringer Landessternwarte Tautenburg,
Sternwarte 5, D-07778 Tautenburg, Germany\\
\email{stecklum@tls-tautenburg.de}\\
\and
Max-Planck-Institut f\"{u}r Astronomie, K\"{o}nigstuhl 17, D-69117 Heidelberg, Germany\\
\email{linz@mpia-hd.mpg.de}\\
}

%
\date{Received date; Accepted date}
%
%
%
\abstract
   {Protostellar jets and outflows are the main outcome of the star formation process, and their analysis can provide us
   with major clues about the ejection and accretion history of young stellar objects (YSOs).}
   {We aim at deriving the main physical properties of massive jets from near-IR (NIR) observations, 
   comparing them to those of a large sample of jets from low-mass YSOs, and relating them to the 
   main features of their driving sources.}
   {We present a NIR imaging (H$_2$ and $K_s$) and low-resolution spectroscopic (0.95-2.50\,$\mu$m) survey of 18 massive jets towards GLIMPSE extended 
   green objects (EGOs), driven by intermediate- and high-mass YSOs, which have bolometric luminosities ($L_{\rm bol}$) 
   between 4$\times$10$^2$ and 1.3$\times$10$^5$\,L$_{\sun}$.}
  {As in low-mass jets, H$_2$ is the primary NIR coolant, detected in all the analysed flows, whereas
the most important ionic tracer is [\ion{Fe}{ii}], detected in half of the sampled jets. 
Our analysis indicates that the emission lines originate from shocks at high temperatures and densities.
No fluorescent emission is detected along the flows, regardless of the source bolometric luminosity.
On average, the physical parameters of these massive jets (i.e. visual extinction, temperature, column density, mass, and
luminosity) have higher values than those measured in their low-mass counterparts.
The morphology of the H$_2$ flows is varied, mostly depending on the complex, dynamic, and inhomogeneous environment in which these massive jets form and propagate.
All flows and jets in our sample are collimated, showing large precession angles. 
Additionally, the presence of both knots and jets suggests that the ejection process is continuous with burst episodes, as in low-mass YSOs. 
We compare the flow H$_2$ luminosity with the source bolometric luminosity confirming the tight correlation between these two quantities. 
Five sources, however, display a lower $L_{H_2}$/$L_{bol}$ efficiency, which might be related to YSO evolution. 
Most important, the inferred $L_{H_2}$ vs. $L_{bol}$ relationship agrees well with the correlation between the momentum flux of the CO outflows 
and the bolometric luminosities of high-mass YSOs indicating that 
outflows from high-mass YSOs are momentum driven, as are their low-mass counterparts.
We also derive a less stringent correlation between the inferred mass of the H$_2$ flows and $L_{bol}$ of the YSOs, indicating that
the mass of the flow depends on the driving source mass.}
{By comparing the physical properties of jets in the NIR, a continuity from low- to high-mass jets is identified. 
Massive jets appear as a scaled-up version of their low-mass counterparts in terms of their physical parameters and origin. 
Nevertheless, there are consistent differences such as a more variegated morphology and, on average, stronger shock conditions, which are likely due to 
the different environment in which high-mass stars form.}
\titlerunning{NIR spectroscopy of massive jets from intermediate- and high-mass YSOs}
\authorrunning{A. Caratti o Garatti et al.}

\maketitle
\section{Introduction}
\label{introduction:sec}

Protostellar jets and outflows are a main outcome of the star formation process from young brown dwarfs to high-mass YSOs~\citep[see e.g.][]{whelan05,arce07,ray07,tan}, 
and are observed over a wide wavelength range~\citep[from X-rays to the radio; e.g.][]{frank}.
These energetic phenomena play a critical role in removing a large fraction of the angular momentum from 
a contracting, rotating core with a magnetised disc, which collimates and accelerates
the flow~\citep[e.g.][]{konigl00,pudritz_PPV,tan}. 
Therefore, they are often used as an indirect tracer of accretion, providing us with fundamental clues 
about the accretion processes and the accretion history of young stellar objects~\citep[e.g.][]{arce07,bally07}. 
They become particularly important in the case of embedded YSOs, in which the accretion process cannot be directly observed.

In this context, bipolar jets trace the ejecta at scales of a few AUs to parsecs from the driving source, 
and they are referred to as the \emph{primary outflows} or jets. Jets, in turn,
accelerate entrained gas of the ambient medium to supersonic velocities, producing large-scale
(tenths to a few parsecs) molecular outflows or \emph{secondary outflows}. 
In this respect, jets are directly related to the
physical processes taking place close to the driving source (e.g. mass loss, mass accretion, etc.). Protostellar jets produce shocks that can be mainly studied 
at optical and IR wavelengths, where the brightest ionic (e.g. [\ion{O}{i}], [\ion{S}{ii}], H$\alpha$, [\ion{Fe}{ii}]) and molecular (e.g. H$_2$)
jet tracers occur~\citep[e.g.][]{reipurth01}. 

In the case of jets from high-mass YSOs 
(HMYSOs\footnote{In this paper we favour the term high-mass young stellar object~\citep[HMYSO; e.g.][]{varricatt} instead of high-mass protostellar object (HMPO), 
because the former also includes young massive stars on the ZAMS.}; $M_*>$8\,M$_\sun$, i.e. $L_{bol} \geq$5$\times$10$^3$\,L$_\sun$),
such observations are strongly limited by large distances (several kpc) and very high visual extinction
($A_{\rm V}$ up to 100\,mag). This makes the detection of high-mass protostellar jets extremely challenging and our knowledge is still limited.


Indeed massive outflows are mostly studied through molecular outflow tracers (e.g. SiO, CO, HCO$^+$, and their isotopologues) 
at submillimetre (submm) and millimetre (mm) wavelengths~\citep[]{beuther,wu,lopez-sepulcre,lopez-sepulcre11,duarte}. 
These surveys provide us with tight correlations between the main 
parameters of the outflows (e.g. power, force, and mass-loss rate) and important source parameters, such as the bolometric luminosity or the envelope mass ($M_{env}$),
over a large range of $L_{\rm bol}$ values up to $\sim$10$^6$\,L$_{\sun}$. These fundamental studies are, however, mostly 
limited to the analysis of the secondary outflow at relatively low spatial resolution, whereas there are just a few  
interferometric studies of single sources that allow us to trace the molecular outflow and the driving source at (sub)arcsecond 
resolution~\citep[see e.g.][]{beuther04,leurini13,sanna,tan}. 

So far, the number of studies of the primary outflow component in HMYSOs are
rare and committed to single objects or regions~\citep[see e.g.][]{marti93,davis04,puga,gredel06,caratti08,martin08}.
This prevents us from properly comparing the ejection properties of these objects with those of their low-mass counterparts. 
Recent NIR surveys in H$_2$ at 2.12\,$\mu$m have dramatically boosted the number of candidate massive jets from
a few to a few hundreds~\citep[e.g.][]{stecklum,varricatt,lee12,lee13}, providing us with a large sample to be studied
in detail. 
With the aim of deriving the main physical properties of massive jets and comparing them with their low-mass counterparts,
we have undertaken a NIR imaging (H$_2$ and continuum emission) and spectroscopic survey (0.95--2.5\,$\mu$m)
of a flux-limited sample of jets from massive YSOs presented in \citet{stecklum}.

This paper is organised as follows.
In Section~\ref{sample:sec} we present the selected sample along with the adopted selection criteria.
In Section~\ref{observations:sec} we report our observations, data reduction, and the ancillary data collected from the literature and public surveys.
Section~\ref{results:sec} describes the results obtained from our imaging and spectroscopy.
A general description of the physical properties of both molecular and ionic components of the flows is provided.
Individual flows are discussed in the Appendix~(see Sect.~\ref{objects:sec}).
Our discussion in Sect.~\ref{discussion:sec} focuses
on the physical processes that produce such flows (Sect.~\ref{discussion1:sec}),
their morphology (Sect.~\ref{discussion2:sec}), their relation
to the driving sources (Sect.~\ref{discussion4:sec}), and comparison with low-mass flows (Sect.~\ref{discussion5:sec}).
Finally, our conclusions are presented in Section \ref{conclusion:sec}.


\section{Sample selection criteria}
\label{sample:sec}

The Spitzer GLIMPSE and GLIMPSE~II surveys~\citep{benjamin,churchwell} have become an important tool to identify HMYSOs. 
Based on these surveys, \citet{cyganowski} and \citet{chen13} selected objects with excess emission 
in the IRAC band 2 images at 4.5\,$\mu$m (called extended green objects or EGOs due to the common 
colour-coding of the 4.5\,$\mu$m band as green in three-colour composite IRAC images at 3.6, 4.5, and 8\,$\mu$m).
Although their exact nature is still under 
debate~\citep[shocked emission in outflows, fluorescent emission or scattered continuum from HMYSOs; see e.g.][]{noriega,debuizer10,takami12},
EGOs seem to be mostly related to massive YSOs.

We first selected $\sim$160 massive outflow candidates by scrutinising the GLIMPSE and GLIMPSE~II surveys~\citep[][]{stecklum}. 
All targets show EGO emission and only half of them are reported in the EGO catalogues of \citet{cyganowski} and \citet{chen13}.
About two thirds of the outflow candidates are associated with CH$_3$OH and/or H$_2$O masers, 
which are typical signposts of very young and luminous objects~\citep[e.g.][]{minier}. 
Additionally, all associated candidate driving sources show emission at 24 and 70\,$\mu$m in the Spitzer MIPS images, and many of them are associated 
with OH masers~\citep[from][]{caswell}, typically coincident with ultra-compact \ion{H}{ii} (UC\ion{H}{ii}) regions or HMYSOs. 
Our subsequent H$_2$ (2.12\,$\mu$m) survey with SofI/NTT confirmed the presence of flows and H$_2$ emission in about half of 
our outflow candidates~\citep[][]{stecklum}. 
Each candidate driving source was identified and selected in the Spitzer IRAC/MIPS images on the basis of its colour as well as its 
location with respect to the outflow lobes. The coordinates of the YSO candidate were then checked in both 2MASS and Herschel images to identify its NIR (when visible) 
and FIR counterparts.

The selection criteria of our spectroscopic sample are therefore based on: 
\textit{a)} clear association of the H$_2$ jets/flows with EGOs; \textit{b)} jets and flows driven by intermediate- or high-mass YSOs ($L_{\rm bol} \sim$10$^2$--10$^5$\,L$_{\sun}$);
\textit{c)} a jet surface brightness $\geq$10$^{-15}$\,erg\,s$^{-1}$\,cm$^{-2}$\,arcsec$^{-2}$ in the H$_2$ continuum-subtracted images, i.e.
bright enough to perform low resolution NIR spectroscopy with SofI/NTT. 

The investigated sample of H$_2$ jets from intermediate- and high-mass YSOs is presented in Table~\ref{sample:tab}, 
which reports for each object: the name of the putative driving source, the flow association with MHOs~\citep[Molecular Hydrogen Emission-Line Object, see][]{davis10} 
and with EGOs in the \citet{cyganowski} catalogue, the driving source coordinates, its bolometric luminosity, and distance.
When available, $L_{\rm bol}$ and distances were retrieved from the literature, as indicated in Table~\ref{sample:tab}.
For five sources $L_{\rm bol}$ values were uncertain or unknown. 
We, therefore, estimated their $L_{\rm bol}$ using the photometric data available for these sources (see Sect.~\ref{SED:sec}).
$L_{\rm bol}$ values in our sample range from 4$\times$10$^2$ to 1.3$\times$10$^5$\,L$_{\sun}$ (see Column\,6 of Table~\ref{sample:tab} and Sect.~\ref{SED:sec}), 
therefore the corresponding zero-age main-sequence (ZAMS) spectral type ranges between B9 and O7. Accordingly, our targets are
intermediate- and high-mass YSOs. Two of them, namely \object{BGPS G014.849-00.992} and \object{GLIMPSE G035.0393-00.4735}, are definitively
intermediate-mass YSOs (see Appendix~\ref{appendixM:sec} and \ref{appendixN:sec}, respectively), whereas \object{IRAS15394-5358} and \object{IRAS16122-5047}
are on the edge of HMYSO classification (see Appendix~\ref{appendixI:sec} and \ref{appendixK:sec}, respectively).

As reported in the table, a few unresolved sources drive more than one jet, indicating the presence of a multiple system. 
In this case the reported bolometric luminosity refers to the whole system.

\begin{table*}
\begin{scriptsize}
\caption[]{ The investigated sample     \label{sample:tab}}
\begin{center}
\begin{tabular}{ccccccccc}
\hline \hline\\[-5pt]
Source &  Associated &   Associated &   $\alpha$(2000.0) & $\delta$(2000.0)  & $L_{\rm bol}$  & D & Ref. & Notes \\
       &     MHO     & EGO & ($hh$:$mm$:$ss$) & ($dd$:$mm$:$ss$) & (10$^4$\,$L_{\sun}$) & (kpc) & &\\
\hline\\[-5pt]
\object{[HSL2000] IRS 1}  & 1701      & G298.26+0.74     & 12:11:47.6&-61:46:19 & 1.6 & 4 & 1,2 & CH$_3$OH masers \\
\object{IRAS12405-6219}  & 1702,1703  &                  & 12:43:31.1&-62:36:13 & 3.6 & 4.4 & 1 & double jet, H$_2$O masers \\
\object{IRAS13481-6124}  &  1700      &                  & 13:51:37.9&-61:39:07 & 5.5 & 3.2 & 1 &  \\
\object{IRAS13484-6100}  & 1704  & G310.15+0.76          & 13:51:57.8&-61:15:49 & 4.3 & 5.4 & 1,3 & CH$_3$OH/H$_2$O masers \\
\object{IRAS14212-6131}  & 1705  & G313.76-0.86          & 14:25:01.5&-61:44:58 &  1.7  & 7.8 & 1 & CH$_3$OH/OH masers \\
\object{SSTGLMC G316.7627-00.0115}  & 1706  &            & 14:44:56.2&-59:47:59 & 0.5 & 2.8 & 4,5 & CH$_3$OH/OH masers \\
\object{Caswell OH 322.158+00.636}  & 1800   &           & 15:18:39.1&-56:38:49 & 13.3 & 4.3 & 4,2 & CH$_3$OH/OH masers \\
\object{IRAS15394-5358}  & 1801,1802  & G326.48+0.70     & 15:43:20.7&-54:07:36 & 0.4 & 1.8 & 1 & double jet, CH$_3$OH masers\\
\object{IRAS15450-5431}  & 1803  & G326.78-0.24          & 15:48:54.4&-54:40:37 & 0.92 & 3.9 & 1 & H$_2$O masers\\
\object{IRAS16122-5047}  & 1920,1921  & G332.35-0.12     & 16:16:05.2&-50:54:36 & 0.4 & 3.1 & 2 & double jet, CH$_3$OH/OH masers  \\
\object{IRAS16547-4247}  & 1900,1901,1902 & G343.12-0.06 & 16:58:17.9&-42:51:55 & 6.6 & 2.8 & 1 &  H$_2$O masers\\
\object{BGPS G014.849-00.992}  & 2307 &                  & 18:21:12.4&-16:30:05 &  0.1 & 2.5 & 4,6,7 & double jet? CH$_3$OH masers \\
\object{GLIMPSE G035.0393-00.4735} & 2429 & G035.04-0.47 & 18:56:58.2&+01:39:34 & 0.04 & 3.4 & 4,7 & \\
\object{NAME G 35.2N}  & 2431 & G035.20-0.74             & 18:58:13.0&+01:40:34 & 3.1 & 2.2 & 1,8 & double jet, CH$_3$OH/OH masers \\
\hline \hline
\end{tabular}

\end{center}
\tablebib{(1)~\citet{lumsden13}; (2)~\citet{urquhart13}; (3)~\citet{faundez}; (4) This work; (5)~\citet{ragan}; 
(6)~\citet{lim}; (7)~Csengeri et al. in prep.; (8)~\citet{zhang09}.
}
\end{scriptsize}
\end{table*}

\section{Observations and data reduction}
\label{observations:sec}

Our observational database on intermediate- and high-mass jets is composed of:
\textit{i)} \emph{ESO/NTT} H$_2$ (2.12\,$\mu$m) and $K_s$ images of the jets;
\textit{ii)} low-resolution NIR spectra of the jets from \emph{ESO/NTT} (to infer
the physical properties of the jets);
\textit{iii)} archival photometric data of their putative driving sources, covering a spectral range from 1\,$\mu$m to 1.1\,mm, 
for the analysis of their spectral energy distribution (SED).


\subsection{Imaging}

Our images are part of a larger H$_2$ survey of massive jets~\citep[see][]{stecklum}. They
were collected at the ESO New Technology Telescope (NTT) with the infrared spectrograph and imaging camera SofI~\citep{moorwood} with a plate scale of
0.288\,$\arcsec$/pix, which provides a 4.9$\arcmin \times$4.9$\arcmin$ field of view (FoV). 
We used a narrow-band filter centred on the H$_2$ line at 2.12\,$\mu$m to detect molecular emission along the flow. 
The observations were obtained by dithering (20$\arcsec$-40$\arcsec$) the telescope to five positions around the target. 
For the faintest targets two or more dithering cycles were adopted. The single frames were then combined in a final mosaic, whose total
exposure time ranges from 1500 to 3600\,s. To remove the continuum emission and detect the H$_2$ line emission,
complementary $K_s$ broad-band images were gathered, adopting the same observational strategy, but with detector integration time (DIT)
values ten times shorter than those of the narrow band filter. Standard dome flat-fields were acquired in both filters.

H$_2$ and continuum images were taken from archival data already published in the literature for two objects, namely \object{IRAS16547-4247}~\citep[from][]{brooks} 
and \object{GLIMPSE G035.0393-00.4735}~\citep[from][]{lee12}. 
For the first object, the data were taken with the infrared imager and spectrograph ISAAC~\citep{ISAAC} at the ESO Very Large Telescope (VLT)
adopting an H$_2$ narrow-band filter and a narrow-band off line continuum filter $K_c$ to remove the continuum from the H$_2$ mosaic. 
For the second target, H$_2$ and $K$ archival data were taken with WFCAM~\citep{WFCAM} at the UK Infrared Telescope (UKIRT).
The details of our imaging observations are provided in Table~\ref{dati:tab} (Columns 2--4).

The data reduction was done using standard \emph{IRAF}\footnote{IRAF (Image Reduction and Analysis Facility) is distributed by the National
Optical Astronomy Observatories, which are operated by AURA, Inc., cooperative agreement with the National Science Foundation.} packages, applying standard procedures 
for sky subtraction, flat-fielding, bad pixel and cosmic ray removal, and image-mosaicking. 
H$_2$ and $K_s$ images of standard stars were acquired to flux calibrate our data. We used K-band stars from 
the Two Micron All Sky Survey~\citep[2MASS,][]{2MASS} for the astrometric calibration of all the images.

\subsection{Spectroscopy}

The NIR low-resolution spectroscopy ($\mathcal R\sim$600, slit width 1$\arcsec$) was acquired with Sofi/NTT during one run (between the 4th
and the 8th of June 2009), using two different grisms (blue, 0.95-1.64\,$\mu$m, and red, 1.53-2.52\,$\mu$m) along the NIR spectrum. 
Each target was observed with the red, or red and blue grisms, with the slit oriented at one or more position angles (P.A.), 
to encompass the entire H$_2$ jet, as well as the driving source. Detailed values are reported in Table~\ref{dati:tab}.
The two slits on \object{NAME G 35.2N} (\textit{hereafter} G\,35.2N) as well as one slit on \object{IRAS13481-6124} (with P.A.=140.7$\degr$) 
are the only observations that were performed with both blue and red grisms.
To perform our spectroscopic measurements, we adopted the usual ABB$'$A$'$ configuration, with a
total integration time of 1200\,s for the red grism and 1800\,s for the blue grism. 
Additional observations of telluric and photometric standard stars were performed to correct for atmospheric and instrumental effects,
and to ensure flux calibration.

\begin{table*}
\begin{scriptsize}
\caption[]{Summary of observations - Imaging and Spectroscopy
    \label{dati:tab}}
\begin{center}
\begin{tabular}{ccccccc}
\hline \hline\\[-5pt]
Object   &  \multicolumn{3}{c}{Imaging}  & \multicolumn{3}{c}{Spectroscopy} \\
         & date (y,m,d) & Band & Integration time (s)   & Wavelength($\mu$m) (grism)&  P.A.($\degr$) & Integration time (s)\\
 \hline\\[-5pt]
\object{[HSL2000] IRS 1} & 2008-Mar-24 & H$_2$, K$_{s}$ & 1500,150 & 1.53-2.52 (red) & 152.5 & 1200\\
\object{IRAS12405-6219}  & 2008-Mar-24 & H$_2$, K$_{s}$ & 1800,200 & 1.53-2.52 (red) & -18.9 & 1200\\
\object{IRAS13481-6124}  & 2009-Jun-06 & H$_2$, K$_{s}$ & 3000,300 & 0.95-2.52 (blue,red) & 27.9 (red),167.9 (red),140.7 (blue+red) & 1800 (blue), 1200 (red)\\
\object{IRAS13484-6100}  & 2008-Mar-24 & H$_2$, K$_{s}$ & 1500,250  & 1.53-2.52 (red) & -17 & 1200\\
\object{IRAS14212-6131}  & 2008-Mar-24 & H$_2$, K$_{s}$ & 1500,250  & 1.53-2.52 (red) & -63.5,55.5  & 1200\\
\object{SSTGLMC G316.7627-00.0115} & 2008-Mar-24 & H$_2$, K$_{s}$ & 1500,250  &1.53-2.52 (red) & 93 & 1200\\
\object{Caswell OH 322.158+00.636} & 2008-Mar-25 & H$_2$, K$_{s}$ &  1500,250  & 1.53-2.52 (red) & 108.3 & 1200\\
\object{IRAS15394-5358}  & 2008-Mar-23 & H$_2$, K$_{s}$ & 1500,250  & 1.53-2.52 (red) & -165.5  & 1200\\
\object{IRAS15450-5431}  & 2008-Mar-23 & H$_2$, K$_{s}$ & 1500,250  & 1.53-2.52 (red) & 171.6,-165.5 & 1200\\
\object{IRAS16122-5047}  & 2008-Mar-23 & H$_2$, K$_{s}$ &  1500,250  &1.53-2.52 (red) & -163.4,-120.3  & 1200\\
\object{IRAS16547-4247}  & $\cdots$ & H$_2$, K$_{c}$$^*$ & $\cdots$ & 1.53-2.52 & 177.6,178.8  & 1200\\
\object{BGPS G014.849-00.992} & 2009-Jun-06 & H$_2$, K$_{s}$ & 1500,250 & 1.53-2.52 (red) & -41.2,-162.7 & 1200\\
\object{GLIMPSE G035.0393-00.4735}  & $\cdots$ & H$_2$, K$^{**}$ &   $\cdots$  &1.53-2.52 (red) & -132.7  & 1200\\
\object{NAME G 35.2N}  & 2009-Jun-08 & H$_2$, K$_{s}$ & 1500,250  & 0.95-2.52 (blue,red) & -158.8,-118 & 1800 (blue), 1200 (red)\\
\hline \hline
\end{tabular}
\end{center}
Notes:$^*$ H$_2$ and K$_{c}$ (narrow-band continuum filter at 2.09\,$\mu$m) ISAAC/VLT images are taken from \citet{brooks}.
$^{**}$ H$_2$ and K WFCAM/UKIRT images are taken from \citet{lee12}.
\end{scriptsize} 
\end{table*}

The data reduction was done using standard \emph{IRAF} tasks.
Each observation was flat fielded, sky subtracted and corrected for the distortion caused by long-slit spectroscopy. 
The atmospheric response was corrected by dividing each spectrum by a telluric standard star, normalised to the blackbody
function at the stellar temperature and corrected for any absorption feature intrinsic to the star.

\subsection{Additional photometry from literature and archival data}

In the context of our investigation, no information is available for five YSOs of our sample (namely \object{G316.762-00.012}, \object{Caswell OH 322.158+00.636}, 
\object{IRAS16122-5047}, \object{BGPS G014.849-00.992} and \object{GLIMPSE G035.0393-00.4735}). 
For these targets, we collected additional photometry from several public surveys to 
construct the SEDs and derive their bolometric luminosities (see also Appendix~\ref{appendixG:sec}, \ref{appendixI:sec}, \ref{appendixK:sec}, \ref{appendixM:sec}, and  \ref{appendixN:sec}, respectively).
The retrieved photometry ranges from 1\,$\mu$m to 1.1\,mm, including data from 2MASS~\citep[$J$, $H$, and $K_s$ bands;][]{2MASS},
the Wide-field Infrared Survey Explorer~\citep[WISE;][]{cutri}, the Galactic Legacy Infrared Mid-Plane Survey
Extraordinaire~\citep[GLIMPSE;][]{benjamin}, AKARI~\citep[][]{akari}, MIPSGAL~\citep[MIPSGAL][]{carey},
MSX Infrared Point Source Catalog~\citep[][]{MSX}, Hi-GAl data~\citep[][]{molinari} from the Herschel data archive,
the ATLASGAL survey~\citep[][]{csengeri}, and the Bolocam Galactic Plane Survey II~\citep[][]{rosolowsky}.

We used Herschel PACS (70\,$\mu$m and 160\,$\mu$m) and SPIRE (250\,$\mu$m, 350\,$\mu$m, and 500\,$\mu$m) data, 
from the bulk reprocessing with SPRG11.1.0, which provided the correct flux calibration for both PACS and SPIRE data. 
Flux densities were estimated by performing aperture photometry on source.
Apertures were set as twice the instrumental full width half maximum (FWHM) at each selected wavelength:
5\farcs7 at 70\,$\mu$m, 12$\arcsec$ at 160\,$\mu$m, 18$\arcsec$ at 250\,$\mu$m, 25$\arcsec$ at 250\,$\mu$m, and 36$\arcsec$ at 500\,$\mu$m.
For the flux calculation, we also subtracted a mean value of the background emission. 
This latter was estimated by measuring the average background emission around each object, avoiding regions contaminated by nearby sources.

\section{Results}
\label{results:sec}

\subsection{H$_2$ imaging}
\label{imaging:sec}

\begin{figure*}
 \centering
   \includegraphics[width=16 cm]{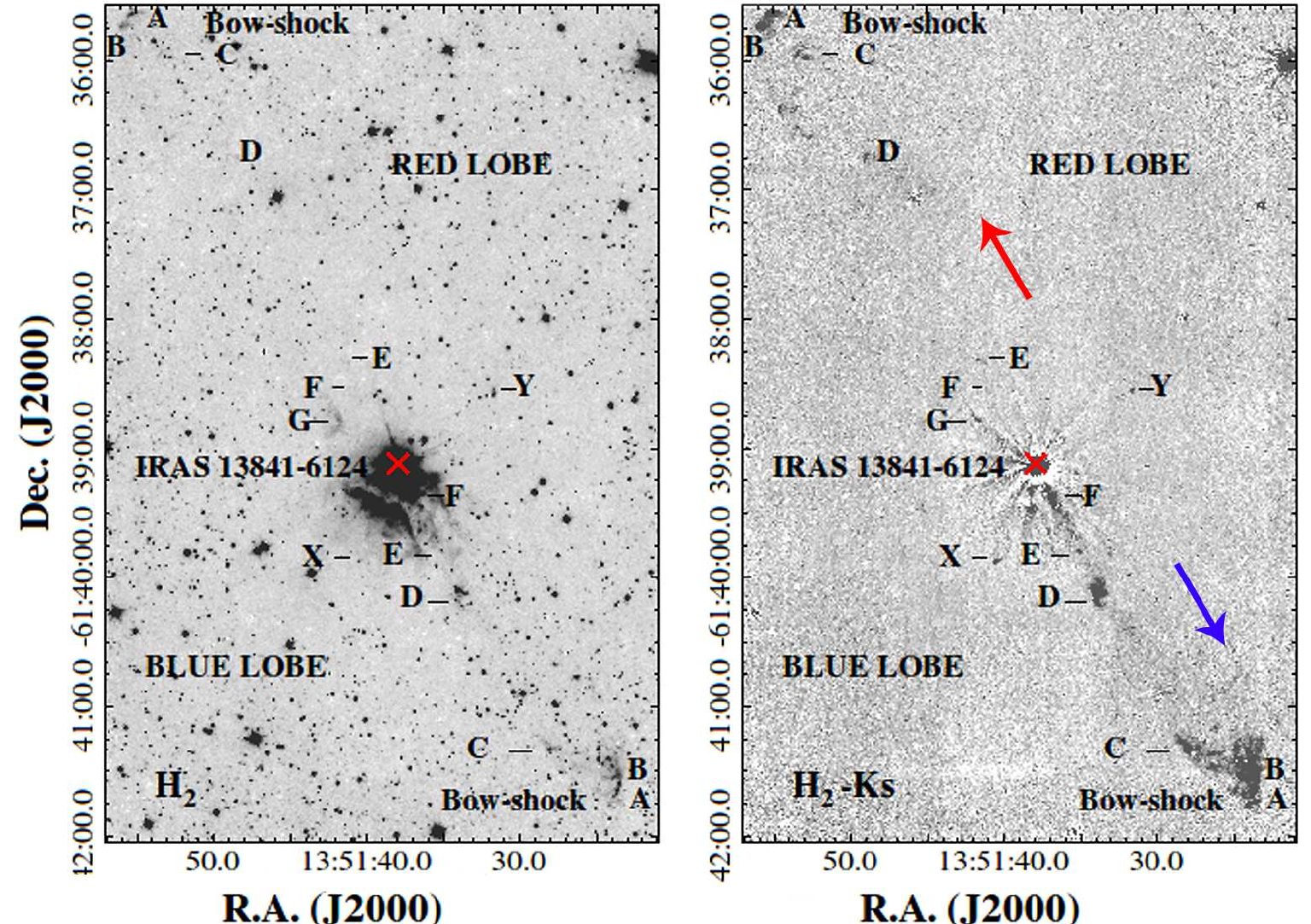}
   \caption{H$_2$ and continuum-subtracted H$_2$ images (\textit{left and right panels}) of the \object{IRAS 13481-6124} jet.
   This jet is one of the few examples of jets with small precession angles and it has asymmetric lobes.
   Source and knots positions are indicated in the figures. Blue and red lobes are also reported according to \citet{kraus10}.
\label{IRAS13481ima:fig}}
\end{figure*}

\begin{figure*}
 \centering
   \includegraphics[width=18 cm]{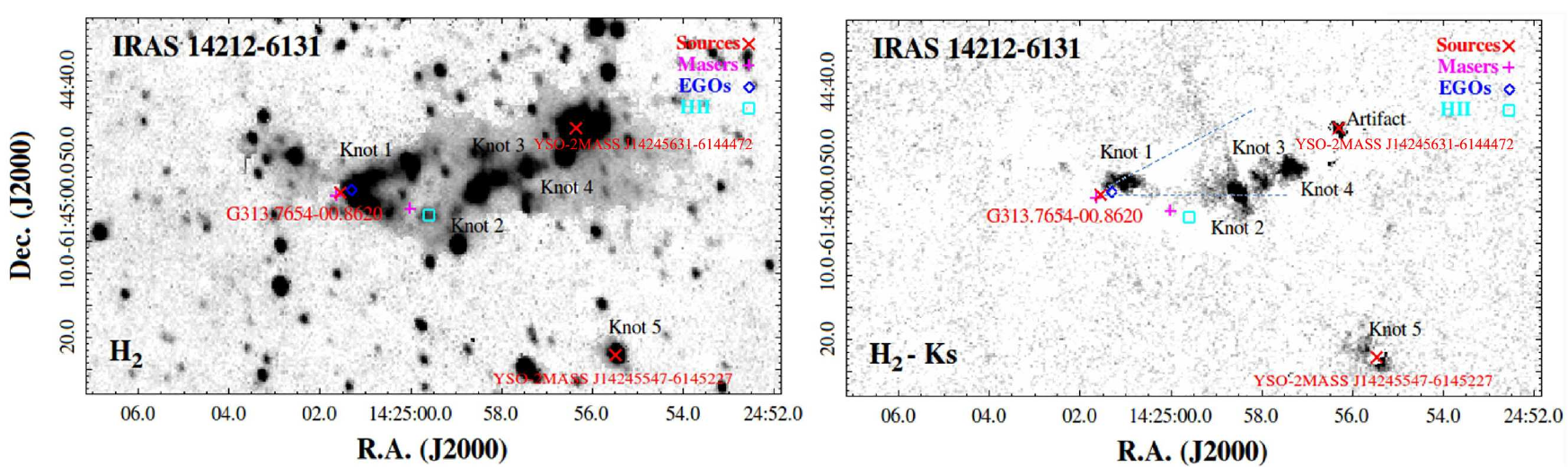}
   \caption{Same as in Fig.~\ref{IRAS13481ima:fig} but for the \object{IRAS14212-6131} flow.
   This object represents a good example of precessing flow with only one detected lobe. The thin dashed lines 
   show the measured precession angle.
   The positions of sources, knots, masers, EGOs, and \ion{H}{ii} regions are indicated in the figures.
   \label{IRAS14212-6131ima:fig}}
\end{figure*}

\begin{figure*}
 \centering
   \includegraphics[width=15 cm]{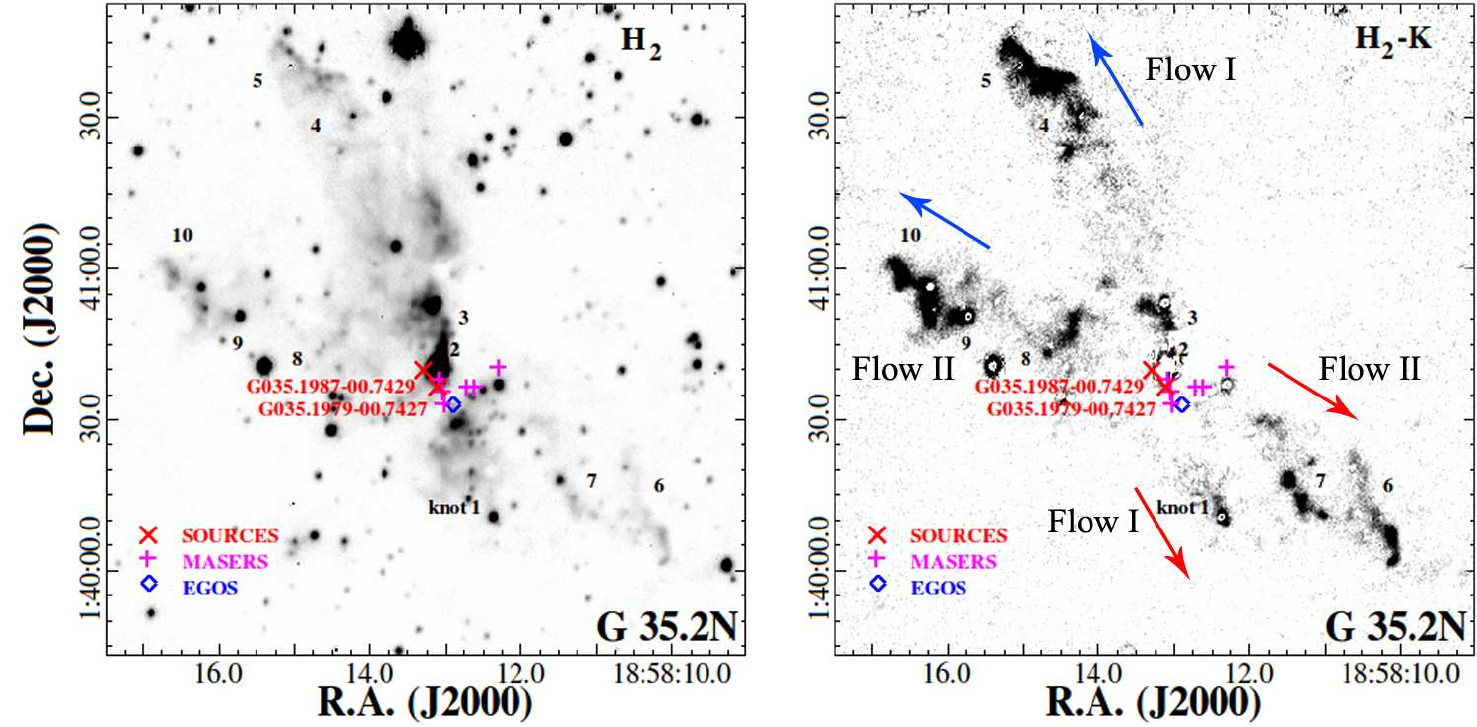} 
   \caption{Same as in Figures~\ref{IRAS13481ima:fig} and \ref{IRAS14212-6131ima:fig} but for the G\,35.2N flows.
   This is one of the best examples of binary jets detected in our sample.
\label{G35ima:fig}}
\end{figure*}

As an example of the complex and different flow morphologies observed in our sample, Figures~\ref{IRAS13481ima:fig}--\ref{G35ima:fig} present
three different examples:
\textit{a)} the \object{IRAS 13484-6124} flow with small precession angle (8$\degr$) and asymmetric lobes (Fig.~\ref{IRAS13481ima:fig});
\textit{b)} the \object{IRAS14212-6131} flow with large precession angle (32$\degr$) and single lobe (Fig.~\ref{IRAS14212-6131ima:fig}); 
\textit{c)} the G\,35.2N flows, one of the best examples of binary precessing jets in the sample (Fig.~\ref{G35ima:fig}).

Left and right panels in the figures show the H$_2$ and continuum-subtracted H$_2$ images of the targets, respectively.
The positions of the sources, detected knots, known masers, EGOs as well as \ion{H}{ii} (HC-, UC- or \ion{H}{ii}) regions reported by SIMBAD
are also labelled in the figures. The images of the remaining jets and a detailed description of the morphology of each object, including the aforementioned targets, 
are given in Sections~\ref{appendixB:sec}--\ref{appendixO:sec} in the Appendix.

Because we do not have any kinematic information on the observed knots, our criterion to associate the H$_2$ emission with a particular 
source relies {\it entirely} on the observed morphology. More precisely, our main criteria are: 
\textit{a)} the presence of bow shocks indicating the position of the driving source; \textit{b)}
the collimation of jets and knots along the flow with respect to the position of the driving source;
\textit{c)} the measured precession angle not exceeding values of 90$\degr$. 

In this paper, we define the current jet position angle (or jet position angle) as the angular offset between the knot closest
to the source and the driving source itself (counterclockwise, north to east). 
In principle, these knots represent the most recent ejecta, therefore they provide the current P.A. of the jet. It is also worth
noting that throughout the paper 
the term precession angle refers to the overall opening angle of the flow, estimated by considering the position
of the driving source and the peak positions of the two most external knots at each side of the jet axis (see dashed lines in Figure~\ref{IRAS14212-6131ima:fig}, right panel).
Strictly speaking, the definition of precession angle should only be applied to wiggling jets, but in some cases we observe jet bending and precession 
(see e.g. Figure~\ref{IRAS16547ima:fig}), jet bending alone (see e.g. Figure~\ref{G332ima:fig}) or even more complex morphologies.
Indeed, in a few objects, the apparent precession might be just a visual effect due to multiple flows.

Often two or more flows are recognised in the observed area, and, in four out of fourteen sources 
(\object{IRAS12405-6219}, \object{IRAS15394-5358}, \object{IRAS16122-5047} and G\,35.2N; Appendix~\ref{appendixC:sec}, \ref{appendixI:sec}, \ref{appendixK:sec}, and \ref{appendixO:sec}), two jets driven by the same source are detected.
Although the sources are not spatially resolved, the detection of two jets indicates the presence of a binary or multiple 
system, as often observed in massive star forming regions~\citep[see e.g.][]{kumar,zinnecker07,varricatt}.
Therefore, the total number of analysed flows in this paper is eighteen.

The H$_2$ flows recognised in our sample typically display large precession angles, up to $\sim$60$\degr$ (see e.g. Fig.~\ref{IRAS14212-6131ima:fig} and
Fig.~\ref{IRAS16547ima:fig}), much larger than those observed in low-mass jets, which are usually well collimated 
or with small precession angles $<$10$\degr$~\citep[see e.g.][]{caratti06,arce07}.
In our sample only four out of eighteen flows have precession angles smaller than 10$\degr$ (see Column\,2 of Table~\ref{phys:tab}; see e.g. Fig.~\ref{IRAS13481ima:fig}
and Fig.~\ref{G316ima:fig}). On the other hand, the jet projected lengths on the sky (Column\,3 of Table~\ref{phys:tab}) range from one tenth up to seven parsecs,
which are values typically observed also in low-mass flows~\citep[e.g. in \object{HH 34} or \object{HH 111}; see e.g.][and references therein]{bally97}. 
H$_2$ bipolar flows are detected in eleven out of the eighteen analysed jets (the number of lobes detected are given in
brackets in Column\,3 of Table~\ref{phys:tab}). 
Seven flows show emission (above a 3$\sigma$ threshold) only from one lobe, as also observed in other surveys of jets from intermediate- 
and high-mass YSOs~\citep[see][]{kumar,varricatt}. This is probably due to the high extinction towards the sources. 
Therefore the undetected lobes are more likely the red-shifted ones, where a higher $A_{\rm V}$ value is expected.
In two out of seven of these targets (i.e. \object{IRAS14212-6131} and \object{GLIMPSE G035.0393-00.4735}) 
there is a marginal detection (below 3$\sigma$) of the second lobe 
(see also Fig.~\ref{IRAS14212-6131ima:fig} and \ref{BS17ima:fig}, and discussion in Appendix~\ref{appendixF:sec} and \ref{appendixN:sec}). 

It is worth noting that the majority of the jets is not easily detectable without subtracting the continuum emission from 
the narrow band images.
Firstly, the extended and bright scattered emission originating from nearby sources or by the driving source itself
may conceal the emission from the jets. Secondly, the observed YSOs are located at low galactic latitudes (i.e. in highly crowded fields),
and at large distances (i.e. the angular extent of the observed H$_2$ knots is often comparable with the measured seeing in our images, $\sim$1$\arcsec$).
Moreover, the morphology of these flows is generally more complex than that of low-mass counterparts, because of their large precession 
angles, and, in some cases, the asymmetric geometry of the two lobes, or the non-detection of one of the lobes.

The jet driven by \object{IRAS 13484-6124} (Figure~\ref{IRAS13481ima:fig}, see also Appendix~\ref{appendixD:sec}) is a clear example 
of a highly collimated bipolar jet in our sample (precession angle $\sim$8$\degr$).
Despite its collimation, the jet is asymmetric, namely the red-lobe is about 1.3 times more extended than the blue lobe. 
A possible explanation might be that the density of the interstellar medium in the blue lobe (SW of the source) is higher than in the red lobe.
As a consequence, blue-shifted knots would have a lower velocity than the red-shifted ones (see also discussion
in Sect.~\ref{discussion2:sec}).

Figure~\ref{IRAS14212-6131ima:fig} provides an example of a highly precessing flow with a single lobe. 
Bright continuum emission (see left panel), depicting the outflow cavity, is detected close to the driving 
source (\object{G313.7654-00.8620}; see also Appendix~\ref{appendixF:sec}), 
along with a more extended continuum, produced by a bright young star (2MASS J14245631-6144472), 
at about 39$\arcsec$ WNW from the source position. 
After subtracting the continuum (Fig.~\ref{IRAS14212-6131ima:fig}, right), one of the two 
precessing lobes (knots 1--4, likely the blue-shifted lobe, precession angle $\sim$32$\degr$) of \object{G313.7654-00.8620} is detected westwards of the source position. 
The second lobe is not clearly detected, likely because of the high extinction towards the target.
More H$_2$ emission, namely Knot 5, is detected towards WSW, likely driven by another YSO 
(i.e. \object{2MASS J14245547-6145227}; see details in Appendix~\ref{appendixF:sec}) marked in Fig.~\ref{IRAS14212-6131ima:fig}.

As illustrated in Figure~\ref{G35ima:fig}, the presence of multiple jets in the same region may also confuse the morphology of the flows.
The observed H$_2$ emission was originally recognised as an hourglass shape with a large opening angle of $\sim$40$\degr$~\citep[][]{lee12},  
interpreted as the outflow cavities of G\,35.2N. On the basis of our H$_2$ imaging and NIR spectroscopic analysis as well as
additional data from the literature, we conclude that this emission belongs to two distinct precessing jets (see Appendix~\ref{appendixO:sec}), 
likely driven by two different HMYSOs, as also suggested by ALMA interferometric observations~\citep[see][]{sanchez}.

\subsection{SEDs of the driving sources}
\label{SED:sec}

\begin{figure}
 \centering
   \includegraphics[width=8.7 cm]{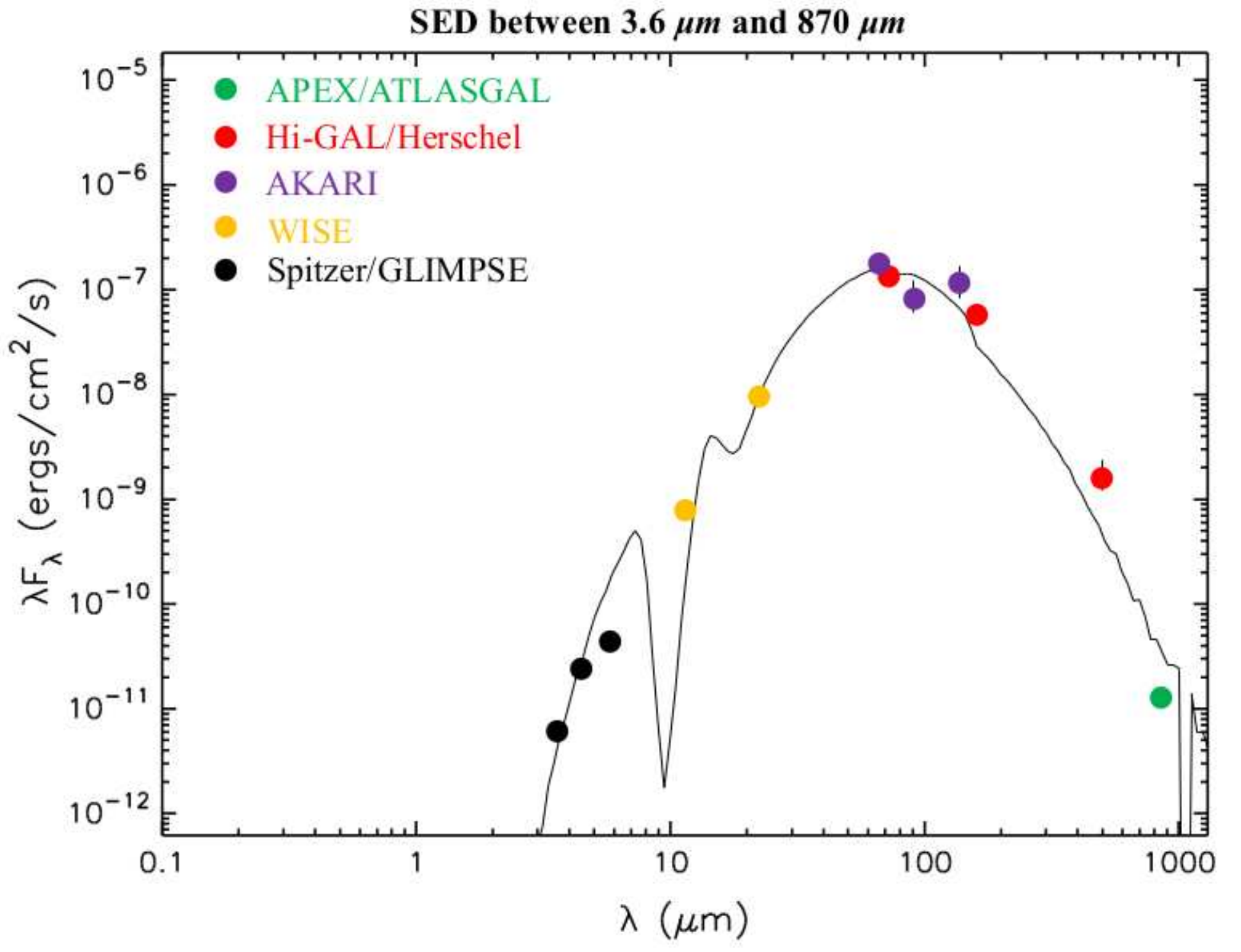} 
   \caption{Spectral energy distribution (SED) of \object{Caswell OH 322.158+00.636} constructed with all photometric data available from the literature,
   namely from 3.6\,$\mu$m to 870\,$\mu$m, and by assuming a distance of 4.2\,kpc to the source~\citep[][]{urquhart13}. 
   The SED was fitted with the radiative transfer model developed by \citet{robitaille07}. The fitting tool is available
   at http://caravan.astro.wisc.edu/protostars/. The solid black line indicates the best-fitting model.
\label{G322SED:fig}}
\end{figure}

From the literature we retrieved estimated distances, SEDs and bolometric luminosities for the majority of the driving sources in our sample (see Table~\ref{sample:tab}).
In five cases, namely \object{G316.762-00.012}, \object{Caswell OH 322.158+00.636}, \object{IRAS16122-5047}, \object{BGPS G014.849-00.992} and
\object{GLIMPSE G035.0393-00.4735}, we construct the SEDs from the collected photometry
to estimate the bolometric luminosity of each source, under the assumption that a single object dominates the luminosity of each region.
The tables listing the available photometry are presented in the Appendix (see Sect.~\ref{objects:sec}).
In each table, we report wavelengths, fluxes and uncertainties as retrieved from the catalogues. No colour correction has been applied to the data.
For each object, we assume the distance reported in Table~\ref{sample:tab} and then
derive its bolometric luminosity by fitting the observed SED  
with the radiative transfer model developed by \citet{robitaille07} (see e.g. Figure~\ref{G322SED:fig}). 
The model assumes a YSO with a circumstellar disc, embedded in an infalling flattened envelope with
outflow cavities. The on-line tool gives the best fit from a large collection of pre-computed model SEDs. 
The fit provides robust estimates of the SED integrated quantities, such as the bolometric luminosity.
Along with the photometric dataset, the employment of this tool requires a range of values for the distance and the foreground extinction. 
For the distance, we adopted the values given in Table~\ref{sample:tab}. For the foreground extinction, we assumed the range of $A_{\rm V}$
values measured along each flow (Table~\ref{phys:tab}).

An example of our SED analysis is presented in Figure~\ref{G322SED:fig}.
The solid black line in Fig.~\ref{G322SED:fig} indicates the best-fitting model for \object{Caswell OH 322.158+00.636}.
Our analysis provides us with an $L_{\rm bol}$ value of $\sim$1.3$\times$10$^5$\,L$_{\sun}$,
roughly corresponding to an O7 ZAMS star of $\sim$30\,M$_{\sun}$. This is the most luminous and massive object in our sample;
a detailed description of this source is given in Appendix~\ref{appendixH:sec}.
The results of the SED analysis and the figures of the four remaining SEDs are given in Appendix~\ref{objects:sec}.

\subsection{Spectroscopy}
\label{spectroscopy:sec}

In Tables \ref{spec_hsl2000:tab}--\ref{spec_G35.2N:tab} (see Appendix, Sect.~\ref{objects:sec}), we report, for each target, the fluxes (uncorrected for the extinction)
of the identified lines together with their vacuum wavelengths. The line fluxes were obtained by fitting the profile with
one, or two Gaussians in case of blending. The associated errors are derived from the RMS of the local baseline multiplied by the line-widths. These are typically
unresolved and therefore comparable with the instrumental profile width. Lines
showing fluxes with a signal-to-noise ratio (S/N) between 2 and 3, as well as those blended, have been labelled. Additional uncertainties
in the fluxes are derived from the absolute calibration ($\approx$ 10\%).

The spectra of the jets are characterised by emission lines from molecular, i.e. H$_2$, and atomic species, mostly
[\ion{Fe}{ii}]. These NIR lines have usually a shock origin and are always observed in protostellar jets~\citep[from both low- and high-mass YSOs;][]{caratti06,caratti08}, 
tracing the jet axis.
Usually H$_2$ or [\ion{Fe}{ii}] lines trace different shock conditions, namely non-dissociative or dissociative shocks. 
The H$_2$ lines are detected in all the observed knots, and they originate from different upper vibrational levels, namely from $v$=1 to $v$=3. 
As expected, the most prominent features are the 1--0\,S(1) at 2.12\,$\mu$m and the 1--0\,Q line series between 2.4 and 2.5\,$\mu$m. 
No H$_2$ line has been detected in the blue part of the spectra (0.95-1.65\,$\mu$m), except in knots 4-5 in G\,35.2N,
where the 2--0\,S(5) transition at 1.08\,$\mu$m is marginally detected (2$<$S/N$<$3). This suggests that the flows are highly reddened.

[\ion{Fe}{ii}] features are detected in 50\% of the flows and in about 30\% of the observed knots (20 out of 65), all showing the bright transition 
at 1.644\,$\mu$m. In some cases (7 out of 20 knots), other fainter lines in the $H$ band, such as the 1.534, 1.600, 1.664, 1.677\,$\mu$m lines, are observed. 
The 1.257\,$\mu$m [\ion{Fe}{ii}] line is detected along the blue-shifted lobe of flow I in G\,35.2N (see Fig.~\ref{G35ima:fig}), 
where also the [\ion{C}{i}] doublet at $\sim$0.98\,$\mu$m is observed (knots 4 and 5). 
This line typically traces the neutral gas beyond the ionisation front~\citep[][]{hollenbach89}, and it has often been detected in shocks
in association with [\ion{Fe}{ii}] emission~\citep[][]{nisini02}.
Notably, we additionally detect two 2P$\rightarrow$4P [\ion{Fe}{ii}] lines in the $K$ band spectral segment of the \object{IRAS14212-6131} source. 
These lines have seldom been observed close to the jet driving sources in the case of low-mass YSOs~\citep[][]{takami,rebeca08}.
They have excitation energies higher than for those in the H band (26\,000 vs. 11\,000\,K), and they are produced in highly dense and excited shocked regions, 
therefore they can be used to investigate the jet/wind-launching regions.

The detection of Br$\gamma$ emission in a few knots (namely knot\,4 in \object{IRAS14212-6131}, knots 4 and 5 in G\,35.2N,
and knot\,1 in \object{IRAS13484-6100}; Appendix~\ref{appendixF:sec}, \ref{appendixO:sec}, and \ref{appendixE:sec}, respectively)
is also noteworthy. The location of these knots is not associated with the position of any HMYSO.
In addition, these features are not spatially extended, 
therefore they are likely produced by shocks rather than emitted by \ion{H}{ii} regions~\citep[e.g.][see also discussion in Sect.~\ref{discussion1:sec}]{rebeca08}.


\begin{figure*}
 \centering
   \includegraphics[width=12 cm]{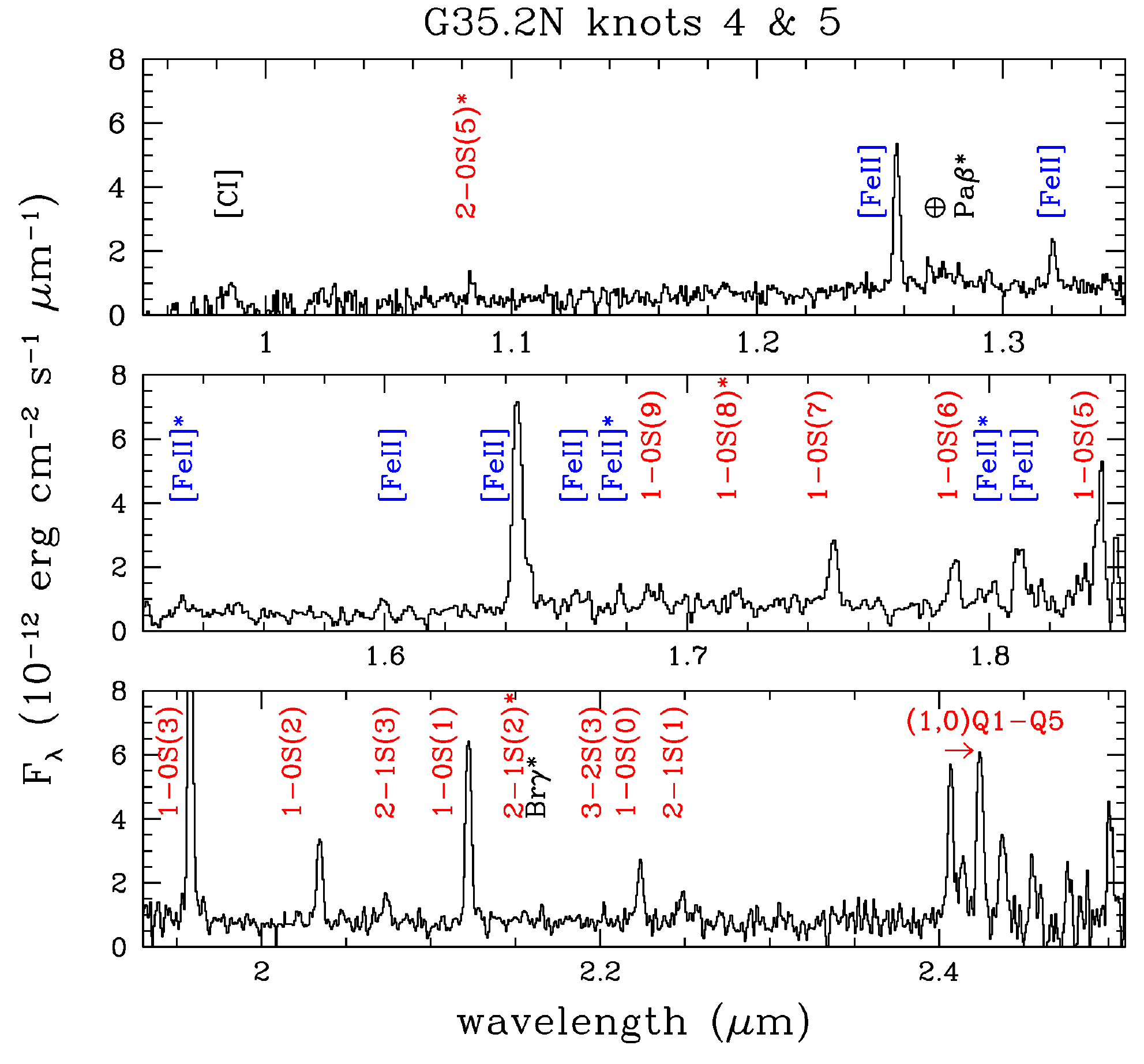}
   \caption{0.95--2.5\,$\mu$m SofI spectrum of knots 4 and 5 in the G35.2N outflows.
   An asterisk near the line identification marks the detections between 2 and 3 sigma. Telluric lines are indicated by the symbol ``$\oplus$''. 
\label{G35.2N_sp:fig}}
\end{figure*}
\begin{figure*}
 \centering
   \includegraphics[width=12 cm]{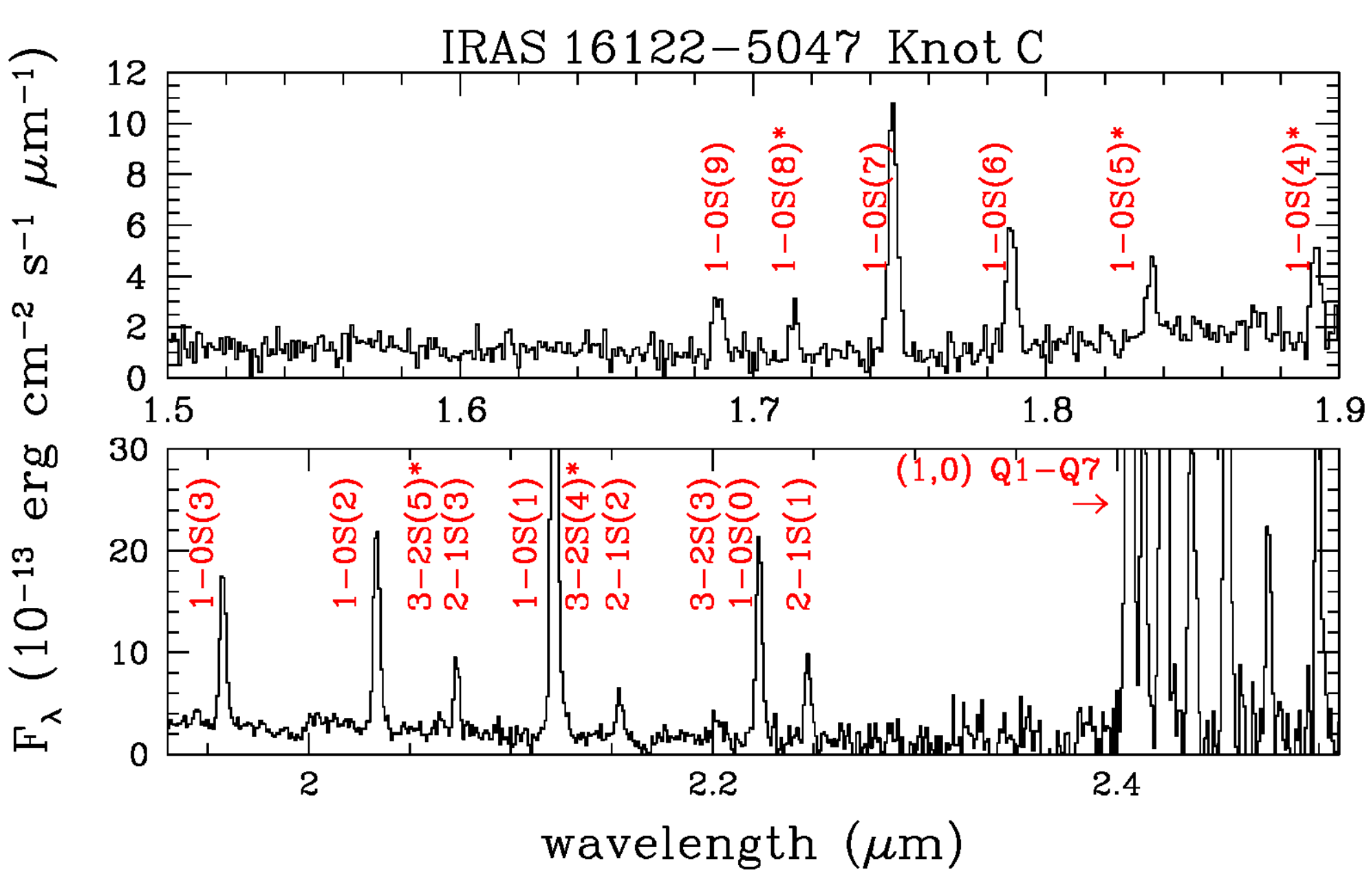}
   \caption{1.5--2.5\,$\mu$m SofI spectrum of Knot C in the \object{IRAS16122-5047} flow. Labels are as in Fig.~\ref{G35.2N_sp:fig}.
\label{IRAS14212_sp:fig}}
\end{figure*}

Our slits also encompass the positions of thirteen out of fourteen driving sources (all targets but \object{Caswell OH 322.158+00.636}).
In most of the cases, the faint NIR continuum from the source comes from emission reflected 
by the outflow cavities or by the circumstellar material. In twelve out of thirteen targets (all but \object{IRAS16547-4247}, 
whose continuum is not detected in our spectra), 
the spectra show steeply rising continua, indicative of highly reddened sources. 
In most of the cases (10 out of 13) the YSOs have also emission lines.
In just one case, i.e. \object{IRAS 15394-5358}, we detect \ion{H}{i} photospheric lines in absorption.
The absence of photospheric lines in the other objects is due to the strong veiling. 
The most prominent emission lines come from shocks along the jet, i.e. H$_2$ (detected in 10 targets) 
and [\ion{Fe}{ii}] (6 out of 10 targets)~\citep[see e.g.][]{nisini02,caratti06}, 
YSO accretion or inner winds, i.e. \ion{H}{i} (6 out of 10 targets)~\citep[see e.g.][]{muzerolle,natta04,caratti12}, or inner disc region, 
namely \ion{Na}{i} and CO (detected in 2 targets)~\citep[see e.g.][]{ilee}. 
These lines are typically observed in HMYSOs~\citep[see e.g.][]{cooper}. Notably, the only fluorescent line, the \ion{Fe}{ii} feature at 1.688\,$\mu$m,
pumped by the Ly\,$\alpha$ line~\citep[][]{lumsden12,caratti13} and frequently detected in the NIR spectra of HMYSOs~\citep[26\% of detection rate in][]{cooper}, 
is observed only in \object{IRAS13481-6124} (see Table~\ref{spec_IRAS13481:tab}), 
clearly one of the least reddened and most evolved HMYSO in our sample. This evidence, along with the high detection rate of jet line tracers on source,
may indicate that, on average, our targets are less evolved than those presented by \citet{cooper}.

\begin{figure*}
 \centering
   \includegraphics[width=12 cm]{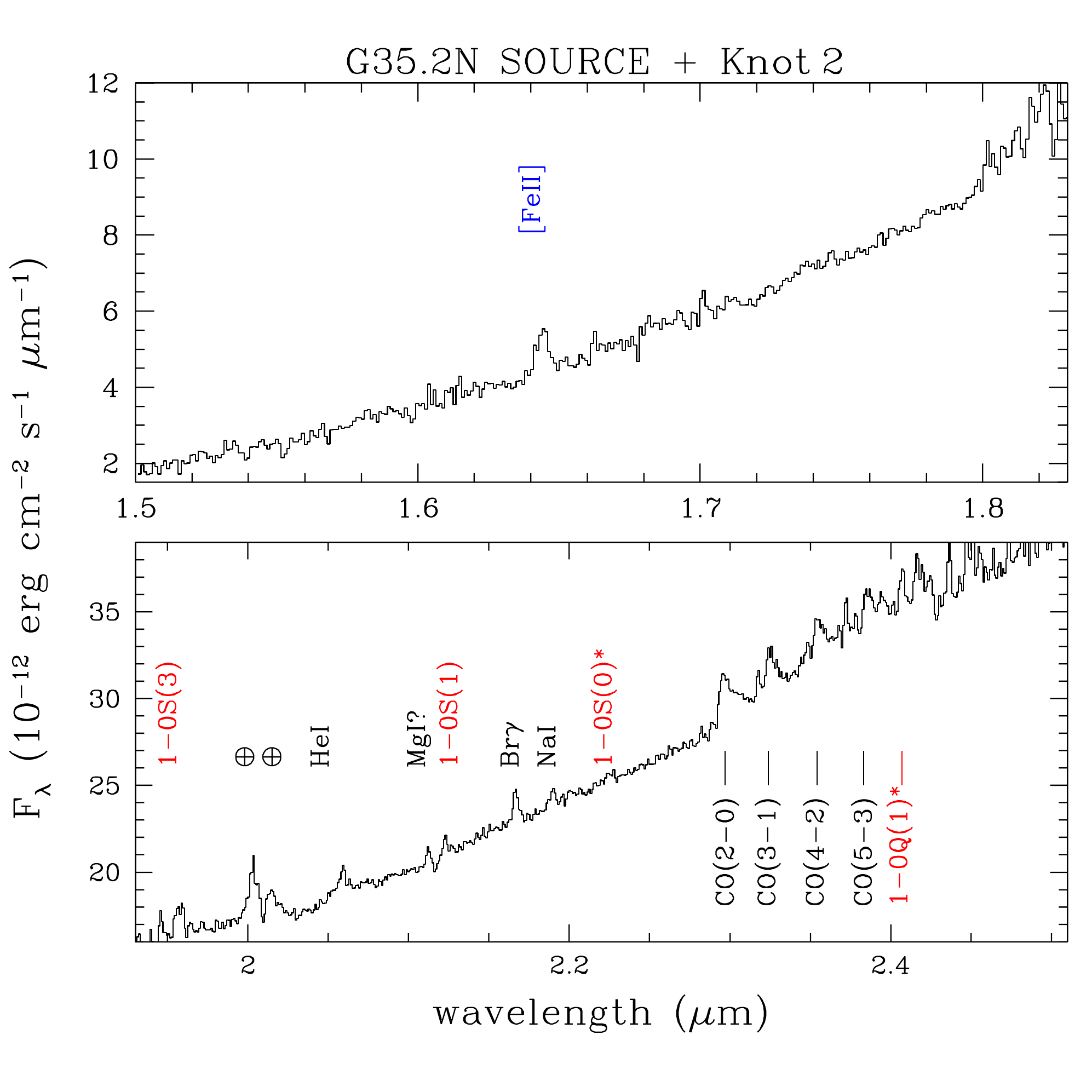}
   \caption{1.5--2.5\,$\mu$m SofI spectrum of G\,35.2N source, which partially includes emission from Knot 2. Labels are as in Fig.~\ref{G35.2N_sp:fig}.
\label{G35.2Nsource_sp:fig}}
\end{figure*}

For the sake of simplicity, we gathered the observed spectra in three groups: \textit{1)} spectra of knots with both ionic and H$_2$ emission;
\textit{2)} spectra of knots with only H$_2$ emission; \textit{3)} YSO spectra.
These groups are summarised in Figure~\ref{G35.2N_sp:fig}, \ref{IRAS14212_sp:fig}, and \ref{G35.2Nsource_sp:fig}, respectively.

Figure~\ref{G35.2N_sp:fig} shows the combined spectrum (0.95--2.5\,$\mu$m) of knots 4 and 5 along the flow I of G\,35.2N, 
positioned NE with respect to the driving source (see also Fig.~\ref{G35ima:fig}). 
The spectrum displays several H$_2$ and [\ion{Fe}{ii}] lines, labelled in red and blue, respectively, as well as faint emission 
from [\ion{C}{i}] and \ion{H}{i}, in black. 
Figure~\ref{IRAS14212_sp:fig} displays the spectrum (1.5--2.5\,$\mu$m) of knot C in the \object{IRAS16122-5047} flow (see also Fig.~\ref{G332ima:fig}). 
The spectrum shows only H$_2$ emission, labelled in red.
Finally, Figure~\ref{G35.2Nsource_sp:fig} gives an example of our YSO spectra, namely 
the 1.5--2.5\,$\mu$m SofI spectrum of G\,35.2N, which partially includes emission from Knot\,2, positioned $\sim$1.5$\arcsec$ NE from the source.
Indeed, in some cases, due to the seeing, it is not possible to disentangle the emission of nearby knots from the YSOs spectra. 
These cases are labelled ``YSO + knot name'' in our tables (see Appendix~\ref{objects:sec}).

\subsubsection{Jet physical parameters from the H$_2$ analysis}
\label{physparam:sec}

The analysis of both H$_2$ and [\ion{Fe}{ii}] lines detected in our spectra 
provides quantitative information about the reddening towards the emitting regions and the excitation conditions
along the jets. In particular, for each knot, we construct H$_2$ ro-vibrational diagrams (Boltzmann plots) to
derive column density, visual extinction, and temperature of the gas.
To achieve it, we apply the same technique as in our previous papers on protostellar jets~\citep[see e.g.][]{caratti06,caratti08}.

\begin{figure}
 \centering
 \includegraphics[width=9.3 cm]{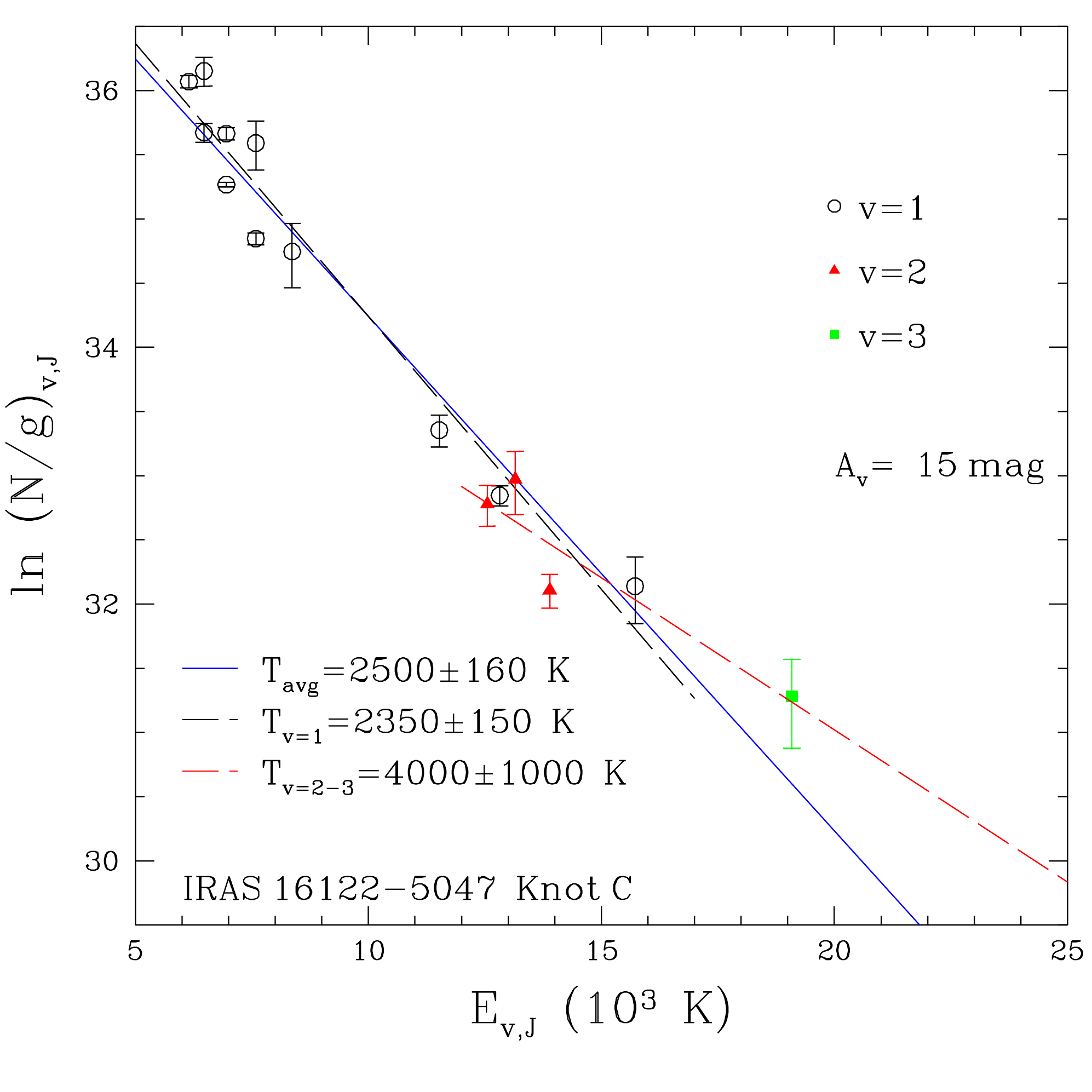}
  \caption{ Ro-vibratational diagram of knot\,C in the \object{IRAS16122-5047} outflow.
   Different symbols indicate lines coming from different vibrational levels, as coded in the upper-right corner of
   the box. The three straight lines represent the best fit through the whole dataset (blue solid line), the v=1 level (black dashed line), 
   and v=2,3 levels (red dashed line), respectively. The corresponding temperatures are also indicated in the lower-left corner of the box.
\label{rot:fig}}
\end{figure}

An example of the ro-vibrational diagram for one of the observed spectra (Knot\,C in the \object{IRAS16122-5047} flow) is presented
in Figure~\ref{rot:fig} (see also the spectrum in Fig.~\ref{IRAS14212_sp:fig}). 
For each observed H$_2$ transition, we plot the natural logarithm of the column density of the molecules ($N_{\rm v,J}$) in the upper energy level
($N_{\rm v,J} = 4\pi I_{\rm v,J} / (A_{\rm v,J} h \nu_{\rm v,J})$, where $I_{\rm v,J}$, $A_{\rm v,J}$, and $\nu_{\rm v,J}$ are
the intensity, Einstein coefficient, and frequency of the transition, respectively) divided by its statistical weight, $g_{\rm v,J}$, 
against its energy level ($E_{\rm v,J}$). To minimise the uncertainties, we only consider unblended lines with S/N$\geq$3.

If the gas is close to the local thermal equilibrium (LTE), then the H$_2$ population follows the Boltzmann distribution 
i.e. $N_{\rm v,J}$/$g_{\rm v,J} \propto \exp(-E_{\rm j}/kT_{\rm ex})$. Data points in the diagram are therefore distributed on a straight line,
and an estimate of the gas excitation temperature ($T_{\rm ex}$) and the average gas column density ($N(H_2)$) can be derived by fitting the data. 
This is the case observed in the majority of the ro-vibrational diagrams of our spectra, where lines with excitation energies lower than
15\,000-16\,000\,K are detected. These are transitions originating from the vibrational levels $v$=1 and $v$=2, usually detected in the $H$ and $K$ bands.
In this case a single temperature, typically between 2000 and 3000\,K, can account for the observed excitation diagrams.

On the other hand, there is not much evidence of temperature stratification in our sample. 
Namely, only four out of 65 knots show transitions $v \geq$3, i.e. 
with excitation energies higher than 16\,000\,K. These high vibrational states are sensitive to higher temperatures,
typically between $\sim$3000 and $\sim$4000\,K~\citep[see e.g.][]{giannini02,giannini04}.
Unfortunately, many of these bright transitions are located in the $J$ band and cannot be detected in highly reddened objects
as in our sample ($A_{\rm V} >$5--10\,mag). Therefore, for these four knots, 
we have to rely on the faint $v$=3 transitions in the $K$ band.
Here, two different temperatures are derived by fitting the $v$=1, and $v$=2 plus $v$=3
vibrational states. Additionally, a single fit through all the lines is obtained, giving a measurement of the average temperature, 
which is then reported in our Tables.

An optimal value of the visual extinction ($A_{\rm V}$) can be inferred by varying the extinction values and maximizing the goodness of each fit.
To deredden the observed line fluxes, we adopt the standard reddening law from \citet{rieke85}.
Because the random scatter of data points about the fit is mainly due to extinction, by varying the extinction values and maximising the goodness of each fit, 
an optimal $A_{\rm V}$ value can be inferred. This method is typically adopted when several lines from the same vibrational level are detected, and, in general,
it is more reliable than using the 1--0\,S(i)/1--0\,Q(i+2) ratios, because the 1--0\,Q(i) transitions lie in a region of the K-band (2.4--2.6\,$\mu$m) with poor atmospheric
transmission~\citep[][]{davis04,giannini04,caratti06}. 

In columns 4, 5, and 6 of Table~\ref{phys:tab} we report the range of average temperatures, 
column densities, and visual extinctions derived in {\it each} flow of the sample. 
The values computed for each individual knot 
are reported in the Appendix~\ref{AppendixA:sec} (Table~\ref{knots:tab}).
Measured temperatures range from 1500 to 3300\,K, with typical values around 2400--2500\,K. The measured visual extinction 
values spread from 1 to 50\,mag, with an average value around 15\,mag. Column density values from this warm component range from 10$^{17}$ to 10$^{20}$\,cm$^{-2}$,
with an average value of $\sim$8$\times$10$^{18}$\,cm$^{-2}$. Notably, the values of these physical parameters are, on average, much
higher than those measured in low-mass jets~\citep[see e.g.][]{caratti06}.

Finally, the analysis of all the ro-vibrational diagrams suggests that the H$_2$ emission from the observed knots
is fully thermalised, i.e. it has a shock-heated origin, independently of the $L_{\rm bol}$ of the YSO. Indeed, fluorescence does not seem to be responsible for
the detected emission along the flows, as suggested by the absence of high vibrational states transitions (namely, v$\ge$6)
and the large values of the 1--0\,S(1)/2--1\,S(1) and 1--0\,S(1)/3--2\,S(3) dereddened ratios~\citep[$\geq$10 and $\geq$100, respectively; see e.g.][]{takami00,davis04,caratti08}.
This result is supported by the absence of fluorescent lines observed in the HYMSOs spectra, except for
\object{IRAS13481-6124} (see Sect.~\ref{spectroscopy:sec}, Appendix~\ref{appendixD:sec}, and Table~\ref{spec_IRAS13481:tab}). 
\citet{stecklum12} report H$_2$ fluorescent emission in the immediate surroundings of this
source (within $\sim$5$\arcsec$), but beyond a few arcseconds from the source, the H$_2$ emission along the jet is strictly thermalised,
as also detected in our SofI spectra, indicating that the circumstellar material around the source screens its FUV emission.
The lack of fluorescent emission along massive jets was already noted in a few previous papers, which analysed NIR spectra from 
individual jets~\citep[see,][]{davis04,gredel06,caratti08}. Nevertheless, this is the first time that such analysis is conducted on a large sample of intermediate-
and high-mass jets.

\begin{table*}
\caption[]{Parameters of the jets of the sample derived from the H$_2$ analysis.
    \label{phys:tab}}
\begin{center}
\begin{tabular}{ccccccc}
\hline \hline\\[-5pt]
Outflow & Precession Angle  & Projected Length$^*$   & $T(H_2)^{**}$ & $N(H_2)^{**}$ & $A_{\rm V}(H_2)^{**}$  &  $L_{H_2}$    \\
        & ($\degr$)  & (pc)  &    (K)      &    (cm$^{-2}$) & (mag)        & (L$_{\sun}$)   \\
\hline\\[-5pt]
\object{[HSL2000] IRS 1}           & 30 & 1.8 (2) & 2400--2500 & (7--8)$\times$10$^{17}$ & 14--15 & 9$\pm$4 \\  
\object{IRAS12405-6219} flow 1  & 14 & 1.6 (1) & 2250--2800 & 2$\times$10$^{19}$ & 35--40 & 12$\pm$4 \\  
\object{IRAS12405-6219} flow 2  & 0 & 0.9 (1) & 2250--2800 & 2$\times$10$^{19}$ & 35--40 & 3$\pm$1 \\ 
\object{IRAS13481-6124}            & 8 & 6.9 (2) & 2220--2940 & 2$\times$10$^{19}$ & 1--11 & 8$\pm$2  \\  
\object{IRAS13484-6100}            & 17 & 0.4 (1) & 2580--2700 & (0.3--1.5)$\times$10$^{19}$  & 15 &  7.2$\pm$4.1 \\
\object{IRAS14212-6131}            & 32 & 1.4 (1) & 2100--2600 & (1--4.5)$\times$10$^{18}$  & 7--12 & 11$\pm$4 \\
\object{SSTGLMC G316.7627-00.0115}           & 0 & 0.5 (2) & 1500--2300 & (0.3--1)$\times$10$^{19}$  & 10--30 & 2$\pm$1 \\
\object{Caswell OH 322.158+00.636} & 19 & 0.2 (1) & 2100--2700 & (0.1--2)$\times$10$^{20}$  & 30--50 & 45$\pm$20 \\
\object{IRAS15394-5358} flow 1  & 42 & 1.7 (2) & 2300--2600 & (0.1--1)$\times$10$^{19}$  & 3--25 & 2.5$\pm$0.8  \\
\object{IRAS15394-5358} flow 2  & 59 & 0.2 (2) & 2500--2600 & (0.9--9)$\times$10$^{18}$  & 20--30 & 0.5$\pm$0.2 \\
\object{IRAS15450-5431}            & 33 & 0.5 (1) & 2100--2800 & (1--1.4)$\times$10$^{18}$  & 5--10 & 2$\pm$1 \\
\object{IRAS16122-5047} flow 1  & 37 & 3.9 (2) & 2150--2900 & (0.7--9)$\times$10$^{18}$  & 8--30 & 2.1$\pm$0.7 \\
\object{IRAS16122-5047} flow 2  & 3 & 5.6 (2) & 2200--2900 & (1--9)$\times$10$^{18}$  & 20--30 &  2.0$\pm$0.7\\
\object{IRAS16547-4247}            & 57 & 3.4 (2) &  1900--2900 & (1--3)$\times$10$^{18}$  & 10--20 & 22$\pm$4 \\
\object{BGPS G014.849-00.992}      & 10 & 0.4 (2) & 1500--2530 & (0.03--1)$\times$10$^{19}$  & 5--30 & 2.1$\pm$0.5 \\
\object{GLIMPSE G035.0393-00.4735} & 22 & 0.74 (2?)  &  1200--2500 & (0.1--5)$\times$10$^{19}$  & 14--30 & 1$\pm$0.4 \\
\object{NAME G 35.2N} flow 1    & 28 & 1.1 (2) &  2500--3300 & (1--3)$\times$10$^{18}$  & 10--20 &  3$\pm$1\\
\object{NAME G 35.2N} flow 2    & 26 & 1.2 (2)  &  2370--3000 & (0.6--3)$\times$10$^{18}$  & 8--10 & 2.0$\pm$0.6  \\
\hline \hline
\end{tabular}
\end{center}
 Notes: $^*$ The projected length value refers to single (1) or both (2) lobes.
$^{**}$ Lowest and highest values of $T$, $N(H_2)$, and $A_{\rm V}$ measured in the
flow are reported.
\end{table*}

\subsubsection{Jet H$_2$ luminosities and masses}
\label{LH2:sec}

Our previous analysis allows us to estimate the H$_2$ luminosity ($L_{H_2}$) from the HMYSO jets.
Given the large number of H$_2$ emission lines in our NIR spectra, we expect that $L_{H_2}$ represents
a significant fraction of the overall energy radiated away during the gas cooling in these jets. Previous studies indicate that 
this is the primary coolant in protostellar jets at NIR wavelengths~\citep[see also][]{caratti06,caratti08}.

To derive $L_{H_2}$, we follow the same method as in \citet{caratti06}. 
First, from our H$_2$ images we identify and associate each knot with its driving source (see Sect.~\ref{imaging:sec} and Sect.~\ref{objects:sec}).
Second, the emitting area and flux of each knot are then evaluated by performing photometry on the flux calibrated continuum-subtracted H$_2$ image.
Aperture photometry is achieved with the task {\em polyphot} in IRAF, after defining each region
within a 3$\sigma$ contour level above the sky background. Identified knots along with their coordinates, 2.12\,$\mu$m fluxes, uncertainties, 
and projected areas in the sky, are reported in the Appendix (Columns 1--6 of Table~\ref{knots:tab}).
Third, the $A_{\rm V}(H_2)$ and $T(H_2)$ values derived from our spectroscopy (Sect.~\ref{physparam:sec} and Table~\ref{knots:tab}) are then used to deredden 
the 2.12\,$\mu$m line flux and to derive the intensities of all H$_2$ lines by applying a radiative code with gas in 
LTE~\citep[see details in][]{caratti06,caratti08}. 

For those knots associated with a particular flow but not encompassed by any slit, 
we adopt an average value of the physical parameters derived from the flow knots.
Then, for each considered knot, our LTE code provides: \textit{a)} the line intensities involving levels with $0 \le v \le 14$ and $0 \le J \le 29$
(E$_{v,J}$ $\le 50\,000\,K)$; \textit{b)} the total H$_2$ intensity; and \textit{c)} the H$_2$ luminosity by assuming the distance to the source 
provided in Table~\ref{sample:tab}.
This process is repeated for all the knots associated with each flow, to estimate $L_{H_2}$ of the flow. 
$L_{H_2}$ values are reported in Column\,7 of Table~\ref{phys:tab}.
The inferred values range from 0.5 to 45\,L$_{\sun}$. This roughly reflects the large $L_{\rm bol}$ range in our sample.
It is worth noting, however, that the inferred $L_{H_2}$ values might be lower limits of the real values, because
our estimate does not take into account part of the emission from the cold H$_2$ component, traced by the pure rotational lines in the mid-IR~\citep[][]{caratti08}, 
that usually have temperatures lower and column densities higher than its NIR counterpart.
Moreover, in some cases, we do not detect any emission from one lobe, likely due to the high extinction.

The final step of our H$_2$ analysis consists in deriving a rough estimate of the mass of the H$_2$ emitting region,
by assuming $M_k = 2 \mu m_H N(H_2)_k A_k$, where $\mu$ is the average atomic weight, $m_{H}$ is the proton mass, $N(H_2)_k$ is the H$_2$ column density, 
and $A_k$ is the area of each given knot (k). The mass of the H$_2$ emitting regions estimated 
for the knots detected in our images are given in the Appendix~\ref{AppendixA:sec} (Column\,10 of Table~\ref{knots:tab}), and range
from 0.01 to 3.7\,M$_{\sun}$. As for the $L_{H_2}$ values, we note that these estimates are derived from the warm H$_2$ component, which has a
column density lower than the cold H$_2$ component~\citep[][]{caratti08}.

\subsubsection{Jet physical parameters from the [\ion{Fe}{ii}] analysis}

For a few knots (5 out of 65), where several [\ion{Fe}{ii}] lines (with $S/N\geq$3) have been observed, $A_{\rm V}$ 
and the electron density ($n_{\rm e}$) can also be inferred \citep[see e.g.][]{nisini02}.
To estimate $A_{\rm V}$ (knots 3 and 4-5 of G\,35.2N), we use the 1.257 and 1.644\,$\mu$m lines, that originate from the same upper level.
Their observed ratios depend only on the differential extinction. Their theoretical intensities are derived from the frequencies and Einstein coefficients
of the transitions, 
taken from \citet{nussbaumer88}. As for the H$_2$ analysis, we adopt the \citet{rieke85} extinction law to correct for the differential extinction and compute $A_{\rm V}$.
The values derived from the [\ion{Fe}{ii}] and H$_2$ lines agree within the uncertainties (see Table~\ref{phys_FeII:tab} and \ref{knots:tab}).

To infer the electron density, we use four different line ratios from the brightest transitions in the $H$ band, which are sensitive to density variations,
namely 1.644\,$\mu$m/1.533\,$\mu$m, 1.644\,$\mu$m/1.600\,$\mu$m, 1.644\,$\mu$m/1.677\,$\mu$m~\citep[][]{nisini02,takami,caratti13}.
For each ratio an estimate of $n_{\rm e}$ is derived and a weighted mean of the various estimates is reported in Column\,3 of Table~\ref{phys_FeII:tab}.

\begin{table}
\caption[]{ Parameters of the jets of the sample derived from the [\ion{Fe}{ii}] line analysis.
    \label{phys_FeII:tab}}
\begin{center}
\begin{tabular}{cccc}
\hline \hline\\[-5pt]
Outflow & Knot & $n_{\rm e}$ & $A_{\rm V}$    \\
        &     &   (cm$^{-3}$) & (mag)      \\
\hline\\[-5pt]
\object{IRAS13481-6124} & B-S$^*$ A Red  & (6$\pm$3)$\times$10$^4$ & $\cdots$\\  
\object{IRAS13484-6100} & knot\,1     & (4$\pm$1)$\times$10$^4$ & $\cdots$  \\
\object{IRAS14212-6131} & knot\,1     & (1$\pm$0.5)$\times$10$^4$ & $\cdots$  \\
\object{NAME G 35.2N}   & knot\,3     & (7$\pm$4)$\times$10$^3$&  24$\pm$4 \\
\object{NAME G 35.2N}   & knots\,4 + 5  & (4$\pm$2)$\times$10$^3$ &  12.5$\pm$2.5 \\
\hline \hline
\end{tabular}
\end{center}
Notes: $^*$ B-S=Bow-shock.
\end{table}

\section{Discussion}
\label{discussion:sec}
First of all, it is worth noting that our sample is composed of YSOs, whose $L_{\rm bol}$ represent late B to late O ZAMS spectral types, with
theoretical stellar masses ($M_*$) ranging from $\sim$6\,M$_{\sun}$ to 25--30\,M$_{\sun}$. 
As described in Sect.~\ref{sample:sec}, these targets were selected on the basis of EGO and H$_2$ emissions, detected in previous surveys.
Therefore our sample is biased towards intermediate and high-mass YSOs with bright jets and flows, and it should not be considered representative of
all intermediate and high-mass YSOs.

\subsection{The shocked nature of the massive H$_2$ jets}
\label{discussion1:sec}

Our spectroscopic survey provides an answer to the important question about the nature of the H$_2$ emission from HMYSOs.
In the sample analysed, the H$_2$ emission clearly originates from shocks at high temperatures (2500--3000\,K) 
and with high column densities (10$^{18}$-10$^{20}$\,cm$^{-2}$), as suggested by our ro-vibrational analysis (see e.g. Fig.~\ref{rot:fig}). 
Our spectroscopic analysis indicates that H$_2$ is the major coolant of these flows at NIR wavelengths.
Most of such emission, 70\% of the analysed knots, is likely produced by non-dissociative C-type shocks, because of the high medium density
and because no ionic emission is detected~\citep[][]{draine80,giannini04}. We might speculate
that in some cases the high visual extinction ($A_{\rm V}>$25--30\,mag) could prevent us from detecting the [\ion{Fe}{ii}] emission line at 1.64\,$\mu$m,
which is the brightest [\ion{Fe}{ii}] line at NIR wavelengths. Nevertheless, it is worth noting that [\ion{Fe}{ii}] is detected in the most reddened knot 
of our sample (Knot\,1 of Caswell OH 322.158+00.636, $A_{\rm V}=50$\,mag; see Table~\ref{knots:tab} and \ref{spec_G322:tab}).

On the other hand, 30\% of the knots show both atomic and molecular emissions, indicating the presence of 
J-type dissociative shocks~\citep[see][]{hollenbach89,gredel,nisini02}. Remarkably, the majority of these knots originate from the most luminous (and likely most powerful)
YSOs in our sample.
In this case, fast shocks dissociate the molecules  and destroy the grains along the flow, ionising the medium and producing the observed [\ion{Fe}{ii}] emission. 
Besides, the observed H$_2$ emission in these knots arises from oblique shocks, where the shock velocities are lower (e.g. wings of the bow-shocks, as in the red-lobe bow-shocks A and B 
of IRAS\,13481-6124), or from reverse shocks, where the molecules are re-forming in the cooling post-shock regions behind the 
dissociative shocks~\citep[see e.g.][]{hollenbach89,nisini02,maccoey04}. Although atomic and molecular emissions are physically detached,
in most of our spectral images both emitting regions are spatially coincident, 
because of the SofI low spatial resolution and because of the large distances involved. 

Additionally, in three out of 65 knots, namely Knot 1 in IRAS 13484-6100, Knots 4+5 in G35.2N, and Knot\,4 in IRAS 14212-6131 
(see Table~\ref{spec_IRAS13484:tab}, \ref{spec_G35.2N:tab}, and \ref{spec_IRAS14212:tab}, respectively),
Br$\gamma$ emission is also detected. Such emission is not associated with any YSO, as in other emanations, and it is rarely detected 
in knots~\citep[e.g.][]{rebeca08}. 
It may arise from photoionisation~\citep[][]{takami00} or from the most extreme shock conditions, as in 
strong J-type shocks, with high shock velocities ($v_s \geq$60\,km\,s$^{-1}$) and high pre-shock densities ($n_H \geq$10${^5}$\,cm$^{-3}$),
which form a UV precursor that dissociate or ionise the pre-shocked gas~\citep[see e.g.][]{hollenbach89,maccoey04}.
The observed emissions are compact (1--2\,arcsec$^2$), so they are more likely originating from a shock. In such a case,
the expected Br$\gamma$ flux can be estimated as a function of $v_s$ and $n_H$~\citep[][]{burton89,fernandes97}:

\begin{equation}
\label{FBrG:eq}
F(Br\gamma)=10^{-14}(\frac{n_H}{10^{5} cm^{-3}})(\frac{v_s}{100 km s^{-1}})\,\,erg s^{-1} cm^{-2} arcsec^{-2}.
\end{equation}

The Br$\gamma$ line fluxes from Knot\,1-IRAS13484-6100 and Knots 4+5-G35.2N, and Knot\,4 in IRAS 14212-6131, corrected for visual extinction (see Table~\ref{knots:tab}), 
are 1.8$\pm$0.5$\times$10$^{-14}$ (emitting area $\sim$2\,arcsec$^{2}$), 5.4$\pm$1.9$\times$10$^{-15}$\,erg\,s$^{-1}$\,cm$^{-2}$ (emitting area $\sim$1\,arcsec$^{2}$), 
and 6.2$\pm$1.2$\times$10$^{-15}$\,erg\,s$^{-1}$\,cm$^{-2}$ (emitting area $\sim$1\,arcsec$^{2}$), respectively.
Such values are compatible with J-type shocks with $v_s \sim$90\,km\,s$^{-1}$ (Knot\,1-IRAS13484-6100), and $v_s \sim$60\,km\,s$^{-1}$ (Knots 4+5-G35.2N
and Knot\,4-IRAS 14212-6131), moving in a pre-shocked medium of $n_H$$\sim$10${^5}$\,cm$^{-3}$. 

Combining imaging and spectral analysis, we note that most of the H$_2$ emission is detected along the flows, tracing the direction of the jet, whereas some emission, 
observed nearby or on source, likely traces possible outflow cavities. In conclusion, the observed shocked emissions originate from different mechanisms, 
namely oblique shocks from non-collimated winds (emission from the outflow cavities nearby the source) and magnetically collimated flows (along the jet).
Conversely, the atomic emission is always detected along the jet axis. The observed flows move in a dense and highly 
inhomogeneous medium, as suggested by the range of observed column densities (10$^{17}$--10$^{20}$\,cm$^{-2}$).
Our analysis indicates that, on average, both visual extinction and gas column density are highest towards the source position and decrease along the flow.

Finally, it is worth asking why no relevant fluorescent H$_2$ emission is detected.
Our previous reasoning demonstrates that the absence of strong UV radiation fields towards the observed jets is the main reason.
Indeed, the HMYSOs in the sample are relatively young and they mostly lack of strong PDRs or \ion{H}{ii} regions in the surroundings.
Although some HMYSOs do possess HC- or UC-\ion{H}{ii} regions, the high visual extinction possibly screens the UV emission from the central
source. Alternatively, in some cases the absence of UV emission might be also abscribed to high accretion 
rates~\citep[$>$10$^{-3}$\,M$_\sun$\,yr$^{-1}$; see e.g.][]{hosokawa}, which inflate the stellar radii up to several hundreds R$_\sun$,
producing lower effective temperatures and stellar UV luminosities.

\subsection{The morphology of the massive jets}
\label{discussion2:sec}

The morphology of the detected flows appear to be quite varied, ranging from a simple straight bipolar geometry to a more complex one, 
as monopolar, highly precessing and with patchy structures. 
Such a variety has been detected in many other surveys of H$_2$ massive flows~\citep[see][]{kumar,stecklum,varricatt,lee13}, as
well as in many low-mass protostellar jets~\citep[see e.g.][]{reipurth01}. As a matter of fact, 
many low-mass YSOs are not born as single, isolated stars, nor they form in a relatively still and homogeneous environment.
As a consequence, straight symmetric jets from low-mass YSOs, such as HH\,212~\citep[][]{zinnecker98}, are not the most common. 
Indeed, several factors play a fundamental role in shaping the jet/flow morphology:
the environment and the dynamical interactions among YSOs; the high medium density and its inhomogeneities; the 
large extinction values; the different evolutionary stages of the YSOs; the presence of multiple driving sources~\citep[see e.g.][]{jochen00_h2,peterson}. 
These elements are largely present in high-mass star forming regions, which have extremely complex, dynamic, 
and inhomogeneous environments~\citep[see e.g.][]{zinnecker07}.
In the following, we analyse each mechanism in the context of our observations.

{\it i) Jet precession}.

HMYSOs preferentially form in clusters or small associations of YSOs. Therefore N-body interactions are likely a 
fundamental ingredient in the evolution of these objects as well as in shaping their flows.
Changes in the flow orientation may be induced by anisotropic accretion from cores or envelopes, or by disc tilting caused by tidal interaction with close 
companions~\citep[][]{bally05}. This is the case for almost all of our jets (with the possible exception of \object{G316.762-00.012}; 
Appendix~\ref{appendixG:sec}),
which show large precession angles indicating the presence of multiple systems (see e.g. Figures~\ref{IRAS14212-6131ima:fig}, \ref{G35ima:fig}, \ref{HSL2000ima:fig}
and \ref{IRAS16547ima:fig}). Notably, the large precession angles observed in our sample will likely produce poorly collimated outflows, as often
observed in HMYSOs. It is worth to emphasise, however, that the mere fact of these jets 
having large precession, bending or showing abrupt changes in their direction does not mean that, {\it per-se}, they are not well collimated, because 
jets are magneto-centrifugally collimated at their base~\citep[see e.g.][]{pudritz_PPV,pudritz12}.
Jets in our sample are well collimated, even if they have large precession angles and, sometimes, a twisted geometry.

{\it ii) Jet multiplicity}. 

In addition to jet precession, jet multiplicity also plays a role in confusing the flow morphology.
In four out of fourteen sources of the sample (i.e. IRAS 12405-6219, IRAS 15394-5338, IRAS16122-5047, and G35.2N; see Figures~\ref{IRAS12405ima:fig},
\ref{G326ima:fig}, \ref{G332ima:fig}, and \ref{G35ima:fig}), or even in five if we include \object{BGPS G014.849-00.992} whose jet multiplicity is uncertain 
(see Sect.~\ref{appendixM:sec} and Fig.~\ref{BS2ima:fig}),
double or multiple jets have been clearly detected. Such detections allow us to infer the presence of multiple systems in spatially unresolved sources.

{\it iii) Jet asymmetry}.

All the observed bipolar jets display some degree of asymmetry in the two lobes, the difference in the lengths of the lobes being the most common.
Pairs of ejected knots are not equally spaced from the source (see e.g. Figures~\ref{IRAS13481ima:fig}, \ref{G35ima:fig}, and \ref{IRAS16547ima:fig}). 
It is worth noting that a highly asymmetric jet will likely produce a highly asymmetric outflow.
These asymmetries are commonly observed in protostellar jets from YSOs with different masses and at different
evolutionary stages~\citep[see e.g.][]{podio11,caratti13,ellerbroek}.
The cause of such asymmetry is still debated, and it can be intrinsic or extrinsic to the source~\citep[see e.g.][]{ferreira06,matsakos}.
In the former case, the bipolar jet is launched at different velocities, possibly due to the asymmetric disc structure 
or magnetic field configuration, whereas, in the latter case, the different medium densities produce different velocities.

{\it iv) Monopolar jets/flows}.

H$_2$ bipolar emission is detected above a 3$\sigma$ threshold in eleven out of eighteen flows, whereas the remaining 7 flows are
apparently monopolar (see Column\,3 of Table~\ref{phys:tab}). However, in two out of seven cases (flows from IRAS 14212-6131 and GLIMPSE G035.0393-00.4735; 
see right panels of Fig.~\ref{IRAS14212-6131ima:fig} and ~\ref{BS17ima:fig}) a marginal
detection (between 2 and 3$\sigma$) of H$_2$ emission from the second lobe is observed. This suggests that the higher visual extinction 
towards one of the lobes can hamper our observations in the NIR regime. 
Nevertheless, the presence of monopolar outflows from high-mass~\citep[][]{zapata,fernandez-lopez} as well as low-mass YSOs~\citep[][]{codella14} 
has been observed in a few sources through submm outflow tracers, such as SiO and CO. The authors propose that 
such a monopolar geometry is due to anisotropic ambient cloud conditions or to an intrinsic asymmetry in the flow.
In addition, the presence of close companion(s) might divert or even inhibit the ejection~\citep[see also][and discussion therein]{reipurth00,murphy}.
Unfortunately, we do not have any (sub)mm observations of theses flows to verify any of these scenarios.
However, the Spitzer/GLIMPSE images show
emission in excess at 4.5\,$\mu$m located 
at the position of the undetected lobe in four out of seven flows (namely IRAS 13484-6100, IRAS 14212-6131, IRAS 15450-5431, GLIMPSE G035.0393-00.4735).
Although the nature of such excess is controversial~\citep[e.g.][]{debuizer10,takami12}, it has been often associated with outflows~\citep[][]{cyganowski,tappe}.

Finally, it is worth noting that the observed flows have both extended jet emission and
knotty or bow-shock structures, as those detected in low-mass jets.
By assuming that these flows originate from discs and that ejection and accretion
are tightly related, our observations would imply that the mass accretion rate in HMYSOs is nearly continuous with intermittent accretion outbursts,
which might be identified with the observed knots and bow-shocks along the flow~\citep[see e.g.][]{pudritz_PPV}. 
Such accretion bursts have been observed in low-mass YSOs during almost all
their evolutionary stages~\citep[][]{caratti11,auard}, and it might be related to disc instabilities and/or dynamical interactions~\citep[see e.g.][]{vorobyov}.
Furthermore, the detection of collimated jets indicates the presence of magnetic fields linked to the star/disc system, 
which collimates and accelerates the flows~\citep[see e.g.][]{pudritz_PPV,pudritz12}.

\subsection{H$_2$ jets vs. sources}
\label{discussion4:sec}

In this section we compare the physical properties derived from the flows with those from their driving sources.
Several works discuss the interplay between the evolutionary properties of protostars and the strength 
of their associated outflows in both low- and high-mass YSOs~\citep[][]{bontemps,beuther,duarte}.
In particular, \citet{beuther} and \citet{duarte} derive a straightforward correlation between the momentum flux of the CO outflows ($F_{CO}$) and the 
bolometric luminosities of HMYSOs ($F_{CO} \propto L_{\rm bol}^{0.6}$) as well as a clear relation between 
the envelope mass and the $L_{\rm bol}$ ($M_{\rm env} \propto L_{\rm bol}^{0.6}$). 
If the source is accretion dominated ($L_{\rm bol} = L_{\rm acc} + L_{*} \sim L_{\rm acc}$)
then the $L_{\rm bol}$ value is indicative of both accretion rate and mass of the source (being $L_{\rm acc} \sim G M_* \dot{M}_{acc}/R_*$).
Therefore, the first relationship suggests that mass loss is mainly driven by the accretion power, which grows as the mass of the 
central source increases. The second correlation provides us with a relationship between the mass accretion rate and the YSO evolution. 
$M_{\rm env}$ is an age indicator and it is expected to decrease as the source accretes matter. 
By assuming that the outflow rates are somehow proportional to the accretion rates, the large $F_{CO}$ values found in HMYSOs 
indicate that their mass accretion rates must also be higher than those found in low-mass YSOs.

Additionally, \citet{caratti06} evinced a tight relation between the 
measured molecular hydrogen flow luminosity ($L_{H_2}$) and the bolometric luminosities of embedded
low-mass YSOs ($L(H_2) \propto L_{\rm bol}^{0.6}$), subsequently extended to a few HMYSOs~\citep[][]{caratti08}.
This correlation holds for very young YSOs (Class\,0 and some Class\,I), where the bolometric luminosity is mostly coincident with the 
accretion luminosity of the object and implies that $\dot{M}_{acc}$ increases with the luminosity (i.e. mass) of the protostar.
Observing several Class I outliers, which have less powerful H$_2$ jets, the authors also show that the jet power decreases as low-mass YSOs evolve.
 
\begin{figure}
 \centering
 \includegraphics[width=9.2 cm]{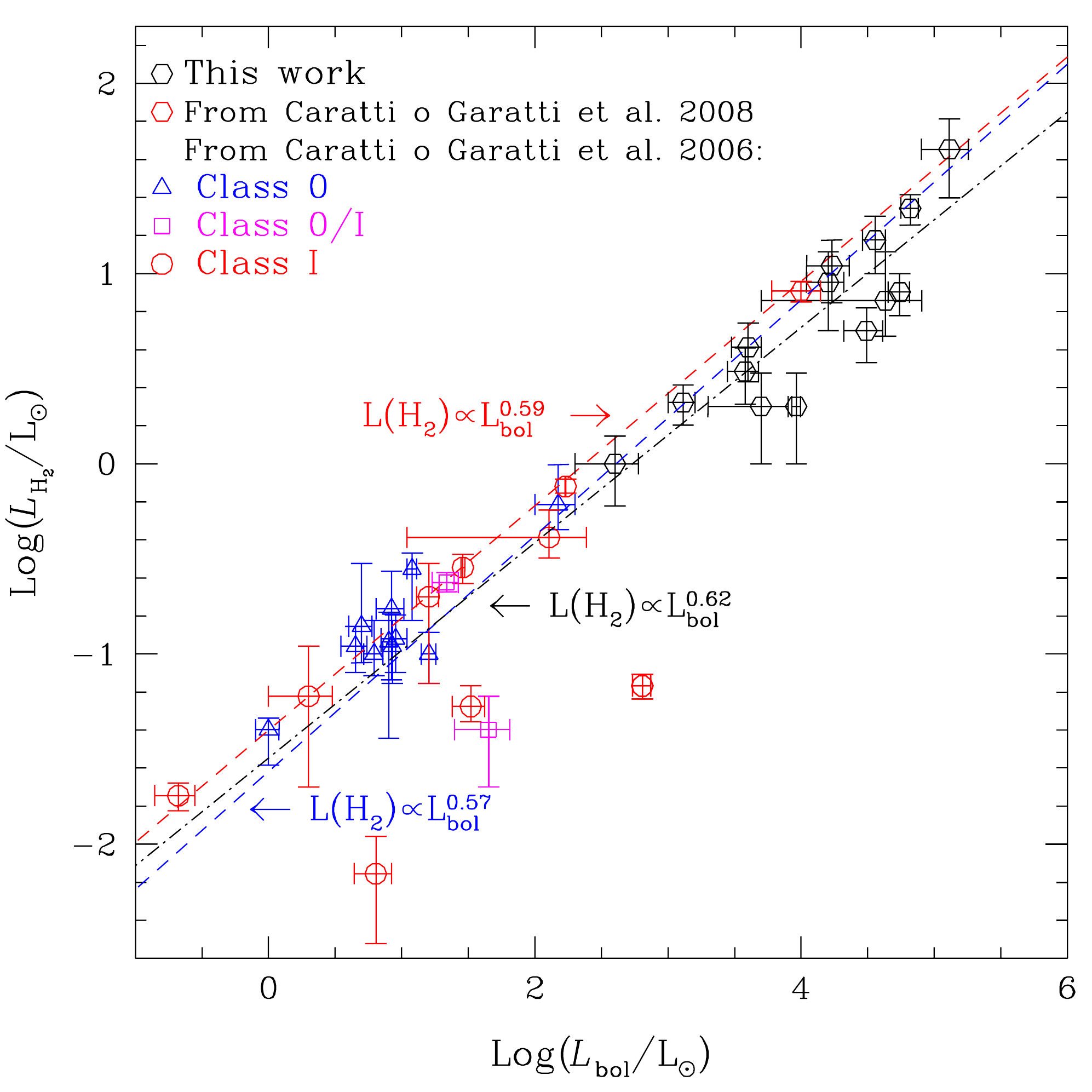}
  \caption{$Log (L_{H_2})$ vs. $Log (L_{\rm bol})$. Results from the current survey (black hexagons) are combined to
  those obtained from previous works~\citep[][]{caratti06,caratti08} as labelled in the upper left corner.
  The red dashed line indicates the previous fit from \citet{caratti06}, which includes only low-mass jets and excludes the four low-mass outliers.
  The blue dashed line shows the best linear fit resulting from the whole sample of HMYSOs. 
  The black dash-dotted line shows the best linear fit resulting from HMYSOs, excluding the five high-mass outliers.
\label{LH2:fig}}
\end{figure}

Indeed, the large H$_2$ luminosity observed in the flows from HMYSOs suggests that high-mass protostars power 
the outflows at a significantly increased accretion rate, as also inferred from their higher $F_{CO}$ values.
This trend is confirmed in our large sample.
In Figure~\ref{LH2:fig} we plot the logarithmic values of flow $L_{H_2}$ vs. source $L_{\rm bol}$, combining the new results of our survey (black hexagons)
with those previously obtained from the HMYSO IRAS\,20126+4104~\citep[red hexagon;][]{caratti08} as well as from low-mass 
YSOs~\citep[red circles, blue triangles, and violet squares indicate Class 0, 0/I and I low-mass YSOs, respectively; from][]{caratti06}.
In the case of multiple flows from unresolved sources, we consider the total $L_{\rm bol}$ of the unresolved cores and the total $L_{H_2}$ from the flows.
The red dashed line indicates the previous fit from \citet{caratti06} with $L(H_2) \propto L_{\rm bol}^{0.6}$, where four low-mass 
outliers were excluded (see Fig.~~\ref{LH2:fig}). The blue dashed line shows the best linear fit resulting from our HMYSO 
sample ($L(H_2) \propto L_{\rm bol}^{0.57}$).

Notably the majority of the analysed HMYSOs (10 out of 15, also including IRAS\,20126+4104 from \citet{caratti08}) agrees well with the relationship
found in low-mass YSOs (red dashed line), whereas five objects are clearly positioned below. This produces a slight shift in the HMYSO fit (blue dashed line).
Interestingly the five outliers (IRAS\,13481-6124, IRAS 14212-6131, IRAS 15450-5431, SSTGLMC G316.7627-00.0115, and G35.2N) are among the most evolved objects in the sample,
and they possess an UC\ion{H}{ii} region. We might therefore suppose that their luminosity is no longer accretion dominated~\citep[see e.g.][]{smith14}.
By excluding these five outliers, we get a better agreement (black dash-dotted line) between the relationships found from low and high-mass YSOs.

Most importantly, the inferred relationship ($L(H_2) \propto L_{\rm bol}^{0.6}$) agrees well with the correlation between the momentum flux of the CO outflows 
($F_{CO}$) and the bolometric luminosities of HMYSOs ($F_{CO} \propto L_{\rm bol}^{0.6}$) from \citet{beuther}. 
This indicates that outflows from HMYSOs are momentum driven, as their low-mass counterparts~\citep[e.g.][]{reipurth01},
as long as the HMYSO $L_{\rm bol}$ is accretion dominated (i.e. before or slightly after that an UC\ion{H}{ii} region is developed).

Several observational studies of massive outflows point to an evolutionary scenario~\citep[see e.g.][]{beuther,beuther-shepherd,shepherd}, 
in which molecular outflows appear to loose their collimation as HMYSOs evolve from B to O spectral types.
During the early B-type stage, the accreting HMYSOs drive collimated outflows, growing further in mass and luminosity, they develop UC\ion{H}{ii}
regions in the late O-type stage, and collimated jets and less collimated winds can coexist producing bipolar outflows with a lower degree of collimation.
Our results show that the HMYSOs in our sample drive collimated, time-variable and highly precessing jets~\citep[e.g.][]{reipurth01}.
Indeed if the massive jets power their CO outflows~\citep[see also][]{davis04,davis07,arce07,caratti08}, 
then the combined action of jet orientation variations and velocity variations may also produce poorly collimated molecular outflows.

\begin{figure}
 \centering
 \includegraphics[width=8 cm]{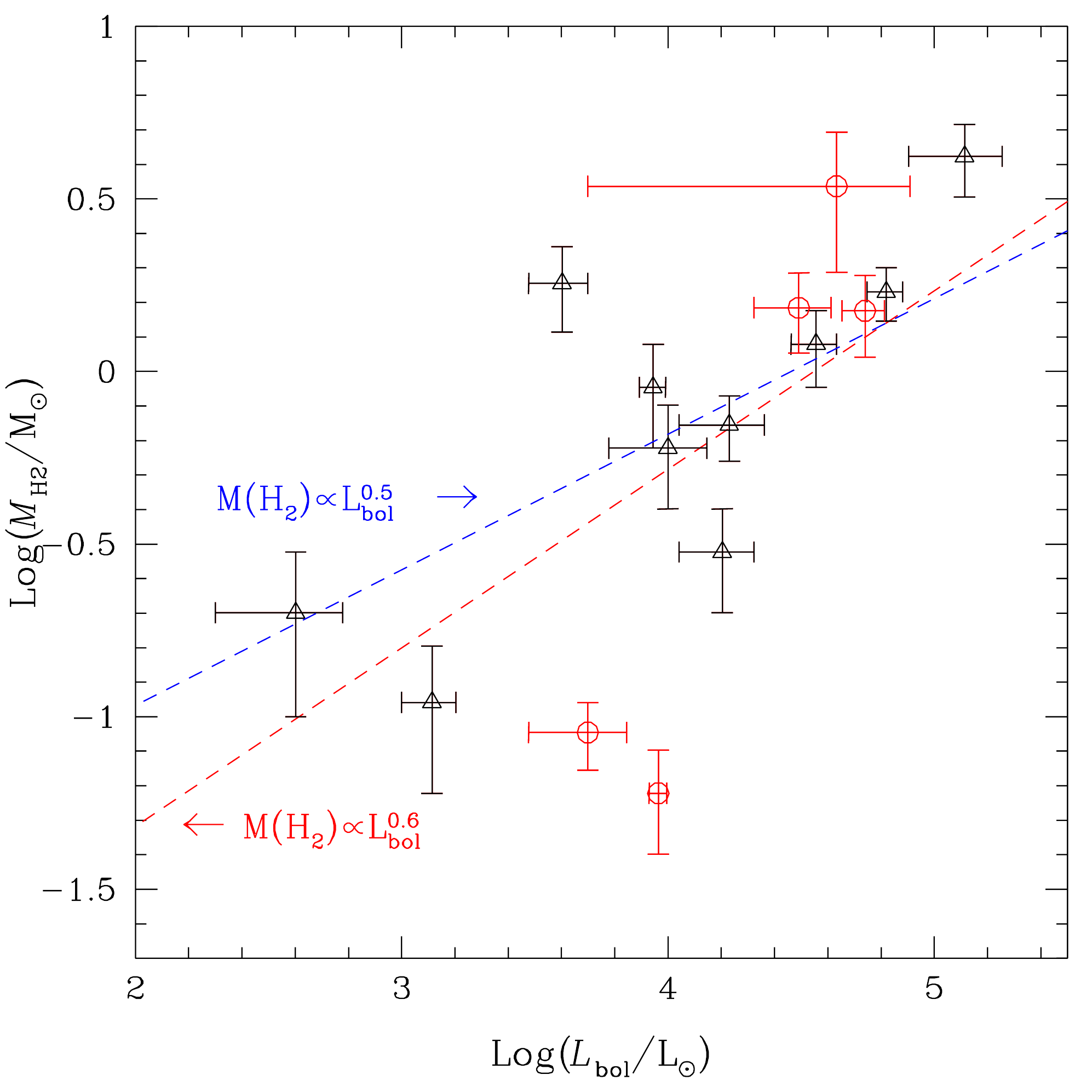}
  \caption{$Log (M(H_2))$\,vs.\,$Log (L_{\rm bol}$) for the sources of the current survey plus IRAS\,20126+4104. The red circles are the
  five outliers of Fig.~\ref{LH2:fig}. The red dashed line represents the best linear fit to the whole sample, whereas 
  the blue dashed line represents the best linear fit to a smaller sample, which excludes the five outliers. 
\label{MH2:fig}}
\end{figure}

Finally, we compare the inferred mass of the H$_2$ jets ($M(H_2)$) with the bolometric luminosities of their driving sources 
(see Fig~\ref{MH2:fig}; the red circles represent the five UC\ion{H}{ii} candidates, i.e. the five HMYSO outliers of Figure~\ref{LH2:fig}).
In principle, $M(H_2)$ should be related to the mass of the driving source, and this can be tested through the $M(H_2)$--$L_{\rm bol}$ relationship.
Despite the large scattering, $M(H_2)$ increases with the $L_{\rm bol}$ of the source, 
indicating that the more massive is the source the more massive is the jet. 
Figure~\ref{MH2:fig} shows the best fit for the whole sample (red dashed line) and for a smaller sample,
which does not include the five UC\ion{H}{ii} candidates (blue dashed line). By excluding the five outliers, the Pearson's coefficient increases from
0.7 to 0.9. By fitting the whole sample, we obtain the same correlation as before, namely $M(H_2) \propto L_{\rm bol}^{0.6}$,
whereas from the reduced sample we get $M(H_2) \propto L_{\rm bol}^{0.5}$, and the uncertainty on the exponent varies from $\sim$40\% to $\sim$30\%, respectively.

\subsection{Comparing jets from low- to high-mass YSOs}
\label{discussion5:sec}

Although we cannot infer the dynamical and kinematic properties of the jets, that will be analysed in a forthcoming paper, 
we can here provide a consistent comparison between the physical properties of jets from low-mass 
($L_{\rm bol} \sim$0.1\,L$_{\sun}$) to high-mass YSOs ($L_{\rm bol} \sim$10$^5$\,L$_{\sun}$) 
by combining results from this and previous works~\citep[][]{caratti06,caratti08}.

As their low-mass counterparts, intermediate and high-mass YSOs possess collimated jets traced by line emission from shocked regions.
High-mass jet tracers in the NIR are those observed in low-mass jets: H$_2$, the major coolant, and [\ion{Fe}{ii}]. At variance with low-mass protostellar jets, 
shocked Br$\gamma$ emission is also detected in a few knots, indicating that here shocks can be much more powerful.
In general, the physical processes, that produce such shocks, are similar, but they mostly take place in a more dense, 
embedded and inhomogeneous medium, therefore the environmental conditions may differ.
As a result, the physical quantities derived from this study are greater than those found in low-mass objects:
higher $A_{V}$ (10--100\,mag), $N(H_2)$ (10$^{18}$--10$^{20}$\,cm$^{-2}$), $T_{H_2}$ (2500--3000\,K), as well as larger $L_{H_2}$ (10--50\,L$_\sun$), 
and $M(H_2)$ (1--5\,M$_\sun$). These two latter quantities are strictly related to the properties of the driving source, 
namely higher ejection rates and more massive flows, most likely linked to the higher accretion rates of the HMYSOs.
At variance with their low-mass counterparts, high-mass jets have more pronounced precession angles (up to 60$\degr$) and more confused morphologies, 
likely indicative of the crowded environment in which they are born. However, their outflows seem to be momentum driven as their low-mass counterparts,
at least in their early evolutionary stage.
Future observations at higher angular resolution and at longer wavelengths might prove the extent to which such dynamical
interactions affect the jet/flow ejection.

To summarise, our analysis confirms the conclusions previously drawn from a limited sample of three jets from B-type stars~\citep[][]{davis04,gredel06,caratti08},
extending them to a larger sample with HMYSO $L_{\rm bol}$ up to $\sim$10$^5$\,L$_{\sun}$:
the observed high-mass protostellar jets are scaled up versions of those from low-mass protostars, albeit with some differences.

\section{Conclusions} 
\label{conclusion:sec}

We present a NIR imaging (H$_2$ and $K_s$) and low-resolution spectroscopic (0.95-2.50\,$\mu$m) survey of 18 massive jets towards EGOs, 
observed towards the fields of 14 intermediate- and high-mass YSOs, which have $L_{\rm bol}$ between 4$\times$10$^2$ and 1.3$\times$10$^5$\,L$_{\sun}$.
We study the morphology of these flows, deriving their physical parameters, and we compare them with the main properties
of the exciting sources, by means of literature data. 
The physical properties of these massive jets are also examined in comparison to their low-mass counterparts. 
The main results of this work are the following:

\begin{enumerate}
\item[-] As in low-mass jets, H$_2$ is the primary NIR coolant, detected in all the analysed knots. 
The most important ionic tracer is [\ion{Fe}{ii}], detected in 50\% of the jets and $\sim$30\% of the analysed knots. 
The majority of these knots originate from the most luminous YSOs, which likely drive the most powerful outflows of the sample.
Br$\gamma$ emission is detected in $\sim$5\% of the knots and it originates from shocks.

\item[-] Our analysis indicates that the observed emission lines originate from shocks at high temperatures and densities.
No fluorescent emission is detected along the flows, independently of the source bolometric luminosity.

\item[-] On average, the physical parameters of these massive jets are greater than those measured in low-mass YSOs.
We measure a high visual extinction towards these jets with values up to 50\,mag. The excitation conditions of the jets 
indicate high H$_2$ column densities up to 10$^{20}$\,cm$^{-2}$ (1 to 4 orders of magnitude higher than the values observed in low-mass jets). On average,
the temperatures traced by the rovibrational lines (v=1--3) in the $H$ and $K$ bands are higher ($\sim$2500\,K) than those typically inferred
in low-mass jets ($\sim$2000\,K). Inferred masses of high-mass knots are much larger than in low-mass YSOs, with values up to a few solar masses for the more massive YSOs.

\item[-] The morphology of the detected H$_2$ flows is heterogeneous, ranging from a simple straight bipolar geometry to a more complex one,
which includes monopolar flows, highly precessing jets with patchy structures, and asymmetric lobes. 
Apart from the observational bias caused by the high visual extinction, such a variety depends on
the complex, dynamic, and inhomogeneous environment in which these massive jets form and propagate.

\item[-] All flows and jets of our sample are collimated, indicating a disc origin. 
Additionally, the presence of both knots and jets might indicate that ejection is both continuous with intermittent bursts. 
By assuming that accretion and ejection are tightly related, this would
imply that mass accretion in intermediate- and high-mass YSOs is nearly continuous with intermittent accretion outbursts, as in low-mass YSOs.

\item[-] We compare the measured flow H$_2$ luminosity (the jet power) with the source bolometric luminosity (assumed representative of the
accretion luminosity), confirming the tight correlation between these two quantities, already found in low-mass protostellar jets~\citep[][]{caratti06}
and in the HMYSO IRAS\,20126+4104~\citep[][]{caratti08}. Five sources, however, display a lower $L_{H_2}$/$L_{bol}$ efficiency, less than one order of magnitude. 
We interpret this behaviour in terms of YSO evolution, i.e. their luminosity is no longer accretion dominated.
Most important, the inferred relationship ($L(H_2) \propto L_{\rm bol}^{0.6}$) agrees well with the correlation between the momentum flux of the CO outflows 
($F_{CO}$) and the bolometric luminosities of HMYSOs ($F_{CO} \propto L_{\rm bol}^{0.6}$) from \citet{beuther}. 
This indicates that outflows from HMYSOs are momentum driven, as their low-mass counterparts.

\item[-] We also derive a less stringent correlation between the inferred mass of the H$_2$ flows and the YSO bolometric luminosity, suggesting that
the mass of the flow depends on the driving source mass.

In conclusion, by comparing the physical properties of jets in the NIR, a continuity from low- to high-mass jets is identified. 
Massive jets appear as a scaled-up version of their low-mass counterparts in terms of their physical parameters and their origin. 
However, there are consistent differences, as a more variegated morphology likely due to the environment, as well as stronger shock conditions
possibly due to more powerful sources. 

\end{enumerate}

\begin{acknowledgements}
We are grateful to Dr. Timea Csengeri for providing us with ATLASGAL fluxes and for her help.
We wish to thank an anonymous referee for useful insights and comments.
This publication makes use of data products from: the Wide-field Infrared Survey Explorer, which is a joint project of 
the University of California, Los Angeles, and the Jet Propulsion Laboratory/California Institute of Technology, 
funded by the National Aeronautics and Space Administration; the 2MASS data, obtained as part of the Two Micron All Sky Survey,
a joint project of the University of Massachusetts and the Infrared Processing and Analysis Center/California Institute of Technology.
This research has also made use of NASA's Astrophysics Data System Bibliographic Services and the SIMBAD database, operated at the CDS, Strasbourg, France.

\end{acknowledgements}

\bibliographystyle{aa}
\bibliography{references}

\Online
\begin{appendix}

\section{Physical parameters of all observed knots}
\label{AppendixA:sec}

\longtab{}{
\begin{table*}
\begin{scriptsize}
\caption[]{Coordinates and H$_2$ (2.122$\mu$m) fluxes, and physical parameters of the knots detected along the investigated flows.
    \label{knots:tab}}
\begin{tabular}{cccccccccc}
\hline \hline
OUTFLOW & KNOT--ID & $\alpha$(2000.0)& $\delta$(2000.0) & $F$(2.12$\mu$m)$\pm$$\Delta$$F$ & Area & $A_{\rm V}$   &   $T(H_2)$  & $N(H_2)$ & $M(H_2)$\\
  &  & ($^{h}$ $^{m}$ $^{s}$) & ($\degr$ $\arcmin$ $\arcsec$) & ($10^{-15}$erg\,s$^{-1}$\,cm$^{-2}$) & (10$^{-10}$~sr) &   (mag)  & (K) & (cm$^{-2}$) & (M$_{\sun}$) \\
\hline \hline\\
\object{[HSL2000] IRS 1}  &  knot\,1            & 12:11:47.4 &-61:46:02.7&  53.4$\pm$1.2 &  1.41  &  14$\pm$2   & 2500$\pm$170  &  8.1$\times$10$^{17}$      &  0.02   \\
\object{[HSL2000] IRS 1}  &  knot\,2            & 12:11:46.6 &-61:45:53.6&  53.4$\pm$1.2 &  8.36  &  15$\pm$2   & 2400$\pm$400  &  7$\times$10$^{17}$     &  0.13   \\
\object{IRAS 12405-6129}   &  YSO               & 12:43:31.5 &-62:36:13.5&  4$\pm$1  &  0.45  &  $\cdots$   & $\cdots$   &  $\cdots$      & $\cdots$   \\
\object{IRAS 12405-6129}   &  knot\,1           & 12:43:31.7 &-62:36:29.3&  3.7$\pm$1.2 &  1.15  &  40$\pm$5   & 2250$\pm$60    &  2.2$\times$10$^{19}$  &  0.6  \\
\object{IRAS 12405-6129}   &  knot\,2           & 12:43:32.1 &-62:36:35.6&  3.7$\pm$1.2 &  0.61  &  35$\pm$5   & 2800$\pm$200   &  2.2$\times$10$^{19}$   & 0.3  \\
\object{IRAS 13481-6124}   &bow-shock\,A Red    & 13:51:55.3 &-61:35:38.5&  48$\pm$3 &  11.44  &  2$\pm$1   & 2740$\pm$160   &  3.2$\times$10$^{17}$   &  0.04  \\
\object{IRAS 13481-6124}   &bow-shock\,B Red    & 13:51:55.9 &-61:35:44.0&  72$\pm$5 &  17.16  &  2$\pm$1   & 2770$\pm$160   &  3.2$\times$10$^{17}$    &  0.06  \\
\object{IRAS 13481-6124}   &bow-shocks\,A/B Blue& 13:51:24.0 &-61:41:41.5&  780$\pm$10 &  186.4  &  1$\pm$1   & 2940$\pm$150   &  3.1$\times$10$^{17}$      & 0.59   \\
\object{IRAS 13481-6124}   &knot\,F\,Blue       & 13:51:36.9 &-61:39:23.9&  170$\pm$10 &  41.48  &  3$\pm$1   & 2220$\pm$140   &  1$\times$10$^{18}$      &  0.42  \\
\object{IRAS 13481-6124}   &knot\,E\,Blue       & 13:51:35.0 &-61:39:48.6&  16$\pm$1 &  3.89  &  2$\pm$1   & 2210$\pm$70   &  1$\times$10$^{18}$     & 0.04   \\
\object{IRAS 13481-6124}   &knot\,D\,Blue       & 13:51:34.2 &-61:40:04.8&  130$\pm$10 &  31.01  &  1$\pm$1   & 2390$\pm$160   &  5.4$\times$10$^{17}$   &  0.17  \\
\object{IRAS 13481-6124}   &  YSO               & 13:51:37.9 &-61:39:07.5&  2$\pm$1 &  1.4  &  $\cdots$   & $\cdots$   &  $\cdots$      & $\cdots$   \\
\object{IRAS 13484-6100}  &  knot\,1            & 13:51:59.6 &-61:15:39.6&  96$\pm$2 &  7.74  &  15$\pm$5   & 2580$\pm$120   &  1.5$\times$10$^{19}$    &  3.4  \\
\object{IRAS 13484-6100}  &  knot\,2            & 13:51:59.8 &-61:15:44.0&  11$\pm$1 &  0.50  &  15$\pm$5   & 2700$\pm$500   &  3.1$\times$10$^{18}$  & 0.04   \\
\object{IRAS 13484-6100}  &  YSO                & 13:51:58.2 &-61:15:42.3&  5$\pm$1 &  0.50  &  15$\pm$5   & 2700$\pm$400   &  3.3$\times$10$^{18}$      & 0.05   \\
\object{IRAS 14212-6131}  &  knots\,1 + YSO     & 14:25:01.1 &-61:44:55.8&  104$\pm$1 &  1.42  &  10$\pm$2   & 2600$\pm$200   & 1.7$\times$10$^{18}$   &  0.15  \\
\object{IRAS 14212-6131}  &  knots\,2 + 3       & 14:24:58.5 &-61:44:57.2&  41$\pm$1 &  1.36  &  12$\pm$2   & 2100$\pm$100   & 4.5$\times$10$^{18}$   &  0.37  \\
\object{IRAS 14212-6131}  &  knot\,4            & 14:24:57.4 &-61:44:53.2&  21$\pm$2 &  1.2  &  7$\pm$2   & 2550$\pm$120   & 1.0$\times$10$^{18}$   &  0.07  \\
\object{IRAS 14212-6131}  &  knots\,5 + YSO     & 14:24:55.4 &-61:45:23.2&  3$\pm$1 &  1.39  &  2$\pm$1   & 2300$\pm$60   &  5$\times$10$^{17}$     &  0.01  \\
\object{SSTGLMC G316.7627-00.0115}  &  knot\,1            & 14:44:57.0 &-59:48:00.8&  8.6$\pm$0.8 &  0.88  &  10$\pm$3   & 1500$\pm$300   &  3.2$\times$10$^{18}$     &  0.03  \\
\object{SSTGLMC G316.7627-00.0115}             &  knot\,2 & 14:44:55.9 &-59:48:00.1&  13$\pm$1 &  0.75  &  30$\pm$5   & 2300$\pm$200   &   1$\times$10$^{19}$     &  0.06  \\
\object{SSTGLMC G316.7627-00.0115}             &  YSO     & 14:44:56.4 &-59:48:00.8&  3$\pm$1 &  0.7  &   $\cdots$   & $\cdots$   &  $\cdots$      & $\cdots$   \\
\object{Caswell OH 322.158+00.636} &  knot\,1   & 15:18:42.2 &-56:38:51.0&  50$\pm$1 &  1.22  &  50$\pm$5   & 2700$\pm$160   &  1.9$\times$10$^{20}$      & 4.2   \\
\object{Caswell OH 322.158+00.636} &  knot\,2   & 15:18:34.3 &-56:38:29.7&  7$\pm$1 &  0.50  &  30$\pm$10   & 2100$\pm$400   &   1.1$\times$10$^{19}$    & 0.1   \\
\object{IRAS 15394-5358}  &  knot\,1            & 15:43:19.2 &-54:07:29.8&  319$\pm$1   &  11.9  &  25$\pm$2   & 2300$\pm$90   &  1.2$\times$10$^{19}$     & 0.44   \\
\object{IRAS 15394-5358}  &  knot\,2            & 15:43:18.5 &-54:07:39.2&  13$\pm$1 &  6.10  &  25$\pm$3   & 2400$\pm$100   &  6.5$\times$10$^{18}$      & 0.12   \\
\object{IRAS 15394-5358}  &  knot\,3            & 15:43:18.2 &-54:07:44.2&  15$\pm$1 &  5.10  &  3$\pm$2   & 2350$\pm$90   &  1$\times$10$^{18}$     &   0.08 \\
\object{IRAS 15394-5358}  &  knot\,4            & 15:43:18.1 &-54:07:33.7&  8$\pm$1 &  7.20  &  20$\pm$5   & 2600$\pm$200   & 9$\times$10$^{17}$ &  0.02  \\
\object{IRAS 15394-5358}  &  knot\,5 + YSO      & 15:43:19.1 &-54:07:39.9&  25$\pm$1 &  1.62  &  30$\pm$10   & 2500$\pm$200   & 8.5$\times$10$^{18}$  & 0.04   \\
\object{IRAS 15450-5431}  &  knot\,1            & 15:48:55.5 &-54:40:29.0&  25$\pm$1 &  1.19  &  6$\pm$2   & 2300$\pm$90   & 1.1$\times$10$^{18}$  & 0.02   \\
\object{IRAS 15450-5431}  &  knot\,2            & 15:48:55.7 &-54:40:28.9&  21$\pm$1 &  1.02  &  5$\pm$1   & 2100$\pm$70   & 1.1$\times$10$^{18}$  & 0.01   \\
\object{IRAS 15450-5431}  &  knot\,3            & 15:48:55.3 &-54:40:12.9&  3.0$\pm$0.5 &  0.50  &  10$\pm$5   & 2800$\pm$150   & 1.1$\times$10$^{18}$  & 0.01   \\
\object{IRAS 15450-5431}  &  knot\,4            & 15:48:55.6 &-54:40:33.1&  1.5$\pm$0.5 &  0.47  &  $\cdots$   & $\cdots$   &  $\cdots$      & $\cdots$   \\
\object{IRAS 15450-5431}  &  YSO                & 15:48:55.6 &-54:40:33.1&  8$\pm$1 &  1.16  &  10$\pm$5   & 2300$\pm$100   & 1.4$\times$10$^{18}$     &  0.02  \\
\object{IRAS 16122-5047}  &  knot\,A            & 16:16:07.1 &-50:54:21.4&  16.4$\pm$0.7 &  1.30  &  20$\pm$8   & 2150$\pm$140   & 3.9$\times$10$^{18}$  & 0.32   \\
\object{IRAS 16122-5047}  &  knot\,B            & 16:16:08.9 &-50:54:03.2&  11.8$\pm$0.8 &  0.88  &  8$\pm$3   & 2500$\pm$100   &   7$\times$10$^{17}$  & 0.04   \\
\object{IRAS 16122-5047}  &  knot\,C            & 16:16:10.3 &-50:53:55.1&  29$\pm$1 &  1.22  &  15$\pm$5   & 2250$\pm$130   & 5$\times$10$^{18}$  & 0.32   \\
\object{IRAS 16122-5047}  &  knot\,D            & 16:16:13.7 &-50:53:36.1&  8.8$\pm$0.8 &  0.95  &  15$\pm$7   & 2200$\pm$200   & 2$\times$10$^{18}$  & 0.08   \\
\object{IRAS 16122-5047}  &  knot\,2            & 16:16:10.6 &-50:52:31.0&  5.5$\pm$0.7 &  0.99  &  30$\pm$10   & 2600$\pm$250   &  3$\times$10$^{18}$  &  0.14   \\
\object{IRAS 16122-5047}  &  knot\,3            & 16:16:10.8 &-50:52:25.5&  8.2$\pm$0.8 &  1.62  &  30$\pm$10   & 2300$\pm$100   &  3.1$\times$10$^{18}$    & 0.23   \\
\object{IRAS 16122-5047}  &  knot\,R1           & 16:16:06.8 &-50:54:34.0&  4.0$\pm$0.7 &  0.95  &  20$\pm$10   & 2900$\pm$500   &  1.1$\times$10$^{18}$    & 0.04   \\
\object{IRAS 16122-5047}  &  YSO + knot\,1      & 16:16:06.9 &-50:54:26.9&  14$\pm$1 &  1.62  &  30$\pm$10   & 2200$\pm$300   &   9$\times$10$^{18}$      &   0.67  \\
\object{IRAS 16547-4247}  &  knot\,A1           & 16:58:17.2 &-42:51:43.0&  50$\pm$1 &  6.20  &  15$\pm$7   & 2000$\pm$650   &  5$\times$10$^{17}$       &  0.1  \\
\object{IRAS 16547-4247}  &  knot\,A2           & 16:58:16.5 &-42:51:38.0&  45$\pm$1 &  7.05  &  20$\pm$5   & 2600$\pm$160   &  3.1$\times$10$^{18}$       &  0.71  \\
\object{IRAS 16547-4247}  &  knot\,A4           & 16:58:16.5 &-42:51:26.0&  41$\pm$1 &  2.40  &  15$\pm$5   & 2200$\pm$100   & 1.3$\times$10$^{18}$         & 0.1   \\
\object{IRAS 16547-4247}  &  knot\,A5           & 16:58:16.5 &-42:51:31.0&  58$\pm$1 &  8.04  &  15$\pm$5   & 1900$\pm$150   & 2.8$\times$10$^{18}$         & 0.64   \\
\object{IRAS 16547-4247}  &  knot\,B1           & 16:58:17.4 &-42:52:16.0&  33$\pm$1 &  4.60  &  10$\pm$5   & 2400$\pm$200   &  1$\times$10$^{17}$      &  0.01  \\
\object{IRAS 16547-4247}  &  knot\,B4           & 16:58:17.0 &-42:52:35.0&  43$\pm$1 &  29.36  &  13$\pm$3   & 2170$\pm$80   &  3$\times$10$^{17}$ &  0.28  \\
\object{BGPS G014.849-00.992} &knots\,1         & 18:21:13.0 &-16:30:17.0&  65$\pm$1 &  4.37  &  30$\pm$10   & 2400$\pm$900   & 3$\times$10$^{18}$ &  0.08  \\
\object{BGPS G014.849-00.992} &knot\,2          & 18:21:12.4 &-16:30:05.0&  21$\pm$1 &  2.06  &  18$\pm$5   & 2440$\pm$90   &  2.1$\times$10$^{18}$ &  0.03  \\
\object{BGPS G014.849-00.992} &knot\,3          & 18:21:10.5 &-16:30:34.0&  13$\pm$1 &  2.16  &  5$\pm$3   & 2530$\pm$80   &    3$\times$10$^{17}$ &  0.004  \\
\object{BGPS G014.849-00.992} &knot\,4          & 18:21:11.7 &-16:29:56.1&  4$\pm$1 &  0.58  &  30$\pm$20   & 1500$\pm$500   &  1$\times$10$^{19}$      &  0.04  \\
\object{BGPS G014.849-00.992} &knot\,5          & 18:21:12.0 &-16:29:33.6&  14$\pm$1 &  1.89  &  20$\pm$8   & 2400$\pm$130   &   4.4$\times$10$^{18}$      &  0.06  \\
\object{GLIMPSE G035.0393-00.4735} & knot\,1    & 18:56:57.0 & 1:39:17.8 &  33$\pm$1  &  10.58  &  14$\pm$4   & 2500$\pm$100   & 1.1$\times$10$^{18}$      &  0.13  \\
\object{GLIMPSE G035.0393-00.4735} &  knot\,2   & 18:56:57.9 & 1:39:29.2 &  4$\pm$1 &  2.13  &  30$\pm$10   & 1200$\pm$500   &   5$\times$10$^{19}$    &   0.12  \\
\object{G35.2N}           &  knot\,1            & 18:58:12.3 & 1:40:10.2 &  8.4$\pm$0.8 &  6.03  &  15$\pm$5   & 2500$\pm$150   &  1$\times$10$^{18}$    &   0.03  \\
\object{G35.2N}           &  knots\,2 + YSO     & 18:58:13.1 & 1:40:39.9 &  2$\pm$0.8 &  0.47  &   $\cdots$   & $\cdots$   &  $\cdots$      & $\cdots$   \\  
\object{G35.2N}           &  knot\,3            & 18:58:13.1 & 1:40:48.8 &  15.9$\pm$0.9 &  10.10  &  20$\pm$8   & 3300$\pm$500   &   3$\times$10$^{18}$   &  0.15  \\    
\object{G35.2N}           &  knots\,4 + 5       & 18:58:14.6 & 1:41:37.1 &  97$\pm$1 &  58.82  &  10$\pm$2   & 2800$\pm$200   &  1.1$\times$10$^{18}$     &  0.31  \\    
\object{G35.2N}           &  knot\,6            & 18:58:10.1 & 1:40:07.0 &  29.7$\pm$0.9 & 23.19  &  10$\pm$4   & 2800$\pm$300   & 1.1$\times$10$^{18}$       &  0.12  \\    
\object{G35.2N}           &  knot\,7            & 18:58:11.3 & 1:40:14.5 &   43$\pm$1 &  16.68  &  10$\pm$5   & 3000$\pm$400   &  2.9$\times$10$^{18}$      &  0.24  \\    
\object{G35.2N}           &  knot\,8-1          & 18:58:14.7 & 1:40:42.9 &  21$\pm$2 &  16.17  &  10$\pm$4   & 2900$\pm$300   &  6$\times$10$^{17}$     &  0.08  \\    
\object{G35.2N}           &  knot\,8-2          & 18:58:14.8 & 1:40:47.4 &  10$\pm$3 &  14.1  &    $\cdots$   & $\cdots$   &  $\cdots$      & $\cdots$   \\    
\object{G35.2N}           &  knot\,9            & 18:58:15.9 & 1:40:50.8 &  22.8$\pm$0.9 &  13.09  &  8$\pm$4   & 2800$\pm$300   &  2$\times$10$^{18}$     &  0.10  \\    
\object{G35.2N}           &  knot\,10           & 18:58:16.6 & 1:40:57.8 &  68$\pm$1 &  39.29  &  8$\pm$2   & 2370$\pm$90   &  2$\times$10$^{18}$   &  0.35  \\    
\hline \hline
\end{tabular}
\end{scriptsize}
\end{table*}
}
\section{Individual objects}
\label{objects:sec}

\subsection{[HSL2000] IRS 1}
\label{appendixB:sec}

\begin{table}
\caption{Observed emission lines in the \object{[HSL2000] IRS 1} jet.\label{spec_hsl2000:tab}}
\begin{tabular}{ccccc}
\hline\\[-5pt]
Species & Term &  $\lambda$($\mu$m) & \multicolumn{2}{c}{$F\pm\Delta~F$(10$^{-15}$erg\,cm$^{-2}$\,s$^{-1}$)}\\
\hline\\[-5pt]
                 &    &       &    knot\,1    &  knot\,2         \\
H$_2$ & 1--0 S(7)     & 1.748 &  1.2$\pm$0.3     & 0.9$\pm$0.3      \\
H$_2$ & 1--0 S(3)     & 1.958 & 2$\pm$1$^*$   & $\cdots$       \\
H$_2$ & 1--0 S(2)     & 2.034 & 3$\pm$0.3       & 1.3$\pm$0.3      \\
H$_2$ & 2--1 S(3)     & 2.073 &0.8$\pm$0.3$^*$& $\cdots$      \\
H$_2$ & 1--0 S(1)     & 2.122 & 5.3$\pm$0.4   & 3.4$\pm$0.4     \\
H$_2$ & 1--0 S(0)     & 2.223 & 1.7$\pm$0.5   & 1.5$\pm$0.5     \\
H$_2$ & 2--1 S(1)     & 2.248 &1.5$\pm$0.6$^*$& 1.5$\pm$0.5     \\
H$_2$ & 1--0 Q(1)     & 2.407 & 7$\pm$2      & 6$\pm$2$^*$        \\
H$_2$ & 1--0 Q(3)     & 2.424 & 7$\pm$2      & 6$\pm$2$^*$        \\
\hline\\[-5pt]
\hline
\end{tabular}
\tablefoot{$^*$ S/N between 2 and 3.}
\end{table}

\begin{figure*}
 \centering
   \includegraphics[width=15 cm]{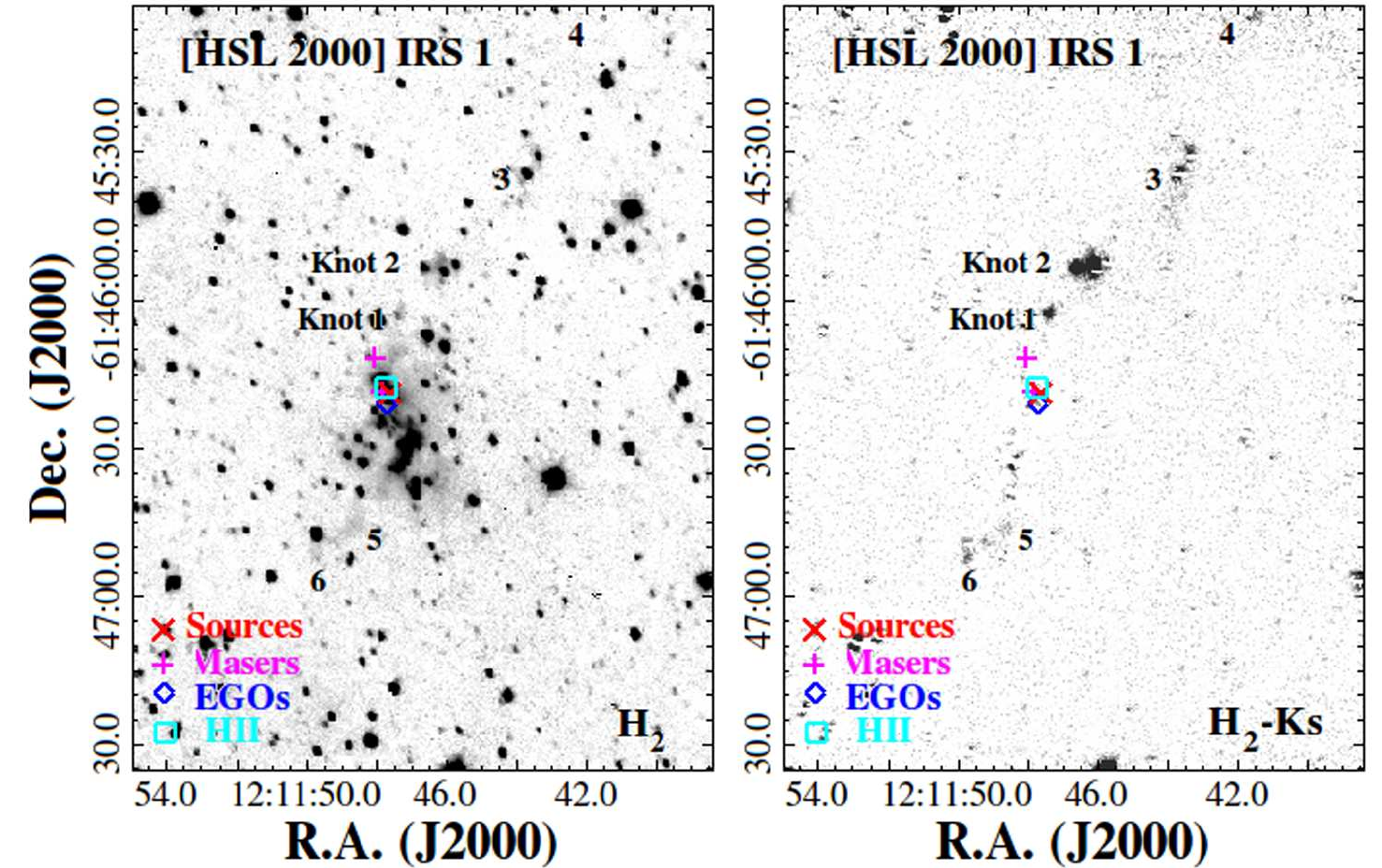} 
   \caption{H$_2$ and continuum-subtracted H$_2$ images (\textit{left and right panels}) of the \object{[HSL2000] IRS 1} outflow.
   The positions of the sources, knots, masers, EGOs, and \ion{H}{ii} regions are indicated in the figures.   
\label{HSL2000ima:fig}}
\end{figure*}

\object{[HSL2000] IRS 1}, coincident with \object{IRAS 12091-6129}, was firstly identified by \citet{henning00} at MIR wavelengths.
The authors also detect a second source, \object{[HSL2000] IRS 2}, located $\sim$28$\arcsec$ westwards.
G298.2622+00.7391 is the dominant source at 8\,$\mu$m.
$L_{\rm bol}$ values from the literature range form 1.6 to 5.2$\times$10$^4$\,L$_{\sun}$~\citep[][]{walsh,henning00,lumsden13}, depending on the adopted distance
(3.8--5.8\,kpc). According to these estimates, the source spectral type ranges from B0.5~\citep[][]{henning00} to O8.5~\citep[][]{walsh}.
Both CH$_3$OH (at 6.67 GHz) and OH maser (at 1.665 GHz) emissions are detected
towards the source~\citep[][]{walsh2}. Although reported in the SIMBAD database, it is not clear whether or not UC\ion{H}{ii} emission is associated with this source.
\citet{walsh2} do not detect any UC\ion{H}{ii} region above 1\,mJy at 8.64 GHz.
Close to the HMYSO, \citet{cyganowski} observed EGO emission (\object{EGO G298.26+0.74}), which was interpreted as scattered emission from the continuum~\citep[][]{takami12}.
Indeed, our image (Fig.~\ref{HSL2000ima:fig}) does not show any H$_2$ emission at the EGO position.
Outflow emission from CO (2--1) and CS (2--1) has been reported~\citep[][]{osterloh,henning00}.
In our image, the H$_2$ jet is well collimated, but, possibly due to the presence of a multiple system, it shows a large precession angle ($\sim$30$\degr$). 
This can be easily inferred by comparing the current P.A. of the jet emerging from the source (P.A.$\sim$10$\degr$) with the P.A. of the most distant 
knot (Knot 4, P.A.$\sim$-20$\degr$). 
This explains why the observed CO outflow~\citep{henning00} has a very small collimation factor~\citep[$R_c$=2.53; see][]{wu}.
The H$_2$ knots are located within the CO bipolar outflow~\citep[see Fig.~1 in][]{henning00}, which shows spatially separated red and blue lobes, located SSE and NNW of the source.
Knots 1, 2, 3, 4 delineate the blue lobe, whereas knots 5, and 6 are located in the red lobe (see Fig.~\ref{HSL2000ima:fig}).
Only H$_2$ emission lines are detected in our spectra (Table~\ref{spec_hsl2000:tab}). 
Our slit encompasses the source position as well, where just a faint rising continuum emission is detected.

\subsection{IRAS12405-6219}
\label{appendixC:sec}

Originally identified as a planetary nebula candidate from its colours~\citep{vandesteene}, it has been later recognised as an HMYSO candidate, 
coincident with an UC\ion{H}{ii} region in the RMS catalogue~\citep[][]{lumsden13}, which reports $L_{\rm bol}$=3.6$\times$10$^4$\,L$_{\sun}$
at a distance of 4.4\,kpc. Location of the object agrees with the position of the NIR source 2MASSJ12433151-6236135 and that of its
MIR counterpart MSX \object{G302.0213+00.2542}. NIR and MIR colours indicate that this is a young and embedded object.
H$_2$O maser emission near the source position has been reported by \citet{suarez}.

\begin{figure*}
 \centering
   \includegraphics[width=15 cm]{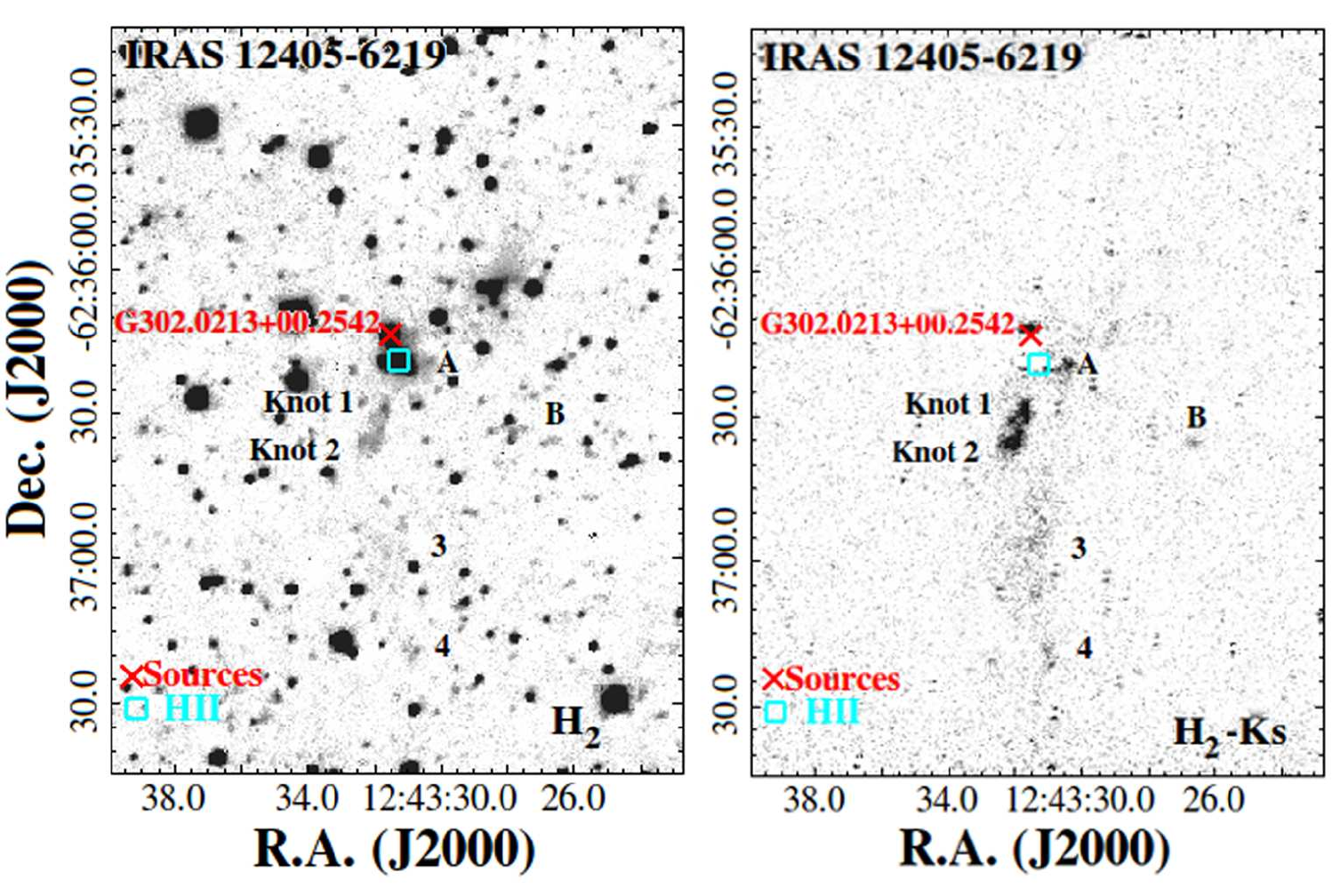} 
   \caption{Same as in Figure~\ref{HSL2000ima:fig} but for the \object{IRAS 12405-6129} flows.   
\label{IRAS12405ima:fig}}
\end{figure*}

Our H$_2$ image (Fig.~\ref{IRAS12405ima:fig}) shows the presence of several knots: knots 1, 2, 3 and 4 are located south of the source, whereas two other knots (A and B) are aligned
with the source position towards WSW. We interpret such a geometry as a combination of two distinct flows: the first flow (knots 1, 2, 3 and 4) is
precessing southwards, with a P.A. ranging $\sim$170$\degr$-184$\degr$, whereas the second flow (knots A and B) is straight with a P.A. of $\sim$236$\degr$.
There is no clear detection of the two opposite lobes, likely the red-shifted ones.
Our spectra of knots 1 and 2 show only H$_2$ emission, and that on the source shows a steeply rising continuum with a bright Br$\gamma$ 
line in emission (Table~\ref{spec_IRAS12405:tab}).

\begin{table}
\caption{Observed emission lines in one of the two \object{IRAS12405-6129} jets. \label{spec_IRAS12405:tab}}
\begin{tabular}{cccccc}
\hline\\[-5pt]
Species & Term &  $\lambda$($\mu$m) & \multicolumn{3}{c}{$F\pm\Delta~F$(10$^{-15}$erg\,cm$^{-2}$\,s$^{-1}$)}\\
\hline\\[-5pt]
                 &    &       &    knot\,1    &  knot\,2     &  YSO$^{**}$      \\
H$_2$ & 1--0 S(3)     & 1.958 & 3$\pm$1       & $\cdots$      & $\cdots$     \\
H$_2$ & 1--0 S(2)     & 2.034 & 3.1$\pm$0.3    & 1.5$\pm$0.4  & $\cdots$   \\
H$_2$ & 2--1 S(3)     & 2.073 &1.2$\pm$0.4    &1.1$\pm$0.4$^*$& $\cdots$    \\
H$_2$ & 1--0 S(1)     & 2.122 & 11.2$\pm$0.5   & 5.1$\pm$0.5 & 4.3$\pm$0.6   \\
\ion{H}{i} & Br$\gamma$&2.166 & $\cdots$      & $\cdots$      &4.7$\pm$0.6   \\
H$_2$ & 1--0 S(0)     & 2.223 & 3.2$\pm$0.6   & 1.2$\pm$0.6$^*$ & 2.8$\pm$0.7   \\
H$_2$ & 1--0 Q(1)     & 2.407 & 16$\pm$3       & 9$\pm$3 & 11$\pm$5$^*$      \\
H$_2$ & 1--0 Q(2)     & 2.413 & 8$\pm$3$^*$   &  $\cdots$    & $\cdots$ \\     
H$_2$ & 1--0 Q(3)     & 2.424 & 21$\pm$3       & 8$\pm$3$^*$ & $\cdots$       \\
H$_2$ & 1--0 Q(5)     & 2.455 & 13$\pm$4       &  $\cdots$   & $\cdots$        \\
\hline\\[-5pt]
\hline
\end{tabular}
\tablefoot{$^*$ S/N between 2 and 3.$^{**}$ YSO = \object{IRAS 12405-6129}, namely \object{2MASS J12433151-6236135}}
\end{table}

\subsection{IRAS 13481-6124}
\label{appendixD:sec}

Located at a distance of 3.1\,kpc~\citep[][]{lumsden13}, \object{IRAS 13481-6124}, spatially 
coincident with the NIR source \object{2MASS J13513785-6139075} (see also Fig~\ref{IRAS13481ima:fig}),
was recognised as an HMYSO candidate by \citet{chan}, and then confirmed through 
the SED modelling by \citet{grave09}, who obtained a stellar mass of $\sim$20\,M$_{\sun}$, O9 spectral type and 6$\times$10$^4$\,yr age.
VLTI interferometric observations~\citep[][]{kraus10} show the presence of a compact dusty disc with a large CO molecular
outflow, driven by a parsec-scale H$_2$ jet~\citep[][]{stecklum12}. Our continuum subtracted H$_2$ image (Fig.~\ref{IRAS13481ima:fig}) 
shows a precessing jet (precession angle $\sim$8$\degr$) extending $\sim$6.9\,pc. The current P.A. of the jet is $\sim$206$\degr$. 
The red lobe is located NE of the source and it is about 
1.3 times more extended than the blue lobe. Both lobes display the same number of knots, whose distances from the source are always
roughly 1.3 times larger in the red-shifted lobe of the flow, likely indicating that the red-shifted jet is moving in a less dense medium.
Two large bow-shocks (A and B) are located at the apex of both lobes, whereas the remaining H$_2$ emanations, from C to G in both lobes, 
have a knotty and jet-like geometry. As evinced from our analysis, visual extinction increases along the jet towards the source position, 
from $A_{\rm V}$=1 to $\sim$13\,mag, whereas both $T(H_2)$ and $N(H_2)$ (i.e. temperature and column density of the shocked excited H$_2$ gas).
decrease (see Table~\ref{knots:tab} and Sect.~\ref{physparam:sec}). As pointed out in Sect.~\ref{physparam:sec}, H$_2$ fluorescent emission 
is confined within few arcseconds from the source~\citep[][]{stecklum12}
and the H$_2$ emission along the jet is strictly thermalised, suggesting that the circumstellar material around the source screens its FUV emission.
Our spectra indicate that H$_2$ emission is present in all knots, whereas [\ion{Fe}{ii}] emission is detected 
only in the red-shifted bow-shocks (see Table~\ref{spec_IRAS13481:tab}).

Apart form those knots belonging to the parsec-scale collimated jet, we detect more H$_2$ knots in our continuum-subtracted H$_2$ image.
These may belong to other flows emanating from the source position (i.e. IRAS 13481-6124 might be then a multiple system) or they might come
from other YSOs. In particular, among the several H$_2$ emanations, two knots (X and Y, see Fig.~\ref{IRAS13481ima:fig}) 
are detected above a 3$\sigma$ threshold and located beyond the extended nebulosity surrounding the source. 
Knot\,X, located SSE of the source, has a bow-shock shape, whose orientation might trace back to the source position.
Knot\,Y is located NW of the source and it does not have a definite shape, therefore no definite direction is recognisable. 
There are no bright YSOs nearby the IRAS source, with the exception of two faint YSO candidates (about 52$\arcsec$ E and 55$\arcsec$ NW of the IRAS position),
which are barely detected at 24\,$\mu$m and at longer wavelengths in the Spitzer/MIPS and Herschel images. Knots X and Y might be then driven by these sources.

Our low resolution spectroscopy on the source (Fig.~\ref{IRAS13481ima:fig}) shows many \ion{H}{i} lines from both Paschen and Brackett series. 
Unfortunately, because the spectral emission
in the $K$ band is completely saturated by the source continuum, we cannot detect the Br$\gamma$ emission on the source, even though it was
previously detected through low-resolution spectroscopy~\citep[][]{beck91}, and at higher spectral and spatial resolution with SINFONI~\citep[][]{stecklum12}.
The spectro-astrometric analysis from these data indicates that the Br$\gamma$ line is spatially extended with its position angle agreeing with that of the jet.
These observations as well as the detection of a collimated parsec-scale jet suggest that the source is still accreting and its disc is not just a passive
disc, as proposed by \citet{kraus10}. IRAS 13481-6124 is among the most evolved and massive objects in our sample, 
but possesses a powerful and collimated jet. Indeed, given the age estimate of this source~\citep[6$\times$10$^4$\,yr;][]{grave09}, 
this might seem a real conundrum.
In principle, the source should be evolved enough and had already developed an UC\ion{H}{ii} region~\citep[see e.g.][]{zinnecker07,tan03}. 
However, there are no observations of any UC\ion{H}{ii} region around it, and ATCA observations at 3 and 6\,cm do not show any radio emission down to 
0.48\,mJy and 1\,mJy, respectively~\citep[][]{urquhart07}. Moreover, by analysing the luminosity-to-mass ratio of IRAS 13481-6124 and other HMYSOs, 
\citet{beltran06} conclude that their HMYSOs are in a pre-ultracompact \ion{H}{ii} phase. 
In conclusion, IRAS 13481-6124 might not be as old as the age estimate suggests.

\begin{landscape}
\begin{table*}Moreover
\begin{scriptsize}
\caption{Observed emission lines in the \object{IRAS 13481-6124} jet. \label{spec_IRAS13481:tab}}
\begin{center}
\begin{tabular}{cccccccccc}
\hline\\[-5pt]
Species & Term &  $\lambda$($\mu$m) & \multicolumn{7}{c}{$F\pm\Delta~F$(10$^{-15}$erg\,cm$^{-2}$\,s$^{-1}$)}\\
\hline\\[-5pt]
                 &                            &       &bow-shock\,A\,Red&bow-shock\,B\,Red&bow-shocks\,A-B\,Blue&knot\,F\,Blue& knot\,E\,Blue&knot\,D\,Blue& YSO$^{**}$\\
\ion{H}{i} & Pa$\gamma$& 1.094                        & $\cdots$      & $\cdots$      & $\cdots$      & $\cdots$      & $\cdots$      & $\cdots$      &130$\pm$40 \\
\ion{H}{i} & Pa$\beta$& 1.282                         & $\cdots$      & $\cdots$      & $\cdots$      & $\cdots$      & $\cdots$      & $\cdots$      &740$\pm$20 \\
\ion{H}{i} & Br19     & 1.526                         & $\cdots$      & $\cdots$      & $\cdots$      & $\cdots$      & $\cdots$      & $\cdots$      &90$\pm$20 \\	 
\ion{H}{i} & Br18     & 1.535                         & $\cdots$      & $\cdots$      & $\cdots$      & $\cdots$      & $\cdots$      & $\cdots$      &200$\pm$20 \\ 
\ion{H}{i} & Br17     & 1.544                         & $\cdots$      & $\cdots$      & $\cdots$      & $\cdots$      & $\cdots$      & $\cdots$      &190$\pm$20 \\
\ion{H}{i} & Br16     & 1.556                         & $\cdots$      & $\cdots$      & $\cdots$      & $\cdots$      & $\cdots$      & $\cdots$      &250$\pm$20 \\		 		 
\ion{H}{i} & Br15     & 1.570                         & $\cdots$      & $\cdots$      & $\cdots$      & $\cdots$      & $\cdots$      & $\cdots$      &280$\pm$20 \\	 
\ion{H}{i} & Br14     & 1.588                         & $\cdots$      & $\cdots$      & $\cdots$      & $\cdots$      & $\cdots$      & $\cdots$      &270$\pm$20 \\	 
{[\ion{Fe}{ii}]} & $a^4\!D_{3/2}-a^4\!F_{7/2}$& 1.600 & 2.9$\pm$0.9   & $\cdots$      & $\cdots$      & $\cdots$      & $\cdots$      & $\cdots$      & $\cdots$ \\ 
\ion{H}{i}                         & Br13     & 1.611 & $\cdots$      & $\cdots$      & $\cdots$      & $\cdots$      & $\cdots$      & $\cdots$      &370$\pm$20 \\
\ion{H}{i}                         & Br12     & 1.641 & $\cdots$      & $\cdots$      & $\cdots$      & $\cdots$      & $\cdots$      & $\cdots$      &480$\pm$20 \\		 		 
{[\ion{Fe}{ii}]} & $a^4\!D_{7/2}-a^4\!F_{9/2}$& 1.644 & 11.5$\pm$0.9  & 3.5$\pm$0.9   & $\cdots$      & 5$\pm$1       & $\cdots$      & $\cdots$      & 60$\pm$20   \\
{[\ion{Fe}{ii}]} & $a^4\!D_{5/2}-a^4\!F_{7/2}$& 1.677 &2.2$\pm$0.9$^*$& $\cdots$      & $\cdots$      & $\cdots$      & $\cdots$      & $\cdots$      &$\cdots$\\ 
\ion{H}{i} & Br11                             & 1.681 & $\cdots$      & $\cdots$      & $\cdots$      & $\cdots$      & $\cdots$      & $\cdots$      &510$\pm$20 \\		 		 
\ion{Fe}{ii}     & $c^4\!F_{9}-z^4\!F_{9}$& 1.688 & $\cdots$      & $\cdots$      & $\cdots$      & $\cdots$      & $\cdots$      & $\cdots$      &270$\pm$20 \\		 		 
H$_2$ & 1--0 S(9) 	                       & 1.688 &1.8$\pm$0.9$^*$& $\cdots$      & 14$\pm$1      & $\cdots$      & 5$\pm$1       & $\cdots$      &$\cdots$\\
H$_2$ & 1--0 S(8) 	                       & 1.715 &1.9$\pm$0.9$^*$& $\cdots$      &   9$\pm$1     & $\cdots$      & $\cdots$      & $\cdots$      & $\cdots$      \\
\ion{H}{i}                          & Br10    & 1.737 & $\cdots$      & $\cdots$      & $\cdots$      & $\cdots$      & $\cdots$      & $\cdots$      &730$\pm$30 \\		 		 
H$_2$ & 1--0 S(7) 	                       & 1.748 & 7$\pm$1       & 4.0$\pm$0.9   &   36$\pm$1    & 4$\pm$1       &  9$\pm$2      & 14$\pm$2      &  $\cdots$     \\
H$_2$ & 1--0 S(6) 	                       & 1.788 & 5$\pm$1       & 2.7$\pm$0.9   & 22$\pm$1      & $\cdots$      &  7$\pm$2      & 8$\pm$2      & $\cdots$   \\
\ion{H}{i} & Br9                              & 1.818 & $\cdots$      & $\cdots$      & $\cdots$      & $\cdots$      & $\cdots$      & $\cdots$      &1300$\pm$100 \\		 		 
H$_2$ & 1--0 S(5) 	                       & 1.836 & $\cdots$      & $\cdots$      & 36$\pm$10     & $\cdots$      & $\cdots$      & $\cdots$      & $\cdots$   \\
H$_2$ & 1--0 S(4) 	                       & 1.892 & $\cdots$      & $\cdots$      & 38$\pm$10     & $\cdots$      & $\cdots$      & $\cdots$      & $\cdots$  \\
H$_2$ & 1--0 S(3) 		               & 1.958 & 25$\pm$8      & $\cdots$      & 73$\pm$10     & 21$\pm$10$^*$ & 33$\pm$1      & $\cdots$      & $\cdots$      \\
H$_2$ & 1--0 S(2)           	               & 2.034 & 9.1$\pm$0.9   & 5$\pm$1       & 35$\pm$1      & 7$\pm$2       & 18$\pm$2      & 15$\pm$2     & $\cdots$   \\
H$_2$ & 2--1 S(3)           	               & 2.073 & $\cdots$      & $\cdots$      & 15$\pm$1      & $\cdots$      & $\cdots$      & 6$\pm$2      & $\cdots$   \\
H$_2$ & 1--0 S(1)           	               & 2.122 & 19$\pm$1      & 12$\pm$1      & 90$\pm$1      & 21$\pm$1      & 47$\pm$2      & 38$\pm$2      &$\cdots$   \\
H$_2$ & 2--1 S(2)           	               & 2.154 & $\cdots$      & $\cdots$      & 8$\pm$1       & $\cdots$      & $\cdots$      & $\cdots$      & $\cdots$      \\
H$_2$ & 3--2 S(3)                             & 2.201 & $\cdots$      & $\cdots$      &  4$\pm$1      & $\cdots$      & $\cdots$      & $\cdots$      & $\cdots$      \\
H$_2$ & 1--0 S(0)           	               & 2.223 &  6$\pm$1      & 3$\pm$1       & 20$\pm$1      & 5$\pm$1       & 13$\pm$2      &10$\pm$3       &$\cdots$   \\
H$_2$ & 2--1 S(1)           	               & 2.248 &  3$\pm$1      & $\cdots$      & 16$\pm$1      & $\cdots$      & 8$\pm$2       & $\cdots$      & $\cdots$   \\
H$_2$ & 1--0 Q(1)           	               & 2.407 & 15$\pm$5      & 10$\pm$5$^*$  & 74$\pm$5      & 18$\pm$5      & 41$\pm$5      & 47$\pm$6      & $\cdots$      \\
H$_2$ & 1--0 Q(2)           	               & 2.413 & $\cdots$      & $\cdots$      & 30$\pm$6      & $\cdots$      & 15$\pm$5	  & $\cdots$	  & $\cdots$	\\
H$_2$ & 1--0 Q(3)           	               & 2.424 & 17$\pm$5      & 11$\pm$5$^*$  & 74$\pm$5      & 20$\pm$5      & 41$\pm$2      & 36$\pm$6      &$\cdots$      \\
H$_2$ & 1--0 Q(4)           	               & 2.437 & $\cdots$      & $\cdots$      & 19$\pm$5      & $\cdots$      & 15$\pm$5	  & $\cdots$	  & $\cdots$	\\
H$_2$ & 1--0 Q(5)           	               & 2.455 & $\cdots$      & $\cdots$      & 47$\pm$5      & 16$\pm$5 	 & 32$\pm$5	  & $\cdots$	  &$\cdots$	\\
H$_2$ & 1--0 Q(6)           	               & 2.476 & $\cdots$      & $\cdots$      & 20$\pm$5      & $\cdots$      & $\cdots$      & $\cdots$	  & $\cdots$	\\
H$_2$ & 1--0 Q(7)           	               & 2.500 & $\cdots$      & $\cdots$      & 30$\pm$5      & $\cdots$      & $\cdots$      & $\cdots$	  &$\cdots$	\\
\hline\\[-5pt]
\hline
\end{tabular}
\tablefoot{$^*$ S/N between 2 and 3. $^{**}$ YSO: \object{IRAS 13481-6124}, namely \object{2MASS J13513785-6139075}. Our $K$ band spectrum is saturated, therefore lines in the K band are not reported.}
\end{center}
\end{scriptsize}
\end{table*}
\end{landscape}

\subsection{IRAS 13484-6100}
\label{appendixE:sec}

IRAS 13484-6100 is a compact GLIMPSE source (G310.1437+00.7598) with a cometary-like \ion{H}{ii} emission, and it is located at a distance of $\sim$5.5\,kpc.
CS emission~\citep[][]{bronfman}, OH~\citep[][]{caswell}, CH$_3$OH~\citep[][]{green12} and H$_2$O~\citep[][]{urquhart09} masers
are detected on the source and in its surroundings.
\citet{faundez} detected the continuum emission from the source at 1.2\,mm, and estimated
$L_{\rm bol}$=7.1$\times$10$^4$\,L$_{\sun}$ and $M_c$=1.1$\times$10$^3$\,M$_{\sun}$ from SED analysis. On the other hand,
the RMS catalogue~\citep[][]{lumsden13} provides two different $L_{\rm bol}$ values of $\sim$5$\times$10$^3$\,L$_{\sun}$ and 4.3$\times$10$^4$\,L$_{\sun}$
for the HMYSO and the \ion{H}{ii} region, respectively.
A bipolar EGO emission (EGO G310.15+0.76) is detected by \citet{cyganowski}, located $\sim$16$\arcsec$ NE of the source position.

\begin{figure*}
 \centering
   \includegraphics[width=9.1 cm]{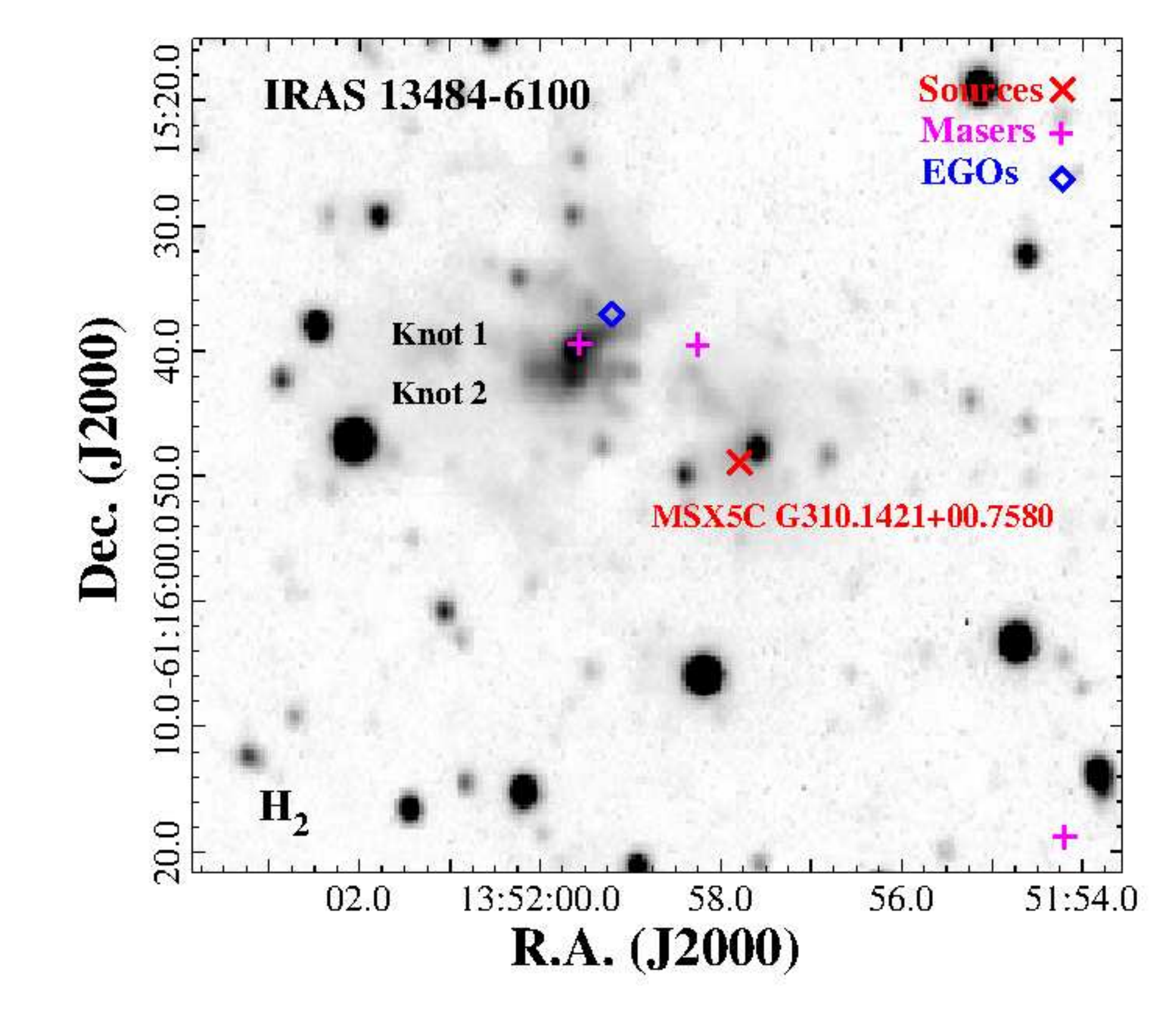} \includegraphics[width=9.1 cm]{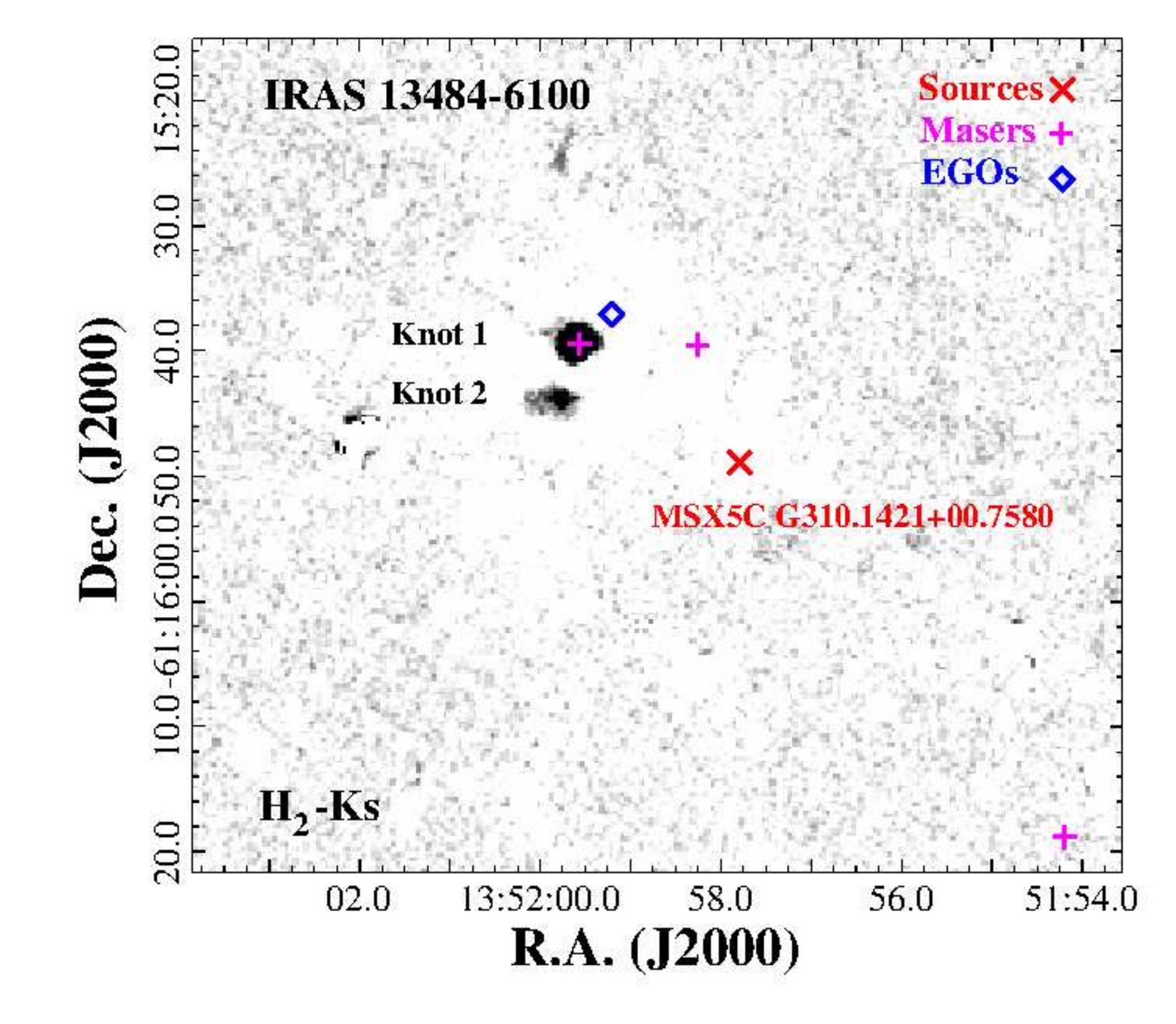}
   \caption{Same as in Figure~\ref{HSL2000ima:fig} but for the \object{IRAS 13484-6100} flow.
   \label{G310ima:fig}}
\end{figure*}

Figure~\ref{G310ima:fig} shows our H$_2$ and continuum-subtracted H$_2$ images of IRAS 13484-6100, and Table~\ref{spec_IRAS13484:tab} lists the emission 
lines detected in the spectra.
In our continuum-subtracted H$_2$ image (Fig.~\ref{G310ima:fig}) the EGO emission partially overlaps with Knot\,1, located at a P.A. of $\sim$53$\degr$
with respect to the source. A second knot (Knot\,2) is observed $\sim$4$\arcsec$ SW of Knot\,1. From our images it is not clear whether or not these emissions
belong to the same flow. Radio continuum emission at 6\,cm is detected towards the Knot\,1 position~\citep[][]{urquhart07}.
Apart from H$_2$ emission, our spectra reveal strong [\ion{Fe}{ii}] lines in both knots as well as towards the source position, which shows a faint raising continuum.
Br$\gamma$ emission from a dissociative shock (see Sect.~\ref{discussion1:sec}) is also detected in Knot\,1 (see Table~\ref{spec_IRAS13484:tab}).

\begin{table*}
\caption{Observed emission lines in the \object{IRAS 13484-6100} jet. \label{spec_IRAS13484:tab}}
\begin{center}
\begin{tabular}{cccccc}
\hline\\[-5pt]
Species & Term &  $\lambda$($\mu$m) & \multicolumn{3}{c}{$F\pm\Delta~F$(10$^{-15}$erg\,cm$^{-2}$\,s$^{-1}$)}\\
\hline\\[-5pt]
                 &    &       &    knot\,1    &  knot\,2       & YSO$^{**}$    \\
{[\ion{Fe}{ii}]} & $a^4\!D_{5/2}-a^4\!F_{9/2}$ & 1.534     & 4$\pm$1  & $\cdots$     & $\cdots$   \\
{[\ion{Fe}{ii}]} & $a^4\!D_{3/2}-a^4\!F_{7/2}$ & 1.600    &  6$\pm$1 & $\cdots$     & $\cdots$ \\
{[\ion{Fe}{ii}]} & $a^4\!D_{7/2}-a^4\!F_{9/2}$ & 1.644    &  39$\pm$1&  9$\pm$1    & 11$\pm$1  \\
{[\ion{Fe}{ii}]} & $a^4\!D_{1/2}-a^4\!F_{5/2}$ & 1.664    & 6$\pm$1  & $\cdots$     & 2.6$\pm$0.8   \\        
{[\ion{Fe}{ii}]} & $a^4\!D_{5/2}-a^4\!F_{7/2}$ & 1.677    & 8$\pm$1  & $\cdots$     & $\cdots$   \\
H$_2$ & 1--0 S(7)     & 1.748 & 11$\pm$1   & 3$\pm$1    & 3$\pm$1	\\
H$_2$ & 1--0 S(6)     & 1.788 & 5$\pm$1    & $\cdots$       &  $\cdots$      \\
{[\ion{Fe}{ii}]} & $a^4\!D_{7/2}-a^4\!F_{7/2}$ &1.810   & 7$\pm$2$^*$ & $\cdots$     & $\cdots$   \\        
               &  +$a^4\!P_{5/2}-a^4\!D_{7/2}$  &1.811  & 14$\pm$2  &$\cdots$ & $\cdots$    \\
H$_2$ & 1--0 S(3)     & 1.958 & 19$\pm$4      & 6$\pm$3$^*$       & 6$\pm$3$^*$   \\
H$_2$ & 1--0 S(2)     & 2.034 & 23$\pm$2   & 5$\pm$2$^*$  & 5$\pm$2$^*$   \\
H$_2$ & 3--2 S(5)     & 2.066 & 3$\pm$1      & $\cdots$       & $\cdots$    \\
H$_2$ & 2--1 S(3)     & 2.073 & 8$\pm$1   & $\cdots$       & $\cdots$   \\
H$_2$ & 1--0 S(1)     & 2.122 & 64$\pm$1  & 11$\pm$1   & 12$\pm$1	\\
\ion{H}{i} & Br$\gamma$&2.166 &  4$\pm$1      & $\cdots$       &  $\cdots$    \\
H$_2$ & 3--2 S(3)     & 2.201 & 3$\pm$1  & $\cdots$   & $\cdots$      \\
H$_2$ & 1--0 S(0)     & 2.223 &  22$\pm$1    & 5$\pm$1    & 6$\pm$1   \\
H$_2$ & 2--1 S(1)     & 2.248 &  9$\pm$1    & $\cdots$   & $\cdots$      \\
H$_2$ & 2--1 S(0)     & 2.355 & 4$\pm$2$^*$ &  $\cdots$       & $\cdots$    \\
H$_2$ & 1--0 Q(1)     & 2.407 & 68$\pm$5      & 11$\pm$5$^*$       & 15$\pm$5      \\
H$_2$ & 1--0 Q(2)     & 2.413 & 29$\pm$5     & $\cdots$   & $\cdots$      \\
H$_2$ & 1--0 Q(3)     & 2.424 & 65$\pm$5      & 11$\pm$5$^*$      & 13$\pm$5$^*$ \\
H$_2$ & 1--0 Q(4)     & 2.437 & 24$\pm$5      & $\cdots$   & $\cdots$      \\
H$_2$ & 1--0 Q(5)     & 2.455 & 48$\pm$5      & $\cdots$   & $\cdots$      \\
H$_2$ & 1--0 Q(6)     & 2.476 & 14$\pm$6$^*$ &  $\cdots$       & $\cdots$    \\
H$_2$ & 1--0 Q(7)     & 2.500 & 35$\pm$10     &  $\cdots$       & $\cdots$    \\
\hline\\[-5pt]
\hline
\end{tabular}
\tablefoot{$^*$ S/N between 2 and 3. $^{**}$ YSO = \object{IRAS 13484-6100}. }
\end{center}
\end{table*}
%

\subsection{IRAS 14212-6131}
\label{appendixF:sec}

Spitzer GLIMPSE images reveal the presence of three IR sources in this region (see Fig.~\ref{IRAS14212-6131ima:fig} and discussion in Sect.~\ref{imaging:sec}). 
The most embedded source (G313.7654-00.8620; $\alpha$(J2000)=14:25:01.53, $\delta$(J2000)=-61:44:57.7) is not 
detected in the NIR and its position agrees with that of the millimetric continuum emission observed by \citet{faundez}. 
The IRAS position is displaced $\sim$15$\arcsec$ westwards of the G313.7654-00.8620 position.
The other two sources are visible at NIR wavelengths (2MASSJ14245631-6144472 and 2MASSJ14245547-6145227) and they are positioned $\sim$38$\arcsec$
WNW and $\sim$40$\arcsec$ SW of the main source, respectively.
The location of the OH maser emission detected by \citet{caswell} agrees with that of G313.7654-00.8620.
Assuming a distance of 4\,kpc, \citet{faundez} derive a bolometric luminosity of $L_{\rm bol}$=1.2$\times$10$^5$\,L$_{\sun}$ and core mass ($M_c$)
of $M_c$=1.1$\times$10$^3$\,M$_{\sun}$ from SED analysis. \citet{lumsden13} assume a distance of 7.8\,kpc and obtain
$L_{\rm bol}$=1.7$\times$10$^4$\,L$_{\sun}$.
G313.7654-00.8620 drives the main flow of the region. 
CS~\citep[][]{bronfman} and CO~\citep[][]{urquhart} emissions have been detected close to the source.
H$_2$ emission was first observed by \citet{debuizer03}. Four CH$_3$OH maser spots are aligned along the flow~\citep[][]{walsh}. 
EGO emission~\citep[EGO G313.76-0.86;][]{cyganowski} lies westwards of the source. 
Our continuum-subtracted H$_2$ image (Fig.~\ref{IRAS14212-6131ima:fig}, right) shows four knots (1--4; suggesting an outflow with a precession angle of $\sim$32$\degr$) 
driven by the central source \object{G313.7654-00.8620}. These knots are located westwards of the source, depicting the blue-shifted lobe as our ISAAC/VLT high
resolution spectra indicate (Caratti o Garatti et al. in preparation). 
From the position of Knot\,1 we infer that the current P.A. of the jet is $\sim$305$\degr$. A faint H$_2$ emission, with signal-to-noise ratio below 3$\sigma$, 
appears eastwards of the source, likely depicting the red-shifted jet. As for other sources in our sample, the high visual extinction possibly prevents us 
from properly detecting one of the two lobes.
A fifth knot (Knot 5) is detected SW from the source and it does not belong to the main flow; it is likely to be associated with a different YSO (2MASSJ14245547-6145227).
The lines detected in our spectra are reported in Table~\ref{spec_IRAS14212:tab}.

\begin{table*}
\caption{Observed emission lines in the \object{IRAS14212-6131} jet. \label{spec_IRAS14212:tab}}
\begin{center}
\begin{tabular}{ccccccc}
\hline\\[-5pt]
Species & Term &  $\lambda$($\mu$m) & \multicolumn{4}{c}{$F\pm\Delta~F$(10$^{-15}$erg\,cm$^{-2}$\,s$^{-1}$)}\\
\hline\\[-5pt]
                 &                             &       &  knot\,1+YSO$^1$  &  knot\,2+3    &   knot\,4     & knot\,5+YSO$^2$       \\
{[\ion{Fe}{ii}]} & $a^4\!D_{5/2}-a^4\!F_{9/2}$ & 1.534 & 6.8$\pm$0.9   & $\cdots$     & $\cdots$      & $\cdots$  \\
{[\ion{Fe}{ii}]} & $a^4\!D_{3/2}-a^4\!F_{7/2}$ & 1.600 & 4.8$\pm$0.8   & $\cdots$     & $\cdots$      & $\cdots$      \\ 
{[\ion{Fe}{ii}]} & $a^4\!D_{7/2}-a^4\!F_{9/2}$ & 1.644 &36.3$\pm$0.6   & 4.5$\pm$0.5   & 12.3$\pm$0.7 &  $\cdots$   \\
{[\ion{Fe}{ii}]} & $a^4\!D_{1/2}-a^4\!F_{5/2}$ & 1.664 & 2.0$\pm$0.6   & $\cdots$      & $\cdots$      & $\cdots$      \\ 
{[\ion{Fe}{ii}]} & $a^4\!D_{5/2}-a^4\!F_{7/2}$ & 1.677 & 4.8$\pm$0.6   & $\cdots$     & $\cdots$      & $\cdots$      \\ 
H$_2$ & 1--0 S(9)                              & 1.688 & $\cdots$      & 1.1$\pm$0.4$^*$& $\cdots$      & 2.2$\pm$0.6     \\ 
{[\ion{Fe}{ii}]} & $a^4\!D_{3/2}-a^4\!F_{5/2}$ & 1.712 & 1.9$\pm$0.6   & $\cdots$     & $\cdots$      & $\cdots$      \\ 
H$_2$ & 1--0 S(8) 	                       & 1.715 &1.6$\pm$0.6$^*$&1.2$\pm$0.4    & $\cdots$      & $\cdots$      \\
H$_2$ & 1--0 S(7) 	                       & 1.748 & 6.7$\pm$0.6   & 4.6$\pm$0.5   & 5.0$\pm$0.7   & 7.0$\pm$0.6   \\
H$_2$ & 1--0 S(6) 	                       & 1.788 & 2.4$\pm$0.6   & 4.0$\pm$0.8   & 2.2$\pm$0.7   & 5.2$\pm$0.7   \\
{[\ion{Fe}{ii}]} & $a^4\!D_{7/2}-a^4\!F_{7/2}$& 1.810 & 6$\pm$3$^*$   & $\cdots$      & 3$\pm$1   & $\cdots$ \\        
               &  +$a^4\!P_{5/2}-a^4\!D_{7/2}$& 1.811 & 14$\pm$4      &$\cdots$       & $\cdots$      & $\cdots$ \\
H$_2$ & 1--0 S(5) 	                       & 1.836 & 10$\pm$4$^*$  &$\cdots$      & 12$\pm$3      & 13$\pm$4       \\
H$_2$ & 1--0 S(3) 		               & 1.958 & 14$\pm$4      &8$\pm$3$^*$    & 8$\pm$3$^*$    & 27$\pm$4         \\
H$_2$ & 1--0 S(2)           	               & 2.034 & 11.7$\pm$0.8  & 11.9$\pm$0.6  & 7.1$\pm$0.6   & 13.7$\pm$0.8   \\
H$_2$ & 2--1 S(3)           	               & 2.073 &  4.5$\pm$0.8  & 4.0$\pm$0.7   & 3.7$\pm$0.6   & 4.3$\pm$0.8   \\
H$_2$ & 1--0 S(1)           	               & 2.122 & 36.4$\pm$0.6  & 41.1$\pm$0.7  & 21.3$\pm$0.6   & 35.7$\pm$0.6  \\
{[\ion{Fe}{ii}]} & $a^2\!P_{3/2}-a^4\!P_{3/2}$& 2.133 &1.7$\pm$0.6$^*$& $\cdots$      & $\cdots$      & $\cdots$ \\     
H$_2$ & 2--1 S(2)           	               & 2.154 & 2.1$\pm$0.6   & 2.5$\pm$0.7   & 1.8$\pm$0.6  & 2.1$\pm$0.6     \\
\ion{H}{i} & Br$\gamma$                       & 2.166 & 2.1$\pm$0.6   & $\cdots$       & 3.1$\pm$0.6 &  $\cdots$    \\
\ion{Na}{i} &                                 & 2.190 &$\cdots$       & $\cdots$      & $\cdots$ &   2.0$\pm$0.6  \\ 
H$_2$ & 3--2 S(3)                             & 2.201 & 2.2$\pm$0.7   & $\cdots$   & $\cdots$      & $\cdots$      \\
H$_2$ & 1--0 S(0)           	               & 2.223 &12.0$\pm$0.7   & 13.0$\pm$0.7   & 5.4$\pm$0.7   & 9.7$\pm$0.7   \\
{[\ion{Fe}{ii}]} & $a^2\!P_{3/2}-a^4\!P_{1/2}$& 2.244 & 3.7$\pm$0.7   & $\cdots$      & $\cdots$      & $\cdots$ \\        
H$_2$ & 2--1 S(1)           	               & 2.248 & 3.7$\pm$0.7   & 7.4$\pm$0.7   & 3.4$\pm$0.7   & 4.2$\pm$0.7   \\
CO & v=2--0                                   & 2.294 & $\cdots$	& $\cdots$    & $\cdots$    &  11$\pm$1         \\ 
CO & v=3--1                                   & 2.323 & $\cdots$	& $\cdots$    & $\cdots$ &  10$\pm$1 \\ 
H$_2$ & 4--3 S(3)           	               & 2.344 &  3$\pm$1      & $\cdots$      & $\cdots$      & $\cdots$      \\
CO & v=4--2                                   & 2.353 & $\cdots$	& $\cdots$ & $\cdots$ & $\cdots$  11$\pm$1   \\ 
H$_2$ & 1--0 Q(1)           	               & 2.407 & 35$\pm$2      & 63$\pm$5      & 20$\pm$3      & 29$\pm$2      \\
H$_2$ & 1--0 Q(2)           	               & 2.413 & 17$\pm$2      & 29$\pm$5      &  7$\pm$3$^*$   & 11$\pm$2      \\
H$_2$ & 1--0 Q(3)           	               & 2.424 & 33$\pm$2      & 62$\pm$5      & 18$\pm$3      & 28$\pm$2      \\
H$_2$ & 1--0 Q(4)           	               & 2.437 & 11$\pm$3      & 21$\pm$7      &  7$\pm$3$^*$   & 9$\pm$3       \\
H$_2$ & 1--0 Q(5)           	               & 2.455 & 23$\pm$5      & 46$\pm$10      &  13$\pm$4   & 24$\pm$5      \\
H$_2$ & 1--0 Q(7)           	               & 2.500 & $\cdots$      & 40$\pm$20$^*$ & $\cdots$      & 25$\pm$10$^*$      \\
\hline\\[-5pt]
\hline
\end{tabular}
\tablefoot{$^*$ S/N between 2 and 3. $^1$ MYSO = \object{2MASSJ 14250114-6144576}. $^2$ YSO = \object{2MASSJ 14245547-6145227}.}
\end{center}
\end{table*}

\subsection{SSTGLMC G316.7627-00.0115}
\label{appendixG:sec}

\begin{figure*}
 \centering
   \includegraphics[width=9.1 cm]{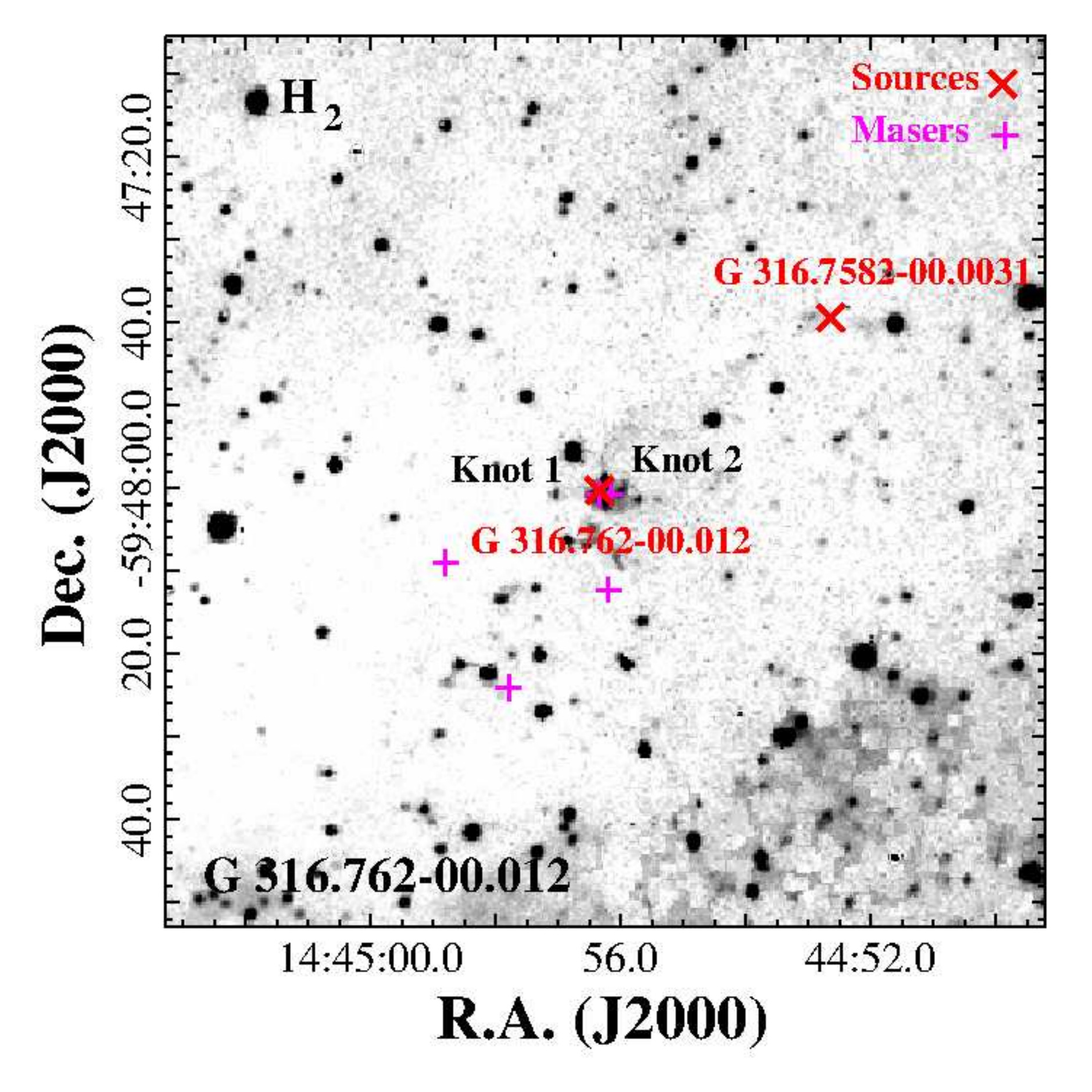} \includegraphics[width=9.1 cm]{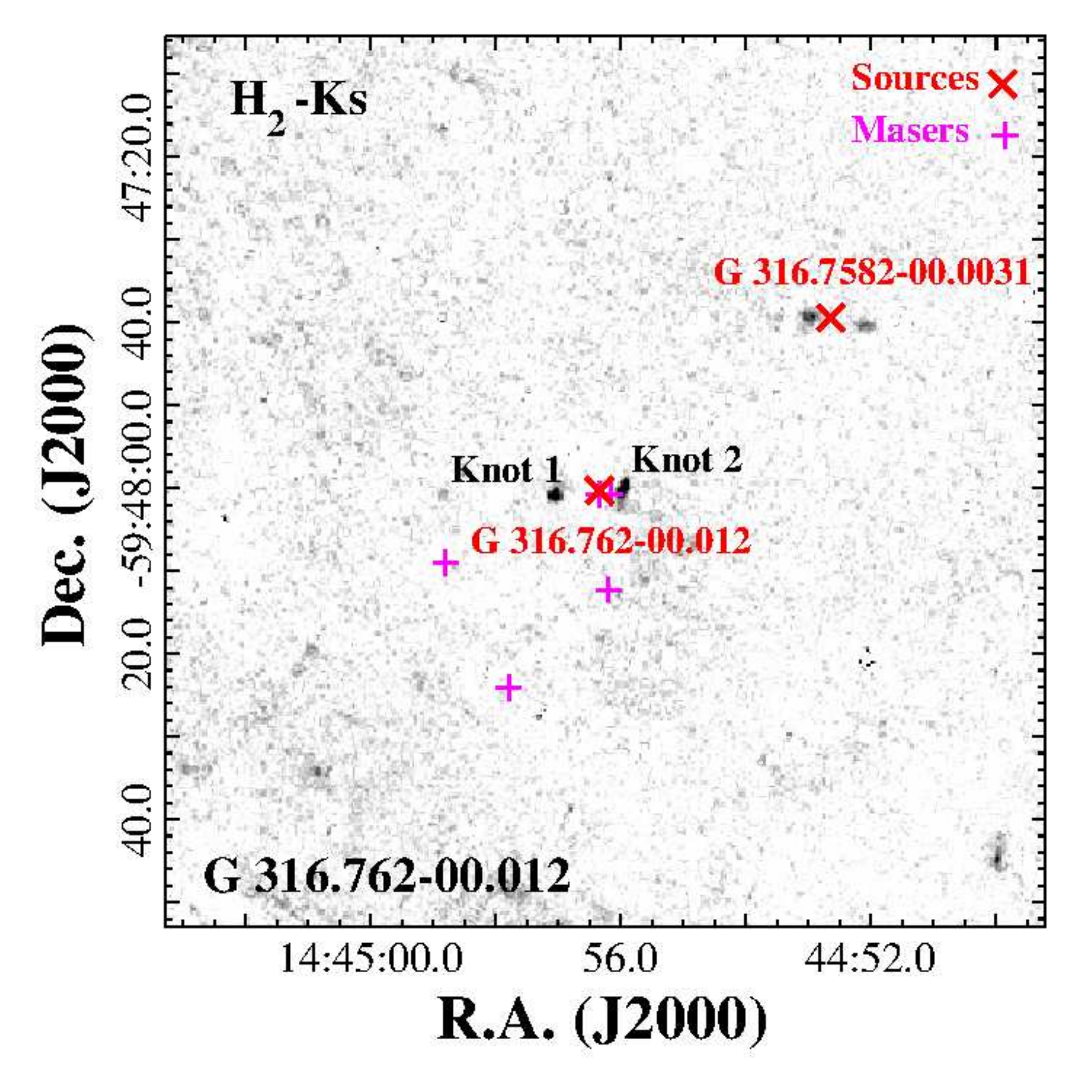}
   \caption{Same as in Figure~\ref{HSL2000ima:fig} but for the \object{SSTGLMC G316.7627-00.0115} flow.
   \label{G316ima:fig}}
\end{figure*}

\begin{figure}
 \centering
   \includegraphics[width=8 cm]{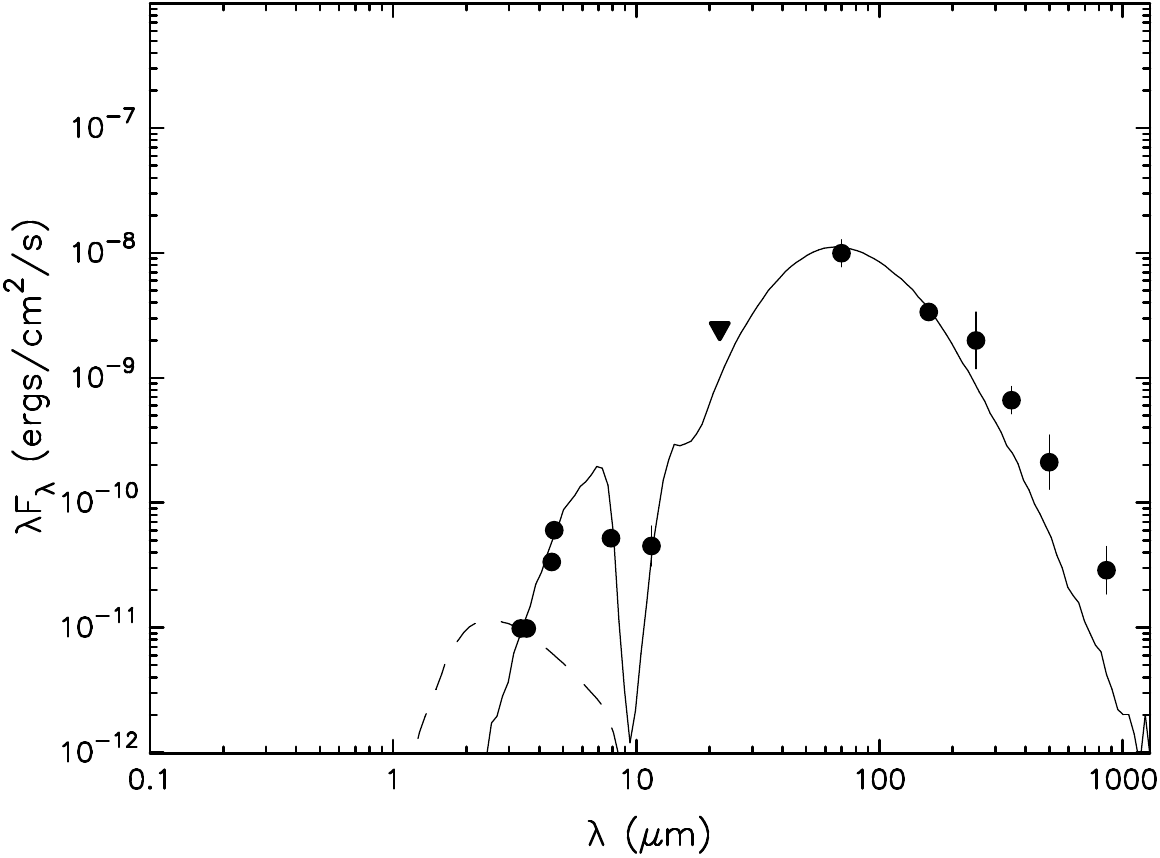} 
   \caption{Spectral energy distribution (SED) of \object{SSTGLMC G316.7627-00.0115} constructed with all photometric data available from the literature,
   namely from 3.6\,$\mu$m to 870\,$\mu$m, and assuming a distance of 2.8\,kpc to the source~\citep[][]{ragan}. 
\label{G316SED:fig}}
\end{figure}

Positioned in the IRAS 14416-5937-B region (d=2.8\,kpc), \object{SSTGLMC G316.7627-00.0115} was first recognised as a massive YSO candidate by \citet{nyman} and
\citet{vig}.
It is located at the south-eastern end of the filamentary IRDC G316.72+0.07. This IRDC is rather pristine and just contains a few weaker mid- and far-infrared 
embedded sources~\citep[][]{ragan}. The IRDC itself attains a high mid-IR extinction contrast (i.e. high column densities)
and contains masses up to 900\,M$_{\sun}$~\citep[][]{linz07}.
\object{SSTGLMC G316.7627-00.0115} does not seem to have a NIR counterpart, although it is quite close ($\sim$2$\arcsec$) 
to \object{2MASSJ 14245547-6145227}, which
was recognised as the NIR counterpart~\citep[][]{vig}.
Two OH maser emissions have been detected close to the source position~\citep{caswell}.
Eastwards and westwards of the source position, we detect two aligned H$_2$ knots (1 and 2) (see Fig.~\ref{G316ima:fig}), 
whose P.A. is $\sim$275$\fdg$.5 or $\sim$95$\fdg$.5, depending on Knot\,2 association with the blue- or red-shifted lobe, respectively.
Only H$_2$ emission lines are detected in the spectra of the two knots (see Table~\ref{spec_G316:tab}).
Our continuum-subtracted H$_2$ image also shows the presence of a second flow $\sim$35$\arcsec$ NE of the source, likely driven by a low-mass YSO.
The photometric data used for our SED analysis are reported in Table~\ref{G316sed:tab}.
Our SED analysis of \object{SSTGLMC G316.7627-00.0115}, shown in Fig.~\ref{G316SED:fig}, provides us with a bolometric 
luminosity of 5$\times$10$^3$\,L$_{\sun}$, at a distance of 2.8\,kpc~\citep[][]{urquhart13}.

\begin{table}
\caption{Observed emission lines in the \object{SSTGLMC G316.7627-00.0115 jet}. \label{spec_G316:tab}}
\begin{tabular}{cccccc}
\hline\\[-5pt]
Species & Term &  $\lambda$($\mu$m) & \multicolumn{3}{c}{$F\pm\Delta~F$(10$^{-15}$erg\,cm$^{-2}$\,s$^{-1}$)}\\
\hline\\[-5pt]
                 &    &       &    knot\,1    &  knot\,2     &  YSO     \\
H$_2$ & 1--0 S(7)     & 1.748 &  $\cdots$     & 0.9$\pm$0.3  &  $\cdots$    \\
H$_2$ & 1--0 S(3)     & 1.958 & 5$\pm$2$^*$   & $\cdots$     & $\cdots$     \\
H$_2$ & 1--0 S(2)     & 2.034 & 2.8$\pm$0.4    & 2.5$\pm$0.4  & $\cdots$   \\
H$_2$ & 2--1 S(3)     & 2.073 & $\cdots$      &0.8$\pm$0.4$^*$& $\cdots$    \\
H$_2$ & 1--0 S(1)     & 2.122 & 8.1$\pm$0.5   & 8.5$\pm$0.5  & 1.0$\pm$0.3   \\
H$_2$ & 1--0 S(0)     & 2.223 & 2.5$\pm$0.6   & 3.2$\pm$0.6  &  $\cdots$    \\
H$_2$ & 1--0 Q(1)     & 2.407 & 9$\pm$2       & 17$\pm$2 &  $\cdots$    \\
H$_2$ & 1--0 Q(2)     & 2.413 & 5$\pm$2$^*$   &  7$\pm$2    & $\cdots$ \\     
H$_2$ & 1--0 Q(3)     & 2.424 & 8$\pm$2       & 17$\pm$2 & $\cdots$       \\
H$_2$ & 1--0 Q(4)     & 2.436 & $\cdots$      & 8$\pm$3$^*$  & $\cdots$        \\
H$_2$ & 1--0 Q(5)     & 2.455 & $\cdots$      & 15$\pm$3   & $\cdots$        \\
\hline\\[-5pt]
\hline
\end{tabular}
\tablefoot{$^*$ S/N between 2 and 3.}
\end{table}

\begin{table}
\caption{\object{SSTGLMC G316.7627-00.0115} available photometry. \label{G316sed:tab}}

\begin{tabular}{ccc}

\hline\\[-5pt]

\multicolumn{1}{c}{$\lambda$}  &  \multicolumn{1}{c}{F$\pm \Delta$F}  & Data \\
\multicolumn{1}{c}{($\mu$m)}   &  \multicolumn{1}{c}{(Jy)}           &  source          \\

\hline\\[-5pt]
3.4 & 0.0109$\pm$0.002  &  WISE            \\ 
3.6 & 0.011$\pm$0.001  &  Spitzer/GLIMPSE \\ 
4.5 & 0.050$\pm$0.08  &  Spitzer/GLIMPSE \\ 
4.6 & 0.09$\pm$0.01   &  WISE            \\ 
8.0 & 0.13$\pm$0.01   &  Spitzer/GLIMPSE \\ 
12 & 0.16$\pm$0.05    &  WISE            \\ 
22 & $<$18.3     &  WISE            \\ 
70 & 240$\pm$60 & Herschel/Hi-GAL  \\ 
160 & 180$\pm$40 & Herschel/Hi-GAL  \\ 
250 & 190$\pm$100     & Herschel/Hi-GAL  \\ 
350& 80$\pm$20       & Herschel/Hi-GAL  \\ 
500& 40$\pm$20        & Herschel/Hi-GAL  \\ 
870& 9.1$\pm$0.4       & APEX/ATLASGAL    \\ 
\hline\\[-5pt]
\hline

\end{tabular}
\tablefoot{These data are used in the SED analysis shown in Fig.~\ref{G316SED:fig}.}

\end{table}

\subsection{Caswell OH 322.158+00.636}
\label{appendixH:sec}
\begin{figure*}
 \centering
   \includegraphics[width=9.1 cm]{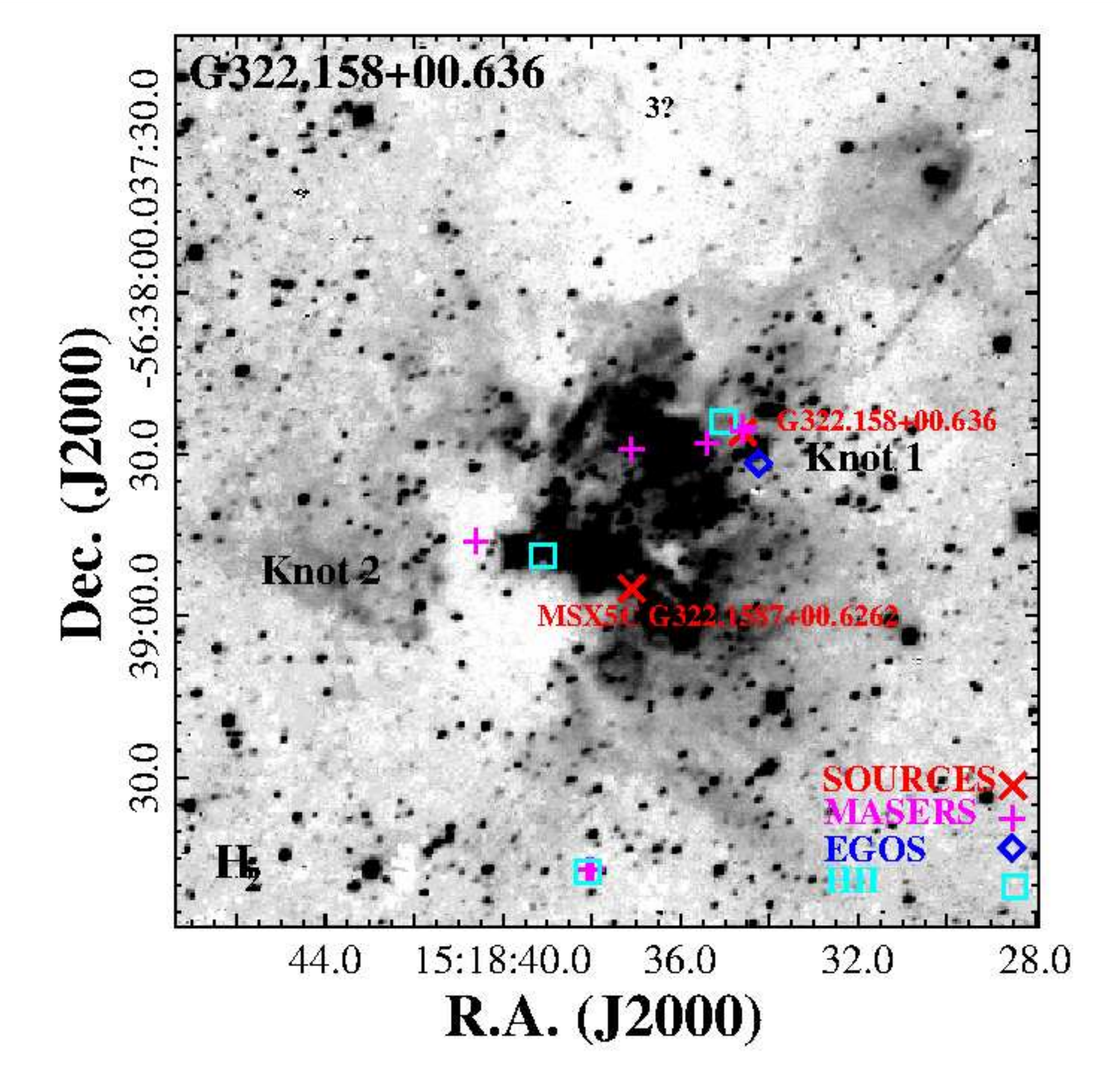} \includegraphics[width=9.1 cm]{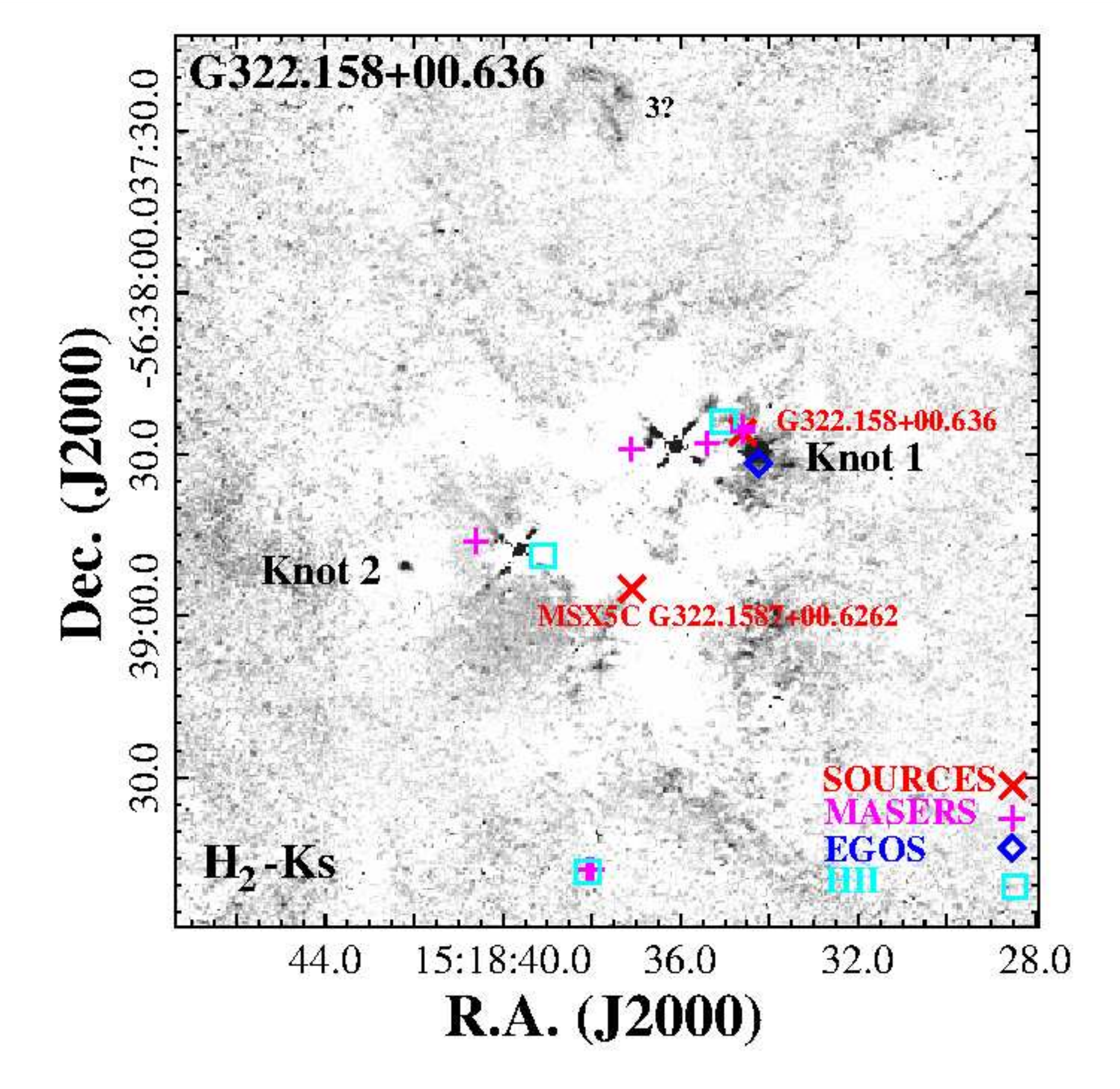}
   \caption{Same as in Figure~\ref{HSL2000ima:fig} but for the \object{Caswell OH 322.158+00.636} flow.
   \label{G322ima:fig}}
\end{figure*}

The location of \object{Caswell OH 322.158+00.636} agrees with that of the MIR source MSX G322.1587+00.6256; no NIR counterpart is detected in the 2MASS
catalogue.
The object is located in a very complex region, dominated at NIR wavelengths by a large nebulosity, possibly illuminated by two bright massive stars~\citep[][]{moises}.
OH and CH$_3$OH masers are detected towards the source position~\citep[][]{caswell,urquhart13}.
In the Spitzer GLIMPSE images the source is saturated at both 24\,$\mu$m and 70\,$\mu$m,
whereas its emission is below the linearity limit in the Herschel PACS images.
A second fainter IR object, MSX5C G322.1587+00.6262, is positioned about 35$\arcsec$ SE of the source.
In our continuum-subtracted H$_2$ image (see Fig.~\ref{G322ima:fig}) we detect a bright knot (Knot\,1) $\sim$5$\arcsec$ SW of the source, 
almost coincident with a very bright EGO
observed in the Spitzer/IRAC images. Our NIR spectroscopy shows both H$_2$ and [\ion{Fe}{ii}] 
emission from this knot (Table~\ref{spec_G322:tab}). The knot position angle is $\sim$210$\degr$. Our continuum-subtracted H$_2$ image also shows 
the presence of a second bright knot, Knot\,2, positioned about 68$\arcsec$ SE of the Knot\,1. 
Knot\,2 does not seem to emanate from MSX G322.1587+00.6256, but more likely from MSX5C G322.1587+00.6262.
The retrieved photometry of MSX G322.1587+00.6256 is reported in Table~\ref{G322sed:tab}.
By assuming a distance of 4.3\,kpc~\citep[][]{urquhart13}, our SED analysis (see Fig.~\ref{G322SED:fig}) provides us with an $L_{\rm bol}$ of $\sim$1.3$\times$10$^5$\,L$_{\sun}$,
roughly corresponding to an O7 ZAMS star with a mass of $\sim$30\,M$_{\sun}$. Therefore this object is the most luminous and massive of our sample.

\begin{table*}
\caption{Observed emission lines in the \object{Caswell OH 322.158+00.636} jet. \label{spec_G322:tab}}
\begin{tabular}{ccccc}
\hline\\[-5pt]
Species & Term &  $\lambda$($\mu$m) & \multicolumn{2}{c}{$F\pm\Delta~F$(10$^{-15}$erg\,cm$^{-2}$\,s$^{-1}$)}\\
\hline\\[-5pt]
                 &    &       &    knot\,1    &  knot\,2         \\
{[\ion{Fe}{ii}]}& $a^4\!D_{7/2}-a^4\!F_{9/2}$ & 1.644   & 2.5$\pm$0.4  & $\cdots$  \\
H$_2$ & 1--0 S(7)     & 1.748   & 2.8$\pm$0.4 &  $\cdots$   	\\
H$_2$ & 1--0 S(3)     & 1.958   & 11$\pm$3     & 5$\pm$2$^*$  \\
H$_2$ & 1--0 S(2)     & 2.034   & 15.8$\pm$0.4 & 3$\pm$1     	 \\
H$_2$ & 2--1 S(3)     & 2.073   & 5.4$\pm$0.4  & $\cdots$       \\
H$_2$ & 1--0 S(1)     & 2.122   & 51.1$\pm$0.4 & 7.1$\pm$0.6 	\\
H$_2$ & 2--1 S(2)     & 2.154   & 4.1$\pm$0.4  & $\cdots$      \\
H$_2$ & 1--0 S(0)     & 2.223   & 19.2$\pm$0.5 & 1.5$\pm$0.5 	\\
H$_2$ & 2--1 S(1)     & 2.248   & 10.6$\pm$0.5 & $\cdots$    	\\
H$_2$ & 3--2 S(1)     & 2.386   & 7$\pm$2      & $\cdots$      \\
H$_2$ & 1--0 Q(1)     & 2.407   & 113$\pm$5    & 10$\pm$2    	\\
H$_2$ & 1--0 Q(2)     & 2.413   &  46$\pm$5    & 5$\pm$2$^*$ 	 \\
H$_2$ & 1--0 Q(3)     & 2.424   & 118$\pm$5    & 10$\pm$2    	 \\
H$_2$ & 1--0 Q(4)     & 2.437   &  42$\pm$5    & $\cdots$       \\
H$_2$ & 1--0(5)       & 2.455   & 71$\pm$10    & 8$\pm$2     	\\
H$_2$ & 1--0(6)       & 2.455   & 32$\pm$10    & $\cdots$    	\\
H$_2$ & 1--0(7)       & 2.455   & 69$\pm$20    & $\cdots$    	\\
\hline\\[-5pt]
\hline
\end{tabular}
\tablefoot{$^*$ S/N between 2 and 3.}
\end{table*}

\begin{table}
\caption{\object{Caswell OH 322.158+00.636} available photometry. \label{G322sed:tab}}

\begin{tabular}{ccc}

\hline\\[-5pt]

\multicolumn{1}{c}{$\lambda$}  &  \multicolumn{1}{c}{F$\pm \Delta$F}  & Data \\
\multicolumn{1}{c}{($\mu$m)}   &  \multicolumn{1}{c}{(Jy)}           &  source          \\

\hline\\[-5pt]
3.6 & 0.0072$\pm$0.0004  &  Spitzer/GLIMPSE \\ 
4.5 & 0.039$\pm$0.003  &  Spitzer/GLIMPSE \\ 
5.8 & 0.087$\pm$0.005   &  Spitzer/GLIMPSE \\ 
12 & 3.2$\pm$0.3    &  WISE            \\ 
22 & 74$\pm$1     &  WISE            \\ 
65 & 4000$\pm$500 & AKARI \\
70 & 3200$\pm$300 & Herschel/Hi-GAL  \\ 
90 & 2800$\pm$1000 & AKARI \\
140 & 5700$\pm$2000 & AKARI \\
160& 3370$\pm$300 & Herschel/Hi-GAL  \\ 
500& 300$\pm$100        & Herschel/Hi-GAL  \\ 
870& 3.8$\pm$0.4       & APEX/ATLASGAL    \\ 
\hline\\[-5pt]
\hline

\end{tabular}

\tablefoot{These data are used in the SED analysis shown in Fig.~\ref{G322SED:fig}.}

\end{table}

\subsection{IRAS 15394-5358}
\label{appendixI:sec}

IRAS 15394-5358 is part of a small cluster of YSOs, where CO~\citep[][]{urquhart} and CS~\citep[][]{bronfman} emissions as well as 
CH$_3$OH~\citep[][]{caswell} and H$_2$O~\citep[][]{scalise} masers are observed.
\citet{faundez} detected continuum emission at 1.2\,mm. By assuming a distance of 2.8\,kpc, they obtained
$L_{\rm bol}$=1.5$\times$10$^4$\,L$_{\sun}$ and $M_c$=8$\times$10$^3$\,M$_{\sun}$ from the SED analysis.
On the other hand, \citet{lumsden13} reported a distance of 1.8\,kpc and derived $L_{\rm bol}$=4$\times$10$^3$\,L$_{\sun}$.

\begin{figure*}
 \centering
   \includegraphics[width=9.1 cm]{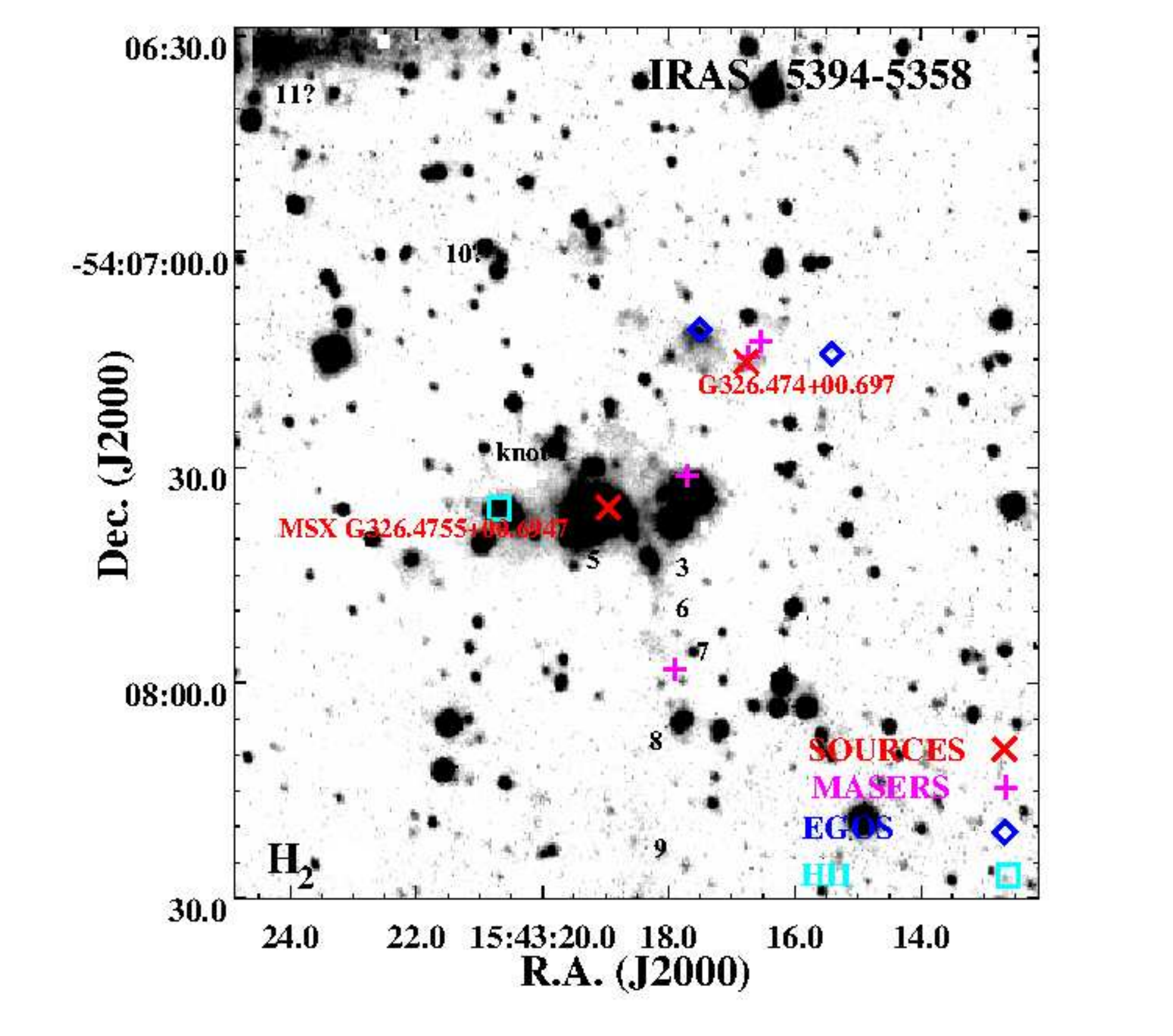} \includegraphics[width=9.1 cm]{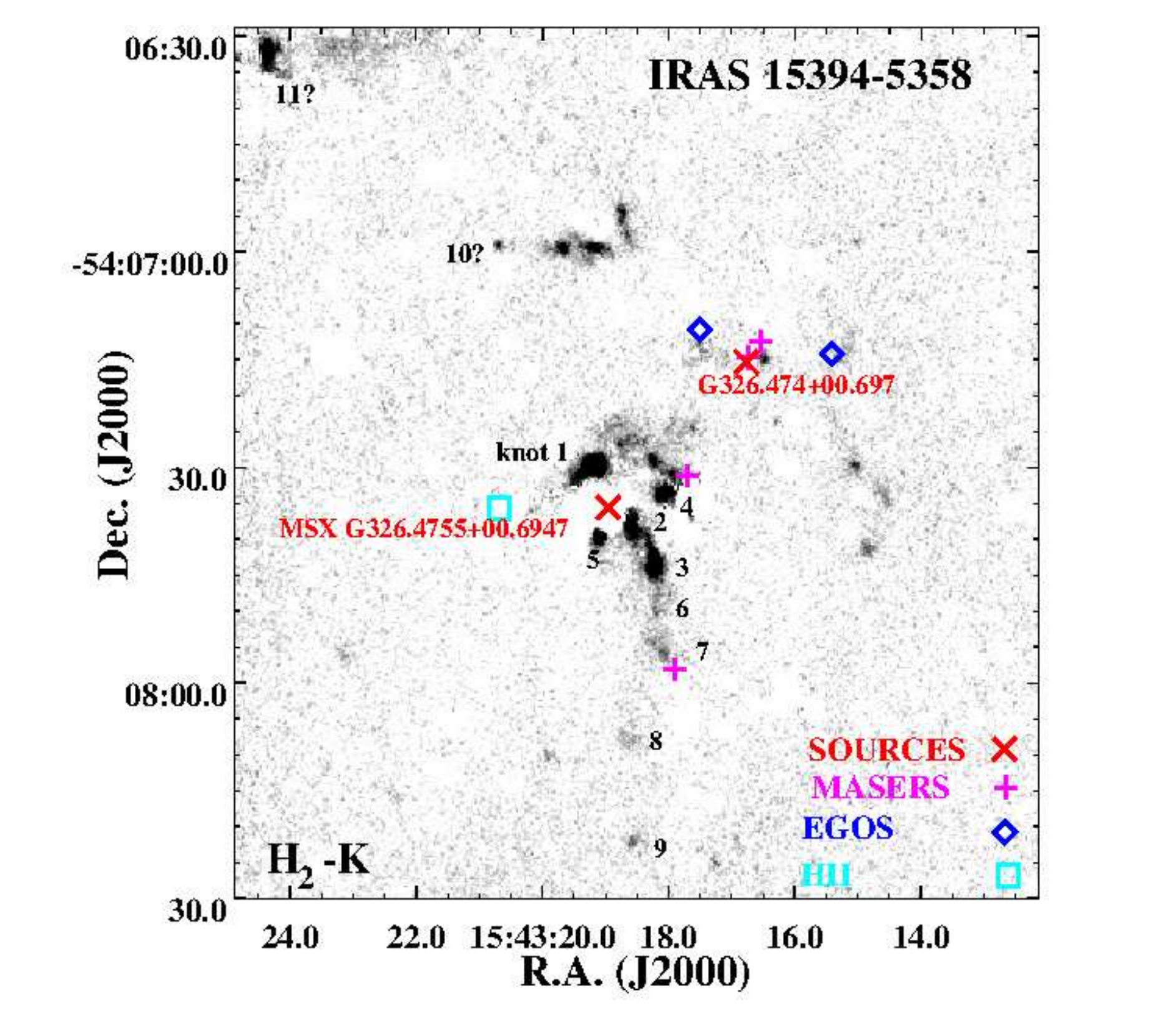}
   \caption{Same as in Figure~\ref{HSL2000ima:fig} but for the \object{IRAS15394-5358} flows.
   \label{G326ima:fig}}
\end{figure*}

Our continuum-subtracted H$_2$ image (Fig.~\ref{G326ima:fig}) shows two bright jets emanating from IRAS 15394-5358, that is surrounded by an extended nebulosity.
There is a relatively bright NIR source at the centre of the nebula (2MASSJ15431897-5407356), and an \ion{H}{ii} region is located 15$\arcsec$ 
eastwards~\citep{bronfman}.
In the $K$ band image, we detect a very faint source ($\alpha$=15:43:18.945, $\delta$=-54:07:37.84, J2000, not detected in 2MASS) $\sim$2$\arcsec$ southwards 
of 2MASSJ15431897-5407356. This might be the second component of the binary system that drives the two flows. The system is not resolved in the Spitzer 
GLIMPSE images, and we do not have any other photometric or colour information to prove the nature of this source.
The first precessing jet (precession angle $\sim$42$\degr$) is composed of knots 2, 3, 6, 7, 8, 9 (the blue-shifted lobe; Caratti o Garatti et al. 
in preparation) and knots 1 and, possibly,
10 and 11 (likely to be the red-shifted lobe).
It has a C shape and it is likely driven by the brightest NIR source (2MASSJ15431897-5407356). The P.A. of the jet is $\sim$207$\degr$. 
The second flow is likely driven by the faintest NIR companion and shows two knots (4 and 5), roughly aligned with the faint NIR source.
Its P.A. is $\sim$300$\degr$, if Knot\,4 is part of the blue-shifted lobe (because of its lowest visual extinction),
or it is $\sim$120$\degr$, if Knot\,4 belongs to the red-shifted lobe.
Our spectral analysis indicates the presence of both H$_2$ and [\ion{Fe}{ii}] emission along the first flow, whereas only H$_2$ emission
is detected along the second flow (see Table~\ref{spec_IRAS15394:tab}).
Finally, in our continuum-subtracted H$_2$ image (Fig.~\ref{G326ima:fig}), a third H$_2$ flow is detected. This flow is likely driven by another 
HMYSO, namely G326.474+0.697~\citep[][]{garay07}, which is detected at FIR wavelengths in both Spitzer and Herschel images.
The flow displays several knots extending NE-SW of the source position. EGO G326.48+0.70~\citep[][]{cyganowski} is associated with this third outflow.
The flow was not encompassed by our slits and it is not discussed further in this paper.

\begin{table*}
\caption{Observed emission lines in the \object{IRAS 15394-5358} jets. \label{spec_IRAS15394:tab}}
\begin{center}
\begin{tabular}{cccccccc}
\hline\\[-5pt]
Species & Term &  $\lambda$($\mu$m) & \multicolumn{5}{c}{$F\pm\Delta~F$(10$^{-15}$erg\,cm$^{-2}$\,s$^{-1}$)}\\
\hline\\[-5pt]
                 &    &       &    knot\,1    &  knot\,2      &   knot\,3     & knot\,4       &  knot\,5+YSO$^{**}$        \\
{[\ion{Fe}{ii}]}& $a^4\!D_{7/2}-a^4\!F_{9/2}$ & 1.644 &0.6$\pm$0.3$^*$& $\cdots$  & 2.6$\pm$0.6   &  $\cdots$   & $\cdots$   \\
H$_2$ & 1--0 S(9)     & 1.688 &0.8$\pm$0.3$^*$& $\cdots$      & $\cdots$      & $\cdots$      & 2.2$\pm$0.6   \\
H$_2$ & 1--0 S(7)     & 1.748 & 2.2$\pm$0.2   & 1.5$\pm$0.5   &  3.9$\pm$0.6  & 1.2$\pm$0.3   & 7.3$\pm$0.7    \\
H$_2$ & 1--0 S(6)     & 1.788 & 1.4$\pm$0.3   &1.0$\pm$0.5$^*$& 2.9$\pm$0.7   & $\cdots$      & 3.3$\pm$0.8   \\
H$_2$ & 1--0 S(3)     & 1.958 & 4$\pm$2$^*$   & $\cdots$      & 6$\pm$3$^*$   & $\cdots$      & 25$\pm$5    \\
H$_2$ & 1--0 S(2)     & 2.034 & 5.7$\pm$0.6   & 5.0$\pm$0.7   & 5.9$\pm$0.7   &  2.4$\pm$0.5  & 12.1$\pm$0.7   \\
H$_2$ & 2--1 S(3)     & 2.073 & 2.6$\pm$0.4   & 2.3$\pm$0.6   & 2.3$\pm$0.6   & $\cdots$      & 6.3$\pm$0.7   \\
H$_2$ & 1--0 S(1)     & 2.122 & 20.9$\pm$0.5  & 13.0$\pm$0.7  & 14.8$\pm$0.7  & 8.2$\pm$0.6   & 32.1$\pm$0.7   \\
H$_2$ & 2--1 S(2)     & 2.154 &1.5$\pm$0.5    & $\cdots$      & $\cdots$      & $\cdots$      & 2.2$\pm$0.7  \\
\ion{H}{i} & Br$\gamma$&2.166 & $\cdots$      & $\cdots$      & $\cdots$      & $\cdots$      & 2.8$\pm$0.7   \\
H$_2$ & 3--2 S(3)     & 2.201 & $\cdots$      & $\cdots$      & $\cdots$      & $\cdots$      & 1.9$\pm$0.7$^*$ \\
H$_2$ & 1--0 S(0)     & 2.223 & 6.2$\pm$0.6   & 5.2$\pm$0.7   & 4.4$\pm$0.7   &  2.7$\pm$0.7  & 7.4$\pm$0.7   \\
H$_2$ & 2--1 S(1)     & 2.248 & 3.7$\pm$0.6   & 3.6$\pm$0.7   & 2.2$\pm$0.7   & $\cdots$      & 4.5$\pm$0.8   \\
H$_2$ & 1--0 Q(1)     & 2.407 & 31$\pm$3      & 21$\pm$3      & 14$\pm$3      & 8$\pm$3$^*$   & 32$\pm$5      \\
H$_2$ & 1--0 Q(2)     & 2.413 & 13$\pm$3      &  8$\pm$3$^*$  &  $\cdots$     & $\cdots$      & 12$\pm$5$^*$       \\
H$_2$ & 1--0 Q(3)     & 2.424 & 32$\pm$4      & 22$\pm$3      & 13$\pm$3      &    12$\pm$3   & 38$\pm$5       \\
H$_2$ & 1--0 Q(4)     & 2.437 & 7$\pm$4$^*$   &  7$\pm$3$^*$  & $\cdots$      & $\cdots$      & 17$\pm$5    \\
H$_2$ & 1--0 Q(5)     & 2.455 & 22$\pm$5      & 17$\pm$5      & 12$\pm$4      & $\cdots$      & 30$\pm$10       \\
\hline\\[-5pt]
\hline
\end{tabular}
\tablefoot{$^*$ S/N between 2 and 3. $^{**}$ YSO = \object{2MASS J15431897-5407356}.}
\end{center}
\end{table*}
%

\subsection{IRAS 15450-5431}
\label{appendixJ:sec}

\begin{figure*}
 \centering
   \includegraphics[width=9.1 cm]{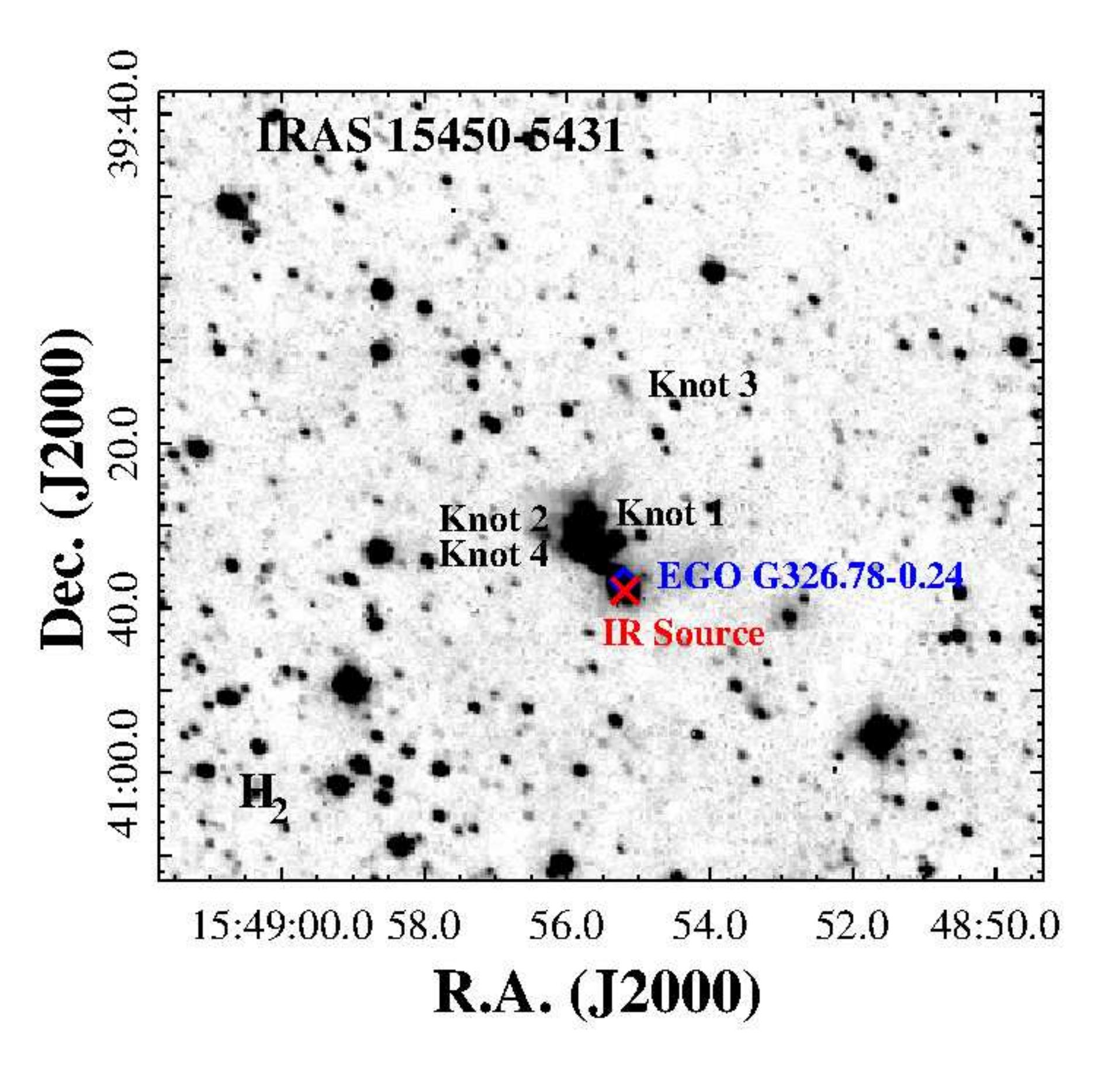} \includegraphics[width=9.1 cm]{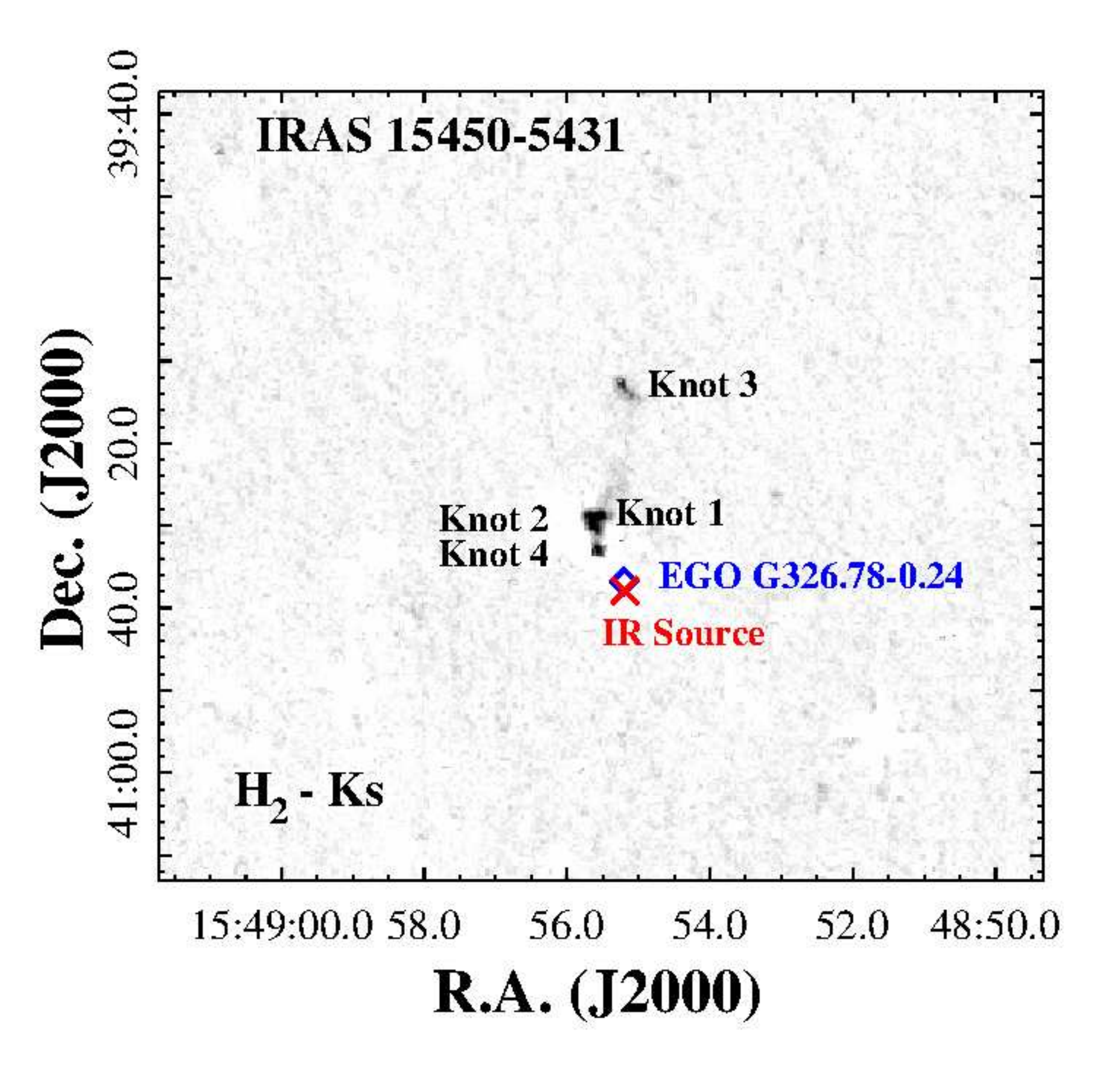}
   \caption{Same as in Figure~\ref{HSL2000ima:fig} but for the \object{IRAS 15450-5431} flow.
   \label{IRAS15450ima:fig}}
\end{figure*}

The location of IRAS 15450-5431 agrees with that of a bright GLIMPSE source (G326.7807-00.2409) in the MIR and with that of 2MASSJ15485523-5440375 in the NIR.
CS~\citep[][]{bronfman} and H$_2$O maser emission~\citep[][]{urquhart09} are reported close to the source location.
The RMS catalogue provides a distance of 3.9\,kpc and a bolometric luminosity of 9$\times$10$^3$\,L$_{\sun}$.
EGO G326.78-0.24~\citep[][]{cyganowski} is located close to the IR source.
Our $K$  band and H$_2$ images (see the H$_2$ image in Fig.~\ref{IRAS15450ima:fig}, left) show a cone-like nebulosity NE of the NIR source, likely depicting an outflow cavity.
A precessing jet (precession angle $\sim$33$\degr$) emerging from this outflow cavity is detected in the 
continuum-subtracted H$_2$ image (Fig.~\ref{IRAS15450ima:fig}, right).
From the position of Knot\,4 we derive a jet P.A. of $\sim$31$\degr$, and only one lobe, likely the blue-shifted one, is detected.
Only H$_2$ emission lines are detected along the flow in our spectra (see Table~\ref{spec_IRAS15450:tab}). 
The on-source spectrum shows a bright rising continuum with puzzling \ion{H}{i} lines from the Brackett series in absorption. 
Indeed there are examples in the literature where photospheric lines could be detected from the scattered light~\citep[e.g.][]{cohen86,neckel95}.

\begin{table*}
\caption{Observed emission lines in the \object{IRAS 15450-5431} jet. \label{spec_IRAS15450:tab}}
\begin{center}
\begin{tabular}{cccccccc}
\hline\\[-5pt]
Species & Term &  $\lambda$($\mu$m) & \multicolumn{5}{c}{$F\pm\Delta~F$(10$^{-15}$erg\,cm$^{-2}$\,s$^{-1}$)}\\
\hline\\[-5pt]
                 &    &       &    knot\,1    &  knot\,2      &   knot\,3     & knot\,4       & YSO$^{**}$        \\
\ion{H}{i} & Br14     & 1.588 & $\cdots$      & $\cdots$      & $\cdots$      & $\cdots$      &-2.1$\pm$0.7 \\		 
\ion{H}{i} & Br13     & 1.611 & $\cdots$      & $\cdots$      & $\cdots$      & $\cdots$      &-3.3$\pm$0.7 \\
\ion{H}{i} & Br12     & 1.641 & $\cdots$      & $\cdots$      & $\cdots$      & $\cdots$      &-3.3$\pm$0.7 \\		 		 
\ion{H}{i} & Br11     & 1.681 & $\cdots$      & $\cdots$      & $\cdots$      & $\cdots$      &-4.6$\pm$0.7 \\		 		 
H$_2$ & 1--0 S(9)     & 1.688 & 1.2$\pm$0.4   &1.0$\pm$0.4$^*$& $\cdots$      & $\cdots$      & $\cdots$    \\
H$_2$ & 1--0 S(8)     & 1.715 &1.1$\pm$0.4$^*$& $\cdots$      & $\cdots$      & $\cdots$      & $\cdots$      \\
\ion{H}{i} & Br10     & 1.737 & $\cdots$      & $\cdots$      & $\cdots$      & $\cdots$      &-6.9$\pm$0.7 \\		 		 
H$_2$ & 1--0 S(7)     & 1.748 & 4.9$\pm$0.4   & 3.2$\pm$0.4   & $\cdots$      & $\cdots$      & 3.0$\pm$0.5    \\
H$_2$ & 1--0 S(6)     & 1.788 & 2.8$\pm$0.4   & 2.3$\pm$0.4   & $\cdots$      & $\cdots$      & 2.1$\pm$0.5   \\
\ion{H}{i} & Br9      & 1.818 & $\cdots$      & $\cdots$      & $\cdots$      & $\cdots$      & -8$\pm$1  \\		 		 
H$_2$ & 2--1 S(8)     & 1.821 &2$\pm$1$^*$    & $\cdots$      & $\cdots$      & $\cdots$      & $\cdots$       \\
H$_2$ & 1--0 S(5)     & 1.836 & 7$\pm$2       & $\cdots$      & $\cdots$      & $\cdots$      & 1.0$\pm$0.3   \\
H$_2$ & 1--0 S(4)     & 1.892 & 7$\pm$2       & $\cdots$      & $\cdots$      & $\cdots$      & 1.0$\pm$0.3   \\
H$_2$ & 1--0 S(3)     & 1.958 & 19$\pm$2      & 16$\pm$2      & $\cdots$      & $\cdots$      & 14$\pm$3    \\
H$_2$ & 1--0 S(2)     & 2.034 & 8.8$\pm$0.5   & 6.2$\pm$0.5   & 1.0$\pm$0.3   & $\cdots$      & 4.8$\pm$0.5   \\
H$_2$ & 2--1 S(3)     & 2.073 & 3.4$\pm$0.5   & 2.2$\pm$0.5   &0.7$\pm$0.3$^*$& $\cdots$      & 2.1$\pm$0.5   \\
H$_2$ & 1--0 S(1)     & 2.122 & 25.0$\pm$0.5  & 20.8$\pm$0.5  & 2.6$\pm$0.4   & 2.2$\pm$0.4   & 32.2$\pm$0.5   \\
H$_2$ & 3--2 S(4)     & 2.124 &1.2$\pm$0.5$^*$& $\cdots$      & $\cdots$      & $\cdots$      & $\cdots$      \\
H$_2$ & 2--1 S(2)     & 2.154 &1.2$\pm$0.5$^*$&1.0$\pm$0.5$^*$& $\cdots$      & $\cdots$      & 1.0$\pm$0.5$^*$  \\
\ion{H}{i} & Br$\gamma$&2.166 & $\cdots$      & $\cdots$      & $\cdots$      & $\cdots$      &-3.5$\pm$0.4   \\
H$_2$ & 3--2 S(3)     & 2.201 &1.0$\pm$0.5$^*$&1.0$\pm$0.5$^*$& $\cdots$      & $\cdots$      & $\cdots$      \\
H$_2$ & 1--0 S(0)     & 2.223 & 6.0$\pm$0.6   & 5.5$\pm$0.6   & $\cdots$      & $\cdots$      & 3.9$\pm$0.5   \\
H$_2$ & 2--1 S(1)     & 2.248 & 3.0$\pm$0.6   & 2.1$\pm$0.6   & 3.0$\pm$0.6   & $\cdots$      & 1.7$\pm$0.5   \\
H$_2$ & 1--0 Q(1)     & 2.407 & 24$\pm$2      & 16$\pm$2      & 3$\pm$1       & 2$\pm$1$^*$   & 14$\pm$2      \\
H$_2$ & 1--0 Q(2)     & 2.413 & 10$\pm$2      &  7$\pm$2      &  2$\pm$1$^*$  & $\cdots$      & 7$\pm$2       \\
H$_2$ & 1--0 Q(3)     & 2.424 & 23$\pm$2      & 17$\pm$2      & 4$\pm$1       & 2$\pm$1$^*$   & 16$\pm$2       \\
H$_2$ & 1--0 Q(4)     & 2.437 & 9$\pm$3      &  6$\pm$3$^*$   & $\cdots$      & $\cdots$      & 6$\pm$2    \\
H$_2$ & 1--0 Q(5)     & 2.455 & 19$\pm$3      & 14$\pm$3      & $\cdots$      & $\cdots$      & 14$\pm$3       \\
H$_2$ & 1--0 Q(7)     & 2.500 & 12$\pm$4      & 12$\pm$4      & $\cdots$      & $\cdots$      & 10$\pm$4$^*$      \\ 
\hline\\[-5pt]
\hline
\end{tabular}
\tablefoot{$^*$ S/N between 2 and 3.$^{**}$ YSO = \object{2MASSJ 15485523-5440375}}
\end{center}
\end{table*}

\subsection{IRAS 16122-5047}
\label{appendixK:sec}

\begin{figure*}
 \centering
   \includegraphics[width=9.1 cm]{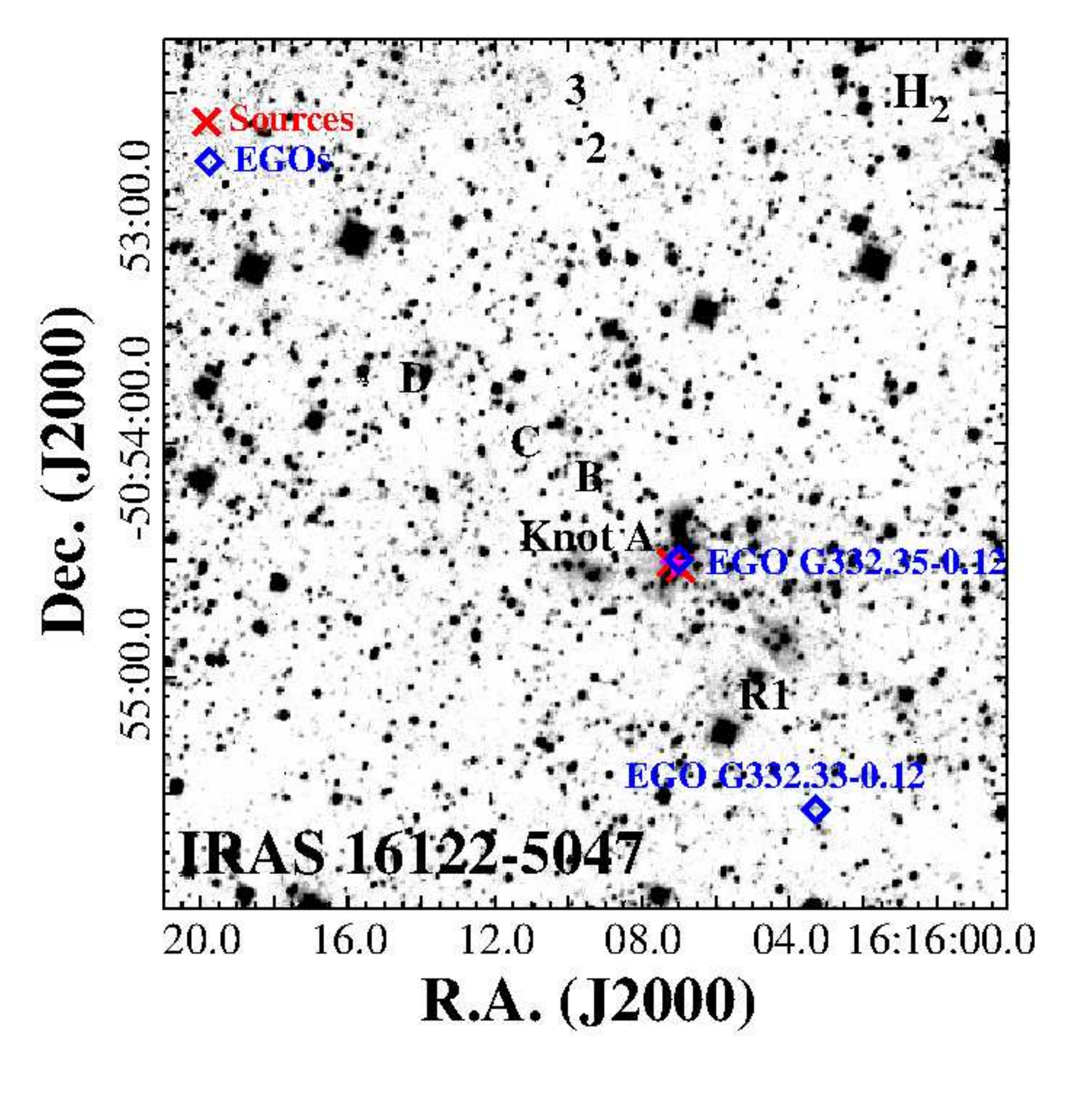} \includegraphics[width=9.1 cm]{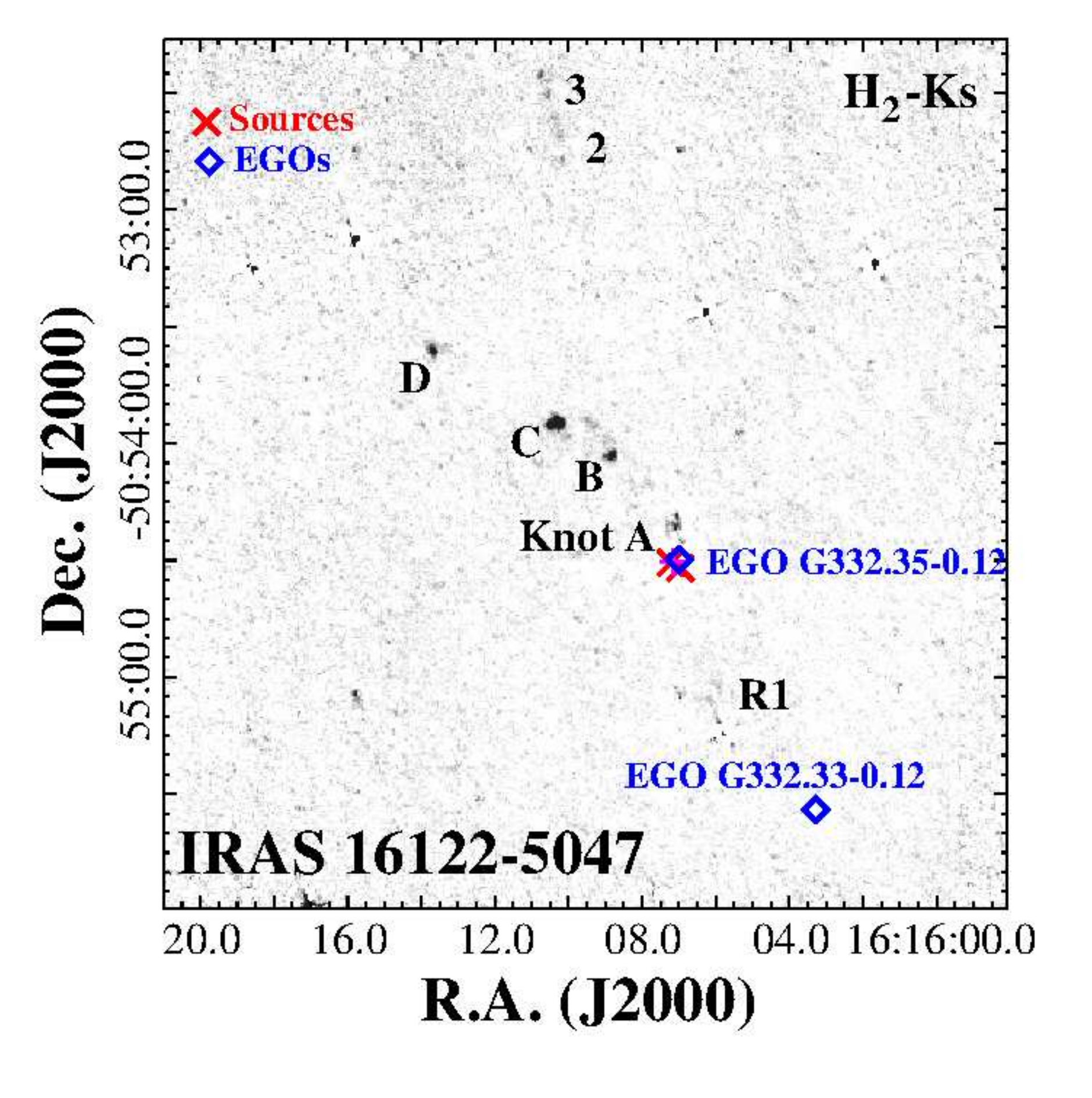}
   \caption{Same as in Figure~\ref{HSL2000ima:fig} but for the \object{IRAS 16122-5047} flows.
   \label{G332ima:fig}}
\end{figure*}

\begin{figure}
 \centering
   \includegraphics[width=8 cm]{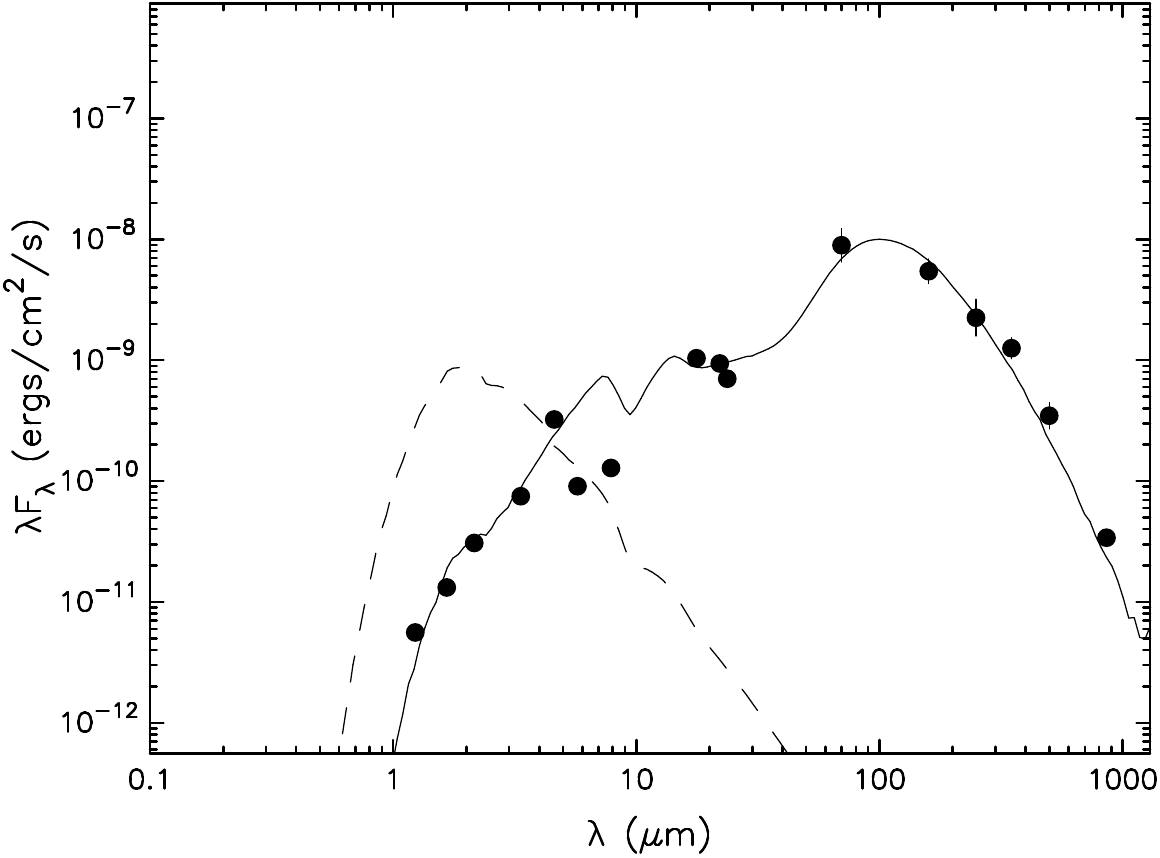} 
   \caption{Spectral energy distribution (SED) of \object{IRAS 16122-5047} constructed with all photometric data available from the literature,
   namely from 3.6\,$\mu$m to 870\,$\mu$m, and assuming a distance of 3.1\,kpc~\citep[][]{urquhart13}. 
\label{G332SED:fig}}
\end{figure}

IRAS 16122-5047 is not well studied.
According to the Spitzer GLIMPSE images, IRAS 16122-5047 seems to be part of a small cluster, composed of two sources, well detected at 24\,$\mu$m,
and a few more fainter objects. The location of the brightest IR source in our field (GLIMPSE G332.3526-00.1154), the only one detected at 70\,$\mu$m with MIPS, agrees with
that of 2MASSJ16160689-5054274.
OH maser emission is detected on source~\citep[][]{caswell} and CH$_3$OH maser emission is detected towards EGO G332.35-0.12~\citep[][]{chen}.
Two EGOs are reported by \citet{cyganowski}, namely EGO G332.35-0.12 close to the NIR source, and EGO G332.33-0.12, about 74$\arcsec$ SW of the 2MASS source.
Both $K$ band and H$_2$ images (see H$_2$ image in Fig.~\ref{G332ima:fig}, left) show some nebulosity around the 2MASS source, that might delineate 
an outflow cavity (towards NNE), 
although the geometry does not appear so clear in our images. The continuum-subtracted H$_2$ image shows two different flows, 
which appear to emanate from the 2MASSJ16160689-5054274 position.
The first flow, which includes knots A, B, C, and D, has a curved shape. Its P.A., measured from Knot\,A position, is $\sim$15$\degr$. 
The second flow includes Knots 1, 2, 3, and it has a straight geometry with a P.A. of $\sim$14$\degr$. There is an additional knot (R1), 
which is located in the opposite direction of both flows. Therefore it is not clear as to which flow it belongs. 
It is likely to be part of the red-shifted lobe of one of the two flows. Only H$_2$ lines are detected in the spectra of the knots (see Table~\ref{spec_IRAS16122:tab}).
Collected photometry is reported in Table~\ref{G332sed:tab}.
Assuming a distance of 3.1\,kpc~\citep[][]{urquhart13}, our SED analysis provides a bolometric luminosity of $\sim$5$\times$10$^3$\,L$_{\sun}$ 
(see Fig.~\ref{G332SED:fig}).

\begin{table*}
\caption{Observed emission lines in the \object{IRAS 16122-5047} jets. \label{spec_IRAS16122:tab}}
\begin{center}
\begin{tabular}{ccccccccccc}
\hline\\[-5pt]
Species & Term &  $\lambda$($\mu$m) & \multicolumn{8}{c}{$F\pm\Delta~F$(10$^{-15}$erg\,cm$^{-2}$\,s$^{-1}$)}\\
\hline\\[-5pt]
                 &    &       &    knot\,A    &  knot\,B      &   knot\,C     & knot\,D       & knot\,2       & knot\,3       & knot\,R1      & YSO$^{**}$ + knot\,1      \\
H$_2$ & 1--0 S(9)     & 1.688 & $\cdots$      & $\cdots$      & 1.2$\pm$0.3   & $\cdots$      & $\cdots$      &$\cdots$       & $\cdots$      & $\cdots$      \\
H$_2$ & 1--0 S(8)     & 1.715 & $\cdots$      & $\cdots$      &0.8$\pm$0.3$^*$& $\cdots$      & $\cdots$      & $\cdots$      & $\cdots$      & $\cdots$      \\
H$_2$ & 1--0 S(7)     & 1.748 & 2.3$\pm$0.4   & 2.8$\pm$0.4   & 3.9$\pm$0.3   & 1.4$\pm$0.3   & $\cdots$      &  $\cdots$     & $\cdots$      & 1.4$\pm$0.3   \\
H$_2$ & 1--0 S(6)     & 1.788 & 1.4$\pm$0.4   & 2.1$\pm$0.5   & 2.4$\pm$0.3   & 1.0$\pm$0.3   &$\cdots$       & $\cdots$      & $\cdots$      & 1.0$\pm$0.3   \\
H$_2$ & 1--0 S(5)     & 1.836 & $\cdots$      & $\cdots$      & 2$\pm$1$^*$   & $\cdots$      & $\cdots$      & $\cdots$      & $\cdots$      & $\cdots$      \\
H$_2$ & 1--0 S(4)     & 1.892 & $\cdots$      & $\cdots$      & 2$\pm$1$^*$   & $\cdots$      & $\cdots$      & $\cdots$      & $\cdots$      & $\cdots$      \\
H$_2$ & 1--0 S(3)     & 1.958 & 3$\pm$1       &  5$\pm$2$^*$  & 6$\pm$1       & $\cdots$      & 2$\pm$1$^*$   & 3$\pm$1       & $\cdots$      & $\cdots$      \\
H$_2$ & 1--0 S(2)     & 2.034 & 4.5$\pm$0.6   & 4.7$\pm$0.5   & 2.3$\pm$0.3   & 3.0$\pm$0.5   & 1.2$\pm$0.4   & 2.2$\pm$0.4   & 2.1$\pm$0.5   & 4.2$\pm$0.5   \\
H$_2$ & 3--2 S(5)     & 2.066 & $\cdots$      &  $\cdots$     &0.9$\pm$0.4$^*$& $\cdots$      & $\cdots$      & $\cdots$      & $\cdots$      & $\cdots$      \\
H$_2$ & 2--1 S(3)     & 2.073 & 1.7$\pm$0.5   & 2.1$\pm$0.6   & 2.3$\pm$0.4   &1.3$\pm$0.5$^*$&  0.9$\pm$0.3  &0.9$\pm$0.4$^*$&1.1$\pm$0.5$^*$& $\cdots$      \\
H$_2$ & 1--0 S(1)     & 2.122 & 16.0$\pm$0.5  & 11.8$\pm$0.6  & 29.0$\pm$0.5  & 8.8$\pm$0.5   & 5.0$\pm$0.5   & 8.1$\pm$0.5   & 4.0$\pm$0.5   & 13.0$\pm$0.5   \\
H$_2$ & 3--2 S(4)     & 2.124 & 1.5$\pm$0.5   & $\cdots$      & 1.4$\pm$0.5$^*$& $\cdots$      & $\cdots$      & $\cdots$      & $\cdots$      & $\cdots$      \\
H$_2$ & 2--1 S(2)     & 2.154 &1.0$\pm$0.5$^*$& $\cdots$      & 2.3$\pm$0.5   & $\cdots$      & $\cdots$      & $\cdots$      & $\cdots$      & $\cdots$      \\
H$_2$ & 3--2 S(3)     & 2.201 & $\cdots$      & $\cdots$      & 1.5$\pm$0.5   & $\cdots$      & $\cdots$      & $\cdots$      & $\cdots$      & $\cdots$      \\
H$_2$ & 1--0 S(0)     & 2.223 & 4.2$\pm$0.6   & 3.5$\pm$0.6   & 8.1$\pm$0.6   & 2.5$\pm$0.6   & $\cdots$      & 2.2$\pm$0.6   & 1.9$\pm$0.6   & 4.4$\pm$0.6   \\
H$_2$ & 2--1 S(1)     & 2.248 & 2.8$\pm$0.6   & 1.9$\pm$0.6   & 1.6$\pm$0.4   &1.6$\pm$0.6$^*$&1.7$\pm$0.6$^*$&1.6$\pm$0.6$^*$& $\cdots$      & 3.6$\pm$0.6   \\
H$_2$ & 1--0 Q(1)     & 2.407 & 26$\pm$2      & 9$\pm$2$^*$   & 41$\pm$2      & 14$\pm$2      & 7$\pm$2       & 12$\pm$2      & 7$\pm$2       & 33$\pm$2      \\
H$_2$ & 1--0 Q(2)     & 2.413 & 13$\pm$2      & 5$\pm$2$^*$   & 18$\pm$2      & 5$\pm$2$^*$   & $\cdots$      & $\cdots$      & $\cdots$      & 18$\pm$2       \\
H$_2$ & 1--0 Q(3)     & 2.424 & 27$\pm$2      & 11$\pm$2      & 42$\pm$2      & 16$\pm$2      & 9$\pm$2       & 14$\pm$2      & 5$\pm$2$^*$   & 40$\pm$2       \\
H$_2$ & 1--0 Q(4)     & 2.437 & 10$\pm$3      & $\cdots$      & 16$\pm$3      & 7$\pm$2       & $\cdots$      & $\cdots$      & $\cdots$      & 20$\pm$5   \\
H$_2$ & 1--0 Q(5)     & 2.455 & 17$\pm$6       & 8$\pm$3$^*$  & 24$\pm$6      & 11$\pm$3      & $\cdots$      & $\cdots$      & $\cdots$      & 34$\pm$6       \\
H$_2$ & 1--0 Q(7)     & 2.500 & 20$\pm$6      & $\cdots$      & $\cdots$      & 12$\pm$4      & $\cdots$      & $\cdots$      & $\cdots$      & 37$\pm$6       \\
\hline\\[-5pt]
\hline
\end{tabular}
\tablefoot{$^*$ S/N between 2 and 3. $^{**}$ YSO = \object{2MASSJ 16160689-5054274}}
\end{center}
\end{table*}

\begin{table}
\caption{\object{IRAS 16122-5047} available photometry. \label{G332sed:tab}}

\begin{tabular}{ccc}

\hline\\[-5pt]

\multicolumn{1}{c}{$\lambda$}  &  \multicolumn{1}{c}{F$\pm \Delta$F}  &  Data \\
\multicolumn{1}{c}{($\mu$m)}   &  \multicolumn{1}{c}{(Jy)}           &  source          \\

\hline\\[-5pt]
1.24 &  0.0023$\pm$0.0002      &  2MASS  \\
1.66 &   0.0073$\pm$0.0005      &  2MASS  \\
2.16 &    0.0222$\pm$0.0004     &  2MASS  \\
3.4 & 0.083$\pm$0.002  &  WISE            \\ 
4.6 & 0.49$\pm$0.01   &  WISE            \\ 
5.8 & 0.176$\pm$0.009   &  Spitzer/GLIMPSE \\ 
8.0 & 0.33$\pm$0.02   &  Spitzer/GLIMPSE \\ 
22 & 6.9$\pm$0.1     &  WISE            \\ 
24 & 5.6$\pm$0.2  &  Spitzer/MIPSGAL \\ 
70 & 220$\pm$70 & Herschel/Hi-GAL  \\ 
160& 300$\pm$70 & Herschel/Hi-GAL  \\ 
250& 200$\pm$70     & Herschel/Hi-GAL  \\ 
350& 150$\pm$30       & Herschel/Hi-GAL  \\ 
500& 60$\pm$15        & Herschel/Hi-GAL  \\ 
870& 9.8$\pm$0.2       & APEX/ATLASGAL    \\ 
\hline\\[-5pt]
\hline

\end{tabular}

\tablefoot{These data are used in the SED analysis shown in Fig.~\ref{G332SED:fig}.}

\end{table}

\subsection{IRAS 16547-4247}
\label{appendixL:sec}

IRAS 16547-4247 or G343.126-0.062 is an isolated massive and dense core located at a distance of 2.9\,kpc~\citep[][]{garay03,rodriguez08}.
Its bolometric luminosity estimates range from 6.2$\times$10$^4$\,L$_{\sun}$~\citep[][]{garay03,faundez}
to 6.6$\times$10$^4$\,L$_{\sun}$~\citep[][who assume a distance of 2.8\,kpc to the source]{lumsden13}.
These values are consistent with an O9 spectral type for a single ZAMS star. The core mass derived from the dust emission at 1.2\,mm
is 1.4$\times$10$^3$\,M$_{\sun}$~\citep[][]{faundez}. The source is not detected in our NIR images or in the Spitzer/IRAC images,
although it is visible as a single object in the Spitzer/MIPS images at 24\,$\mu$m.
At high spatial resolution, Submillimeter Array (SMA) observations of the dust continuum at 1.3\,mm~\citep[][]{franco} trace a more complex
structure, indicating that there might be at least two objects.
H$_2$O masers were detected towards the source position~\citep[][]{batchelor}. SMA observations at a higher spatial resolution show
that water maser emission is present in several distinct parts of the region~\citep[][]{franco}. 
In particular, a compact structure located at the disc position and perpendicular to the radio jet 
shows a velocity gradient possibly arising from Keplerian motion, which would imply a mass of $\sim$30\,M$_{\sun}$ for the central star(s).
The source drives a thermal radio jet, extending 0.14\,pc with a P.A. of $\sim$163$\degr$~\citep[][]{garay03,rodriguez05}.
A parsec-scale H$_2$ flow (MHO 1900--1902) well aligned with the radio jet was observed by \citet{brooks}.
More recently, \citet{garay07} observed a well collimated CO bipolar outflow roughly oriented in
the north–south direction~\citep[the blue-shifted lobe lies southwards; see e.g.][]{garay07a}, close to the plane of the sky (i=84$\degr \pm$2$\degr$).
EGO emission (EGO G343.12-0.06) has been also detected close to the source~\citep[][]{cyganowski}. As for the majority of the EGOs, such emission
can be due to scattered continuum in the HMYSO outflow cavities~\citep{takami12}.

\begin{figure*}
 \centering
   \includegraphics[width=9.1 cm]{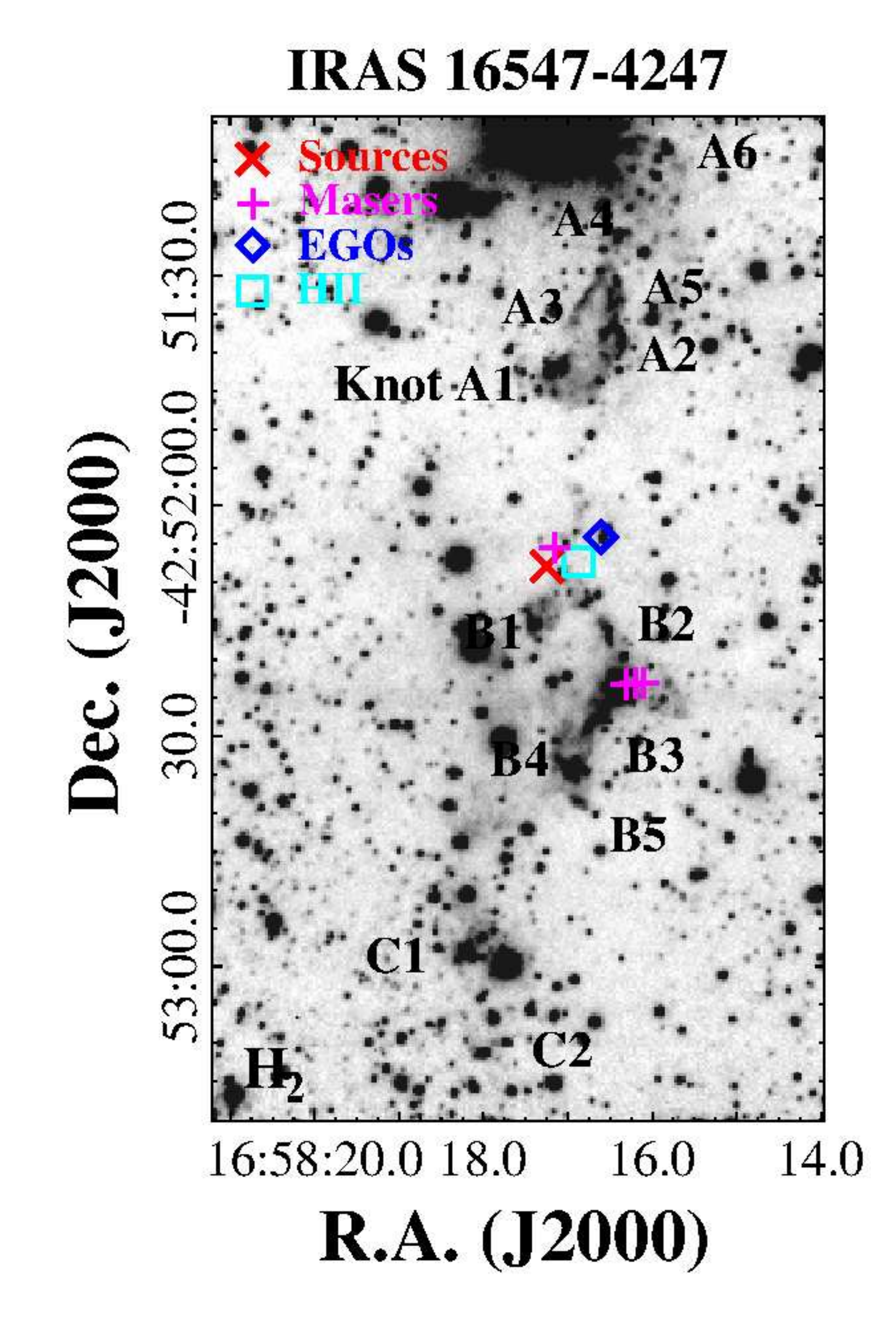} \includegraphics[width=9.1 cm]{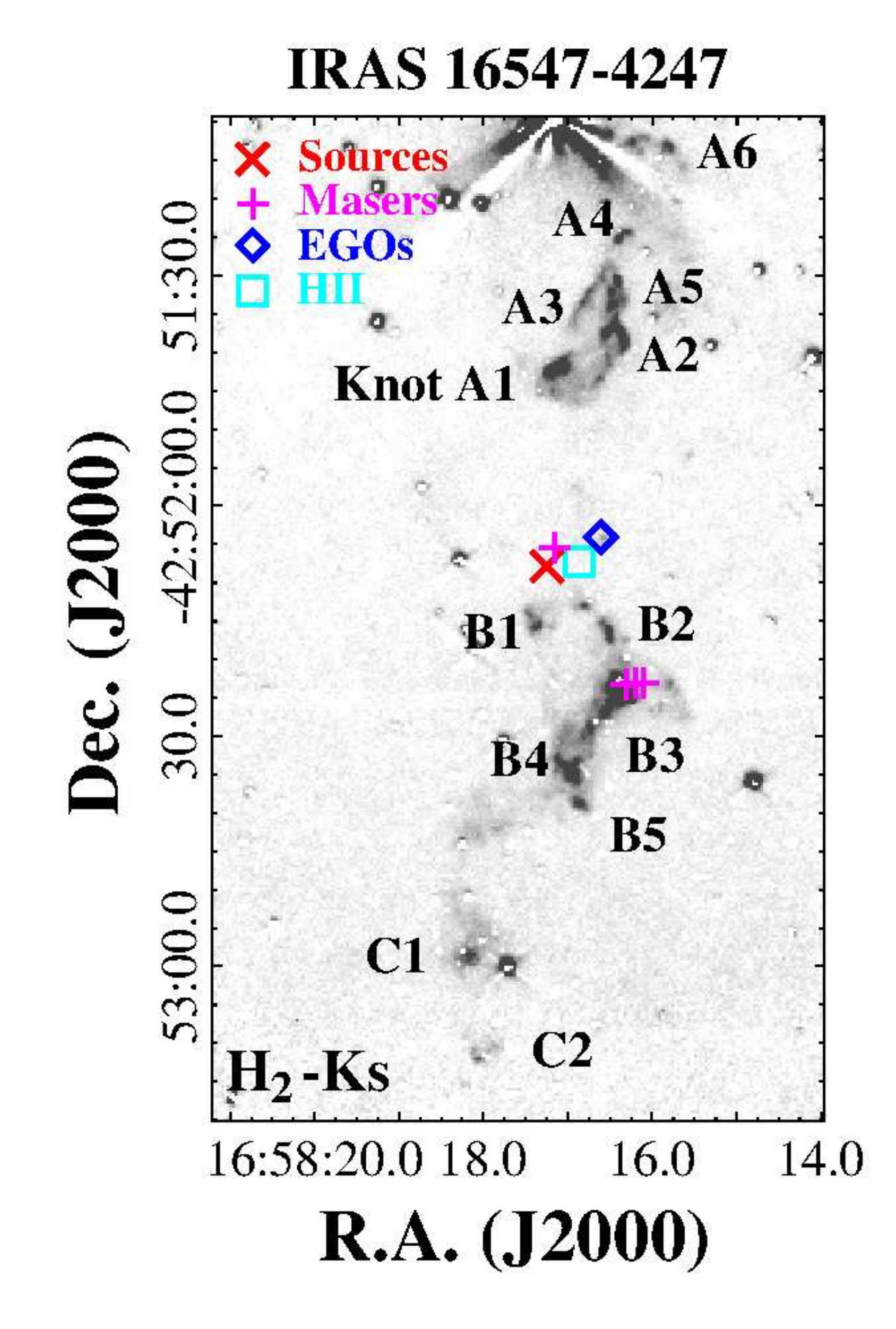}
   \caption{Same as in Figure~\ref{HSL2000ima:fig} but for the \object{IRAS 16547-4247} flow.
   \label{IRAS16547ima:fig}}
\end{figure*}

Figure~\ref{IRAS16547ima:fig} shows the H$_2$ and continuum-subtracted H$_2$ images from ISAAC. These data were originally presented by \citet{brooks}.
We reduced and analysed the data again, finding similar results (e.g. calibrated fluxes are identical, within the errors).
We also keep the knot nomenclature same as in \citet{brooks}. In addition, we detect another knot (named A6), which is also visible in their Figure\,1.
This is likely the terminal bow-shock of the red-shifted lobe.
Because of to the high visual extinction, no H$_2$ emission is detected near the source position, therefore the current jet P.A. is not measurable.
We assume a P.A. of 177$\degr \pm$4$\degr$, according to the observed radio jet emission~\citep[][]{franco}. 
The extension of the whole jet, including both lobes, is about 3.4\,pc at a distance of 2.9\,kpc. The jet shows one of the
largest precession angle in our sample, i.e. 57$\degr$ computed from knots B1 and B2), and the jet axis P.A. drifts from $\sim$161$\degr$ to 218$\degr$.
Our sepectroscopic analysis reveals both H$_2$ and [\ion{Fe}{ii}] emission lines in the analysed knots (see Table~\ref{spec_IRAS16547:tab}).

\begin{table*}
\caption{Observed emission lines in the \object{IRAS 16547-4247} jet. \label{spec_IRAS16547:tab}}
\begin{center}
\begin{tabular}{ccccccccc}
\hline\\[-5pt]
Species & Term &  $\lambda$($\mu$m) & \multicolumn{6}{c}{$F\pm\Delta~F$(10$^{-15}$erg\,cm$^{-2}$\,s$^{-1}$)}\\
\hline\\[-5pt]
                 &                             &       &    knot\,A1   &  knot\,A2     &   knot\,A4    & knot\,A5      & knot\,B1      & knot\,B4\\
{[\ion{Fe}{ii}]} & $a^4\!D_{3/2}-a^4\!F_{7/2}$ & 1.600 & $\cdots$      &0.5$\pm$0.2$^*$& $\cdots$      & $\cdots$      & $\cdots$      &0.4$\pm$0.2$^*$\\ 
{[\ion{Fe}{ii}]} & $a^4\!D_{7/2}-a^4\!F_{9/2}$ & 1.644 & 1.3$\pm$0.3   & 2.9$\pm$0.2   & 1.4$\pm$0.2   & 2.6$\pm$0.2   & $\cdots$      & 1.2$\pm$0.2   \\
{[\ion{Fe}{ii}]} & $a^4\!D_{5/2}-a^4\!F_{7/2}$ & 1.677 & $\cdots$      &0.5$\pm$0.2$^*$& $\cdots$      & $\cdots$      & $\cdots$      &0.5$\pm$0.2$^*$\\ 
H$_2$ & 1--0 S(9) 	                       & 1.688 & $\cdots$      & 0.7$\pm$0.2   & $\cdots$      & $\cdots$      & $\cdots$      &0.4$\pm$0.2$^*$\\
H$_2$ & 1--0 S(8) 	                       & 1.715 & $\cdots$      &0.5$\pm$0.2$^*$& $\cdots$      & $\cdots$      & $\cdots$      & $\cdots$      \\
H$_2$ & 1--0 S(7) 	                       & 1.748 & $\cdots$      & 2.3$\pm$0.2   & 1.0$\pm$0.2   & 1.8$\pm$0.2   & 0.4$\pm$0.1   &  $\cdots$     \\
H$_2$ & 1--0 S(6) 	                       & 1.788 & $\cdots$      & 1.4$\pm$0.3   &0.6$\pm$0.3$^*$& 1.0$\pm$0.3   &0.2$\pm$0.1$^*$& 1.3$\pm$0.2   \\
H$_2$ & 1--0 S(5) 	                       & 1.836 & $\cdots$      &1.6$\pm$0.8$^*$& $\cdots$      & $\cdots$      & $\cdots$      & 1.0$\pm$0.3   \\
H$_2$ & 1--0 S(3) 		               & 1.958 & $\cdots$      & 4.0$\pm$1.0   & $\cdots$      & 4.0$\pm$1.0   & 1.2$\pm$0.4   & $\cdots$      \\
H$_2$ & 2--1 S(4)           	               & 2.004 & $\cdots$      & 1.0$\pm$0.3   & $\cdots$      &0.6$\pm$0.3$^*$& $\cdots$      &   5$\pm$1     \\
H$_2$ & 1--0 S(2)           	               & 2.034 & 1.0$\pm$0.3   & 6.6$\pm$0.3   & 2.3$\pm$0.3   & 4.2$\pm$0.3   & 0.8$\pm$0.2   & 3.0$\pm$0.3   \\
H$_2$ & 3--2 S(5)           	               & 2.066 & $\cdots$      &0.6$\pm$0.3$^*$& $\cdots$      & $\cdots$      & $\cdots$      &0.6$\pm$0.3$^*$\\
H$_2$ & 2--1 S(3)           	               & 2.073 & $\cdots$      & 2.0$\pm$0.3   & 0.9$\pm$0.3   & 1.4$\pm$0.3   & $\cdots$      & 0.9$\pm$0.3   \\
H$_2$ & 1--0 S(1)           	               & 2.122 & 2.4$\pm$0.3   & 21.1$\pm$0.3  & 6.4$\pm$0.4   & 14.6$\pm$0.3  & 2.0$\pm$0.3   & 9.4$\pm$0.3   \\
H$_2$ & 2--1 S(2)           	               & 2.154 & $\cdots$      & 1.5$\pm$0.4   &1.1$\pm$0.4$^*$& $\cdots$      & $\cdots$      & $\cdots$      \\
H$_2$ & 3--2 S(3)                              & 2.201 & $\cdots$      & 0.9$\pm$0.3   & $\cdots$      & $\cdots$      & $\cdots$      & $\cdots$      \\
H$_2$ & 1--0 S(0)           	               & 2.223 &0.8$\pm$0.3$^*$& 7.0$\pm$0.4   & 2.3$\pm$0.4   & 4.1$\pm$0.4   &0.8$\pm$0.3$^*$& 2.7$\pm$0.3   \\
H$_2$ & 2--1 S(1)           	               & 2.248 & $\cdots$      & 3.2$\pm$0.4   & 1.6$\pm$0.4   & 1.4$\pm$0.4   & $\cdots$      & 1.0$\pm$0.3   \\
H$_2$ & 3--2 S(2)           	               & 2.286 & $\cdots$      & 1.7$\pm$0.5   & $\cdots$      & $\cdots$      & $\cdots$      & $\cdots$      \\
H$_2$ & 4--3 S(3)           	               & 2.344 & $\cdots$      & 1.9$\pm$0.5   & $\cdots$      & $\cdots$      & $\cdots$      & $\cdots$      \\
H$_2$ & 2--1 S(0)           	               & 2.355 & $\cdots$      & 1.8$\pm$0.5   & $\cdots$      & $\cdots$      & $\cdots$      & $\cdots$      \\
H$_2$ & 3--2 S(1)           	               & 2.386 & $\cdots$      & 1.8$\pm$0.5   & $\cdots$      & $\cdots$      & $\cdots$      & $\cdots$      \\ 
H$_2$ & 1--0 Q(1)           	               & 2.407 & 3$\pm$1       & 34$\pm$2      & 11$\pm$2      & 23$\pm$2      & 2$\pm$1$^*$   & 10$\pm$1      \\
H$_2$ & 1--0 Q(2)           	               & 2.413 & $\cdots$      & 15$\pm$2      &  5$\pm$2$^*$   & 12$\pm$2      & $\cdots$      & 4$\pm$1       \\
H$_2$ & 1--0 Q(3)           	               & 2.424 &2$\pm$1$^*$    & 33$\pm$2      & 10$\pm$2      & 24$\pm$2      & 2$\pm$1$^*$   & 9$\pm$1       \\
H$_2$ & 1--0 Q(4)           	               & 2.437 & $\cdots$      & 12$\pm$3      &  4$\pm$2$^*$   & 9$\pm$2       & $\cdots$      & 4$\pm$2$^*$   \\
H$_2$ & 1--0 Q(5)           	               & 2.455 & $\cdots$      & 20$\pm$5      &  8$\pm$3$^*$   & 12$\pm$4      & $\cdots$      & 7$\pm$2       \\
\hline\\[-5pt]
\hline
\end{tabular}
\tablefoot{$^*$ S/N between 2 and 3.}
\end{center}
\end{table*}
\clearpage 

\subsection{BGPS G014.849-00.992}
\label{appendixM:sec}

\begin{figure*}
 \centering
   \includegraphics[width=9.1 cm]{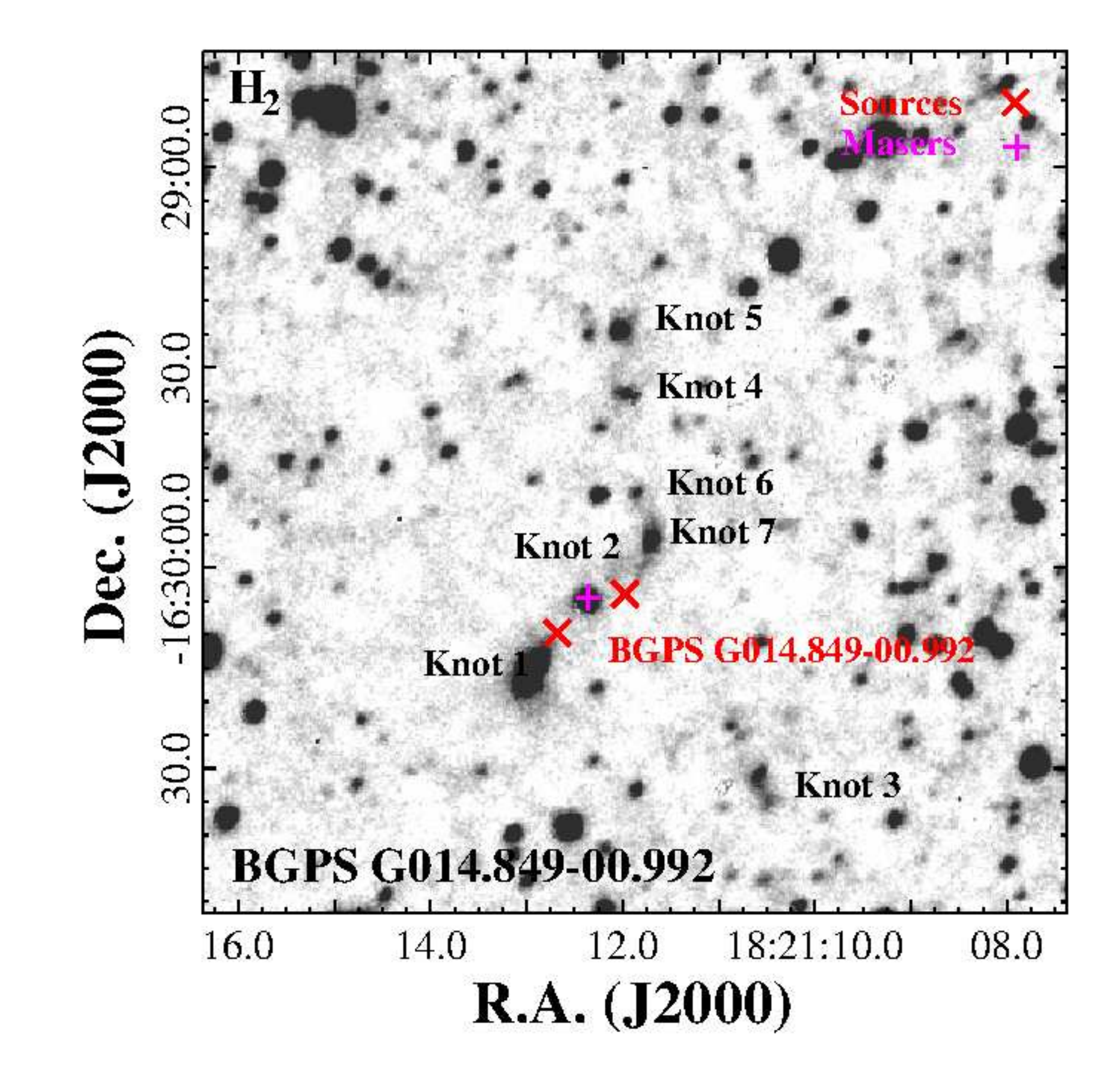} \includegraphics[width=9.1 cm]{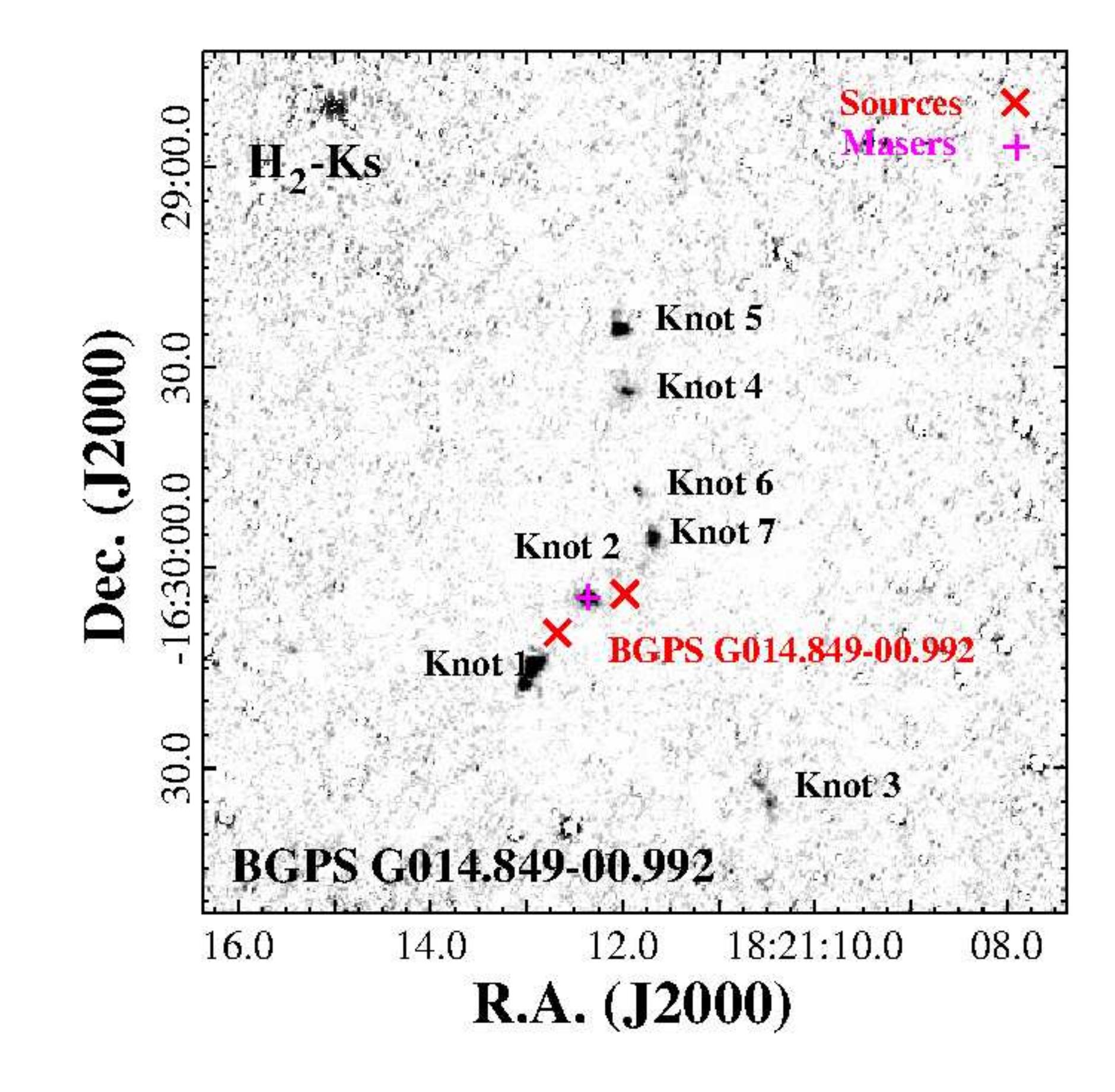}
   \caption{Same as in Figure~\ref{HSL2000ima:fig} but for the \object{BGPS G014.849-00.992} flow.
   \label{BS2ima:fig}}
\end{figure*}

\begin{figure}
 \centering
   \includegraphics[width=8 cm]{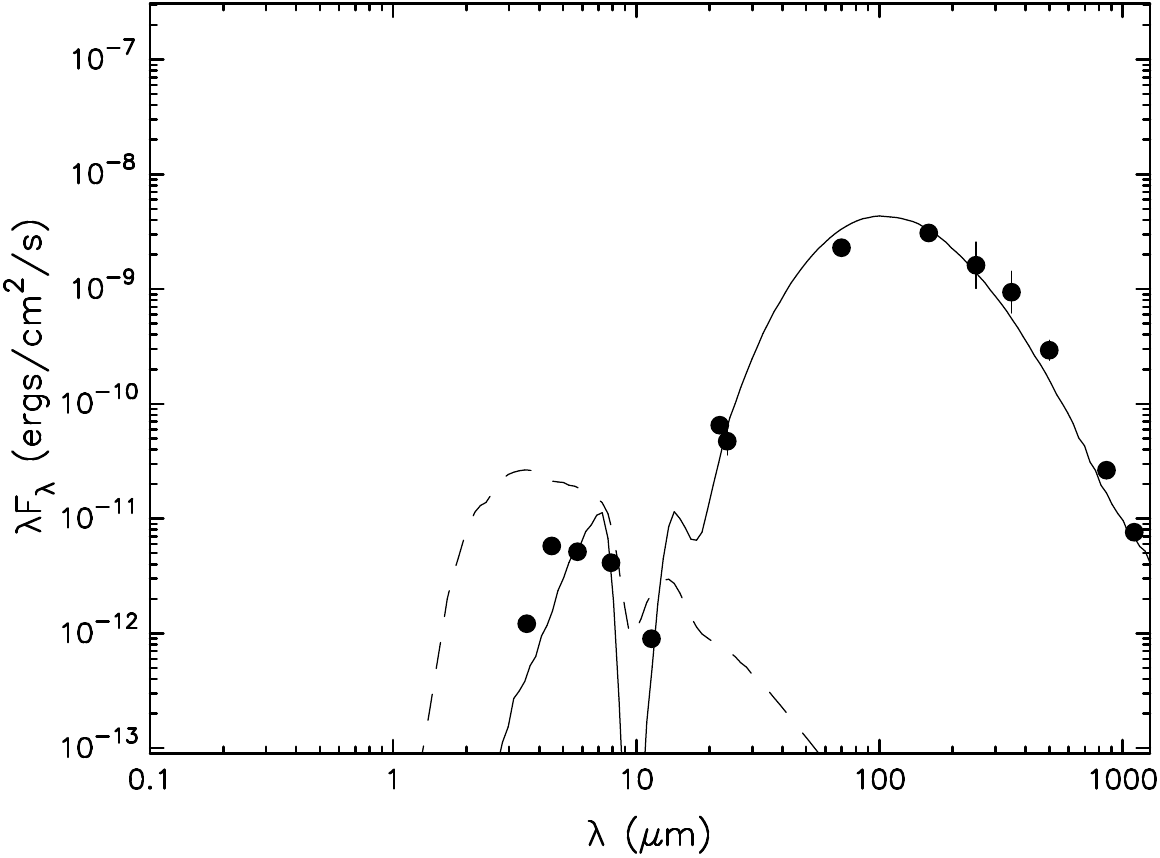} 
   \caption{Spectral energy distribution (SED) of \object{BGPS G014.849-00.992} constructed with all photometric data available from the literature,
   namely from 3.6\,$\mu$m to 1.1\,mm, and assuming a distance of 2.5\,kpc to the source~\citep[][]{lim}. 
\label{G14SED:fig}}
\end{figure}

BGPS G014.849-00.992 ($\alpha$=18:21:12.922, $\delta$=-16:30:10.2, J2000) was first reported in the Bolocam Galactic Plane Survey (BGPS) 
catalogue~\citep[][]{rosolowsky}. A second source (G014.8516-00.9890), located $\sim$12$\arcsec$ NW of BGPS G014.849-00.992, is also detected
in the Spitzer GLIMPSE images at 24\,$\mu$m (see Fig~\ref{BS2ima:fig}). This source is not spatially resolved in the Spitzer/MIPS images at 70\,$\mu$m.
In Herschel PACS images at 70\,$\mu$m, the two sources are partially resolved, and BGPS G014.849-00.992 is the brightest source at 70\,$\mu$m whereas the second
source is barely detected, indicating that BGPS G014.849-00.992 is the dominant source at FIR wavelengths.
\citet{lim} report the detection of strong multiple H$_2$O maser emissions along the flow as well as CH$_3$OH maser emission 
close to the location of the BGPS source ($\sim$10$\arcsec$).
None of the sources is detected in our NIR images. Two flows are likely detected in our continuum-subtracted H$_2$ image (Fig.~\ref{BS2ima:fig}, right). 
The first flow is driven by BGPS G014.849-00.992, located close to Knot 1, and it is composed of knots 1, 2 and, possibly, 7. 
Its P.A. is $\sim$324$\degr$ or $\sim$144$\degr$, depending on whether or not Knot\,2 belongs to the blue-shifted lobe.
There is a second curved flow, composed of knots 4, 5, 6, and, possibly, 7. 
It is slightly misaligned with respect to the second IR source, therefore the latter might not be the driving source. It is worth noting, however, 
that no additional IR sources are detected in the FoV. Our NIR spectra show only H$_2$ emission along the flows (see Table~\ref{spec_BS2:tab}).
Collected photometry is reported in Table~\ref{G14sed:tab}. Figure~\ref{G14SED:fig} shows the SED analysis of BGPS G014.849-00.992.
By assuming a distance of 2.5\,kpc~(Csengeri et al. in prep.), best fitting model indicates an $L_{\rm bol}$ value 
of 1.3$\times$10$^3$\,L$_{\sun}$ and a central mass of $\sim$8\,M$_{\sun}$, namely at the edge of the HMYSO regime.

\begin{table*}
\caption{Observed emission lines in the \object{BGPS G014.849-00.992} jet. \label{spec_BS2:tab}}
\begin{center}
\begin{tabular}{cccccccc}
\hline\\[-5pt]
Species & Term &  $\lambda$($\mu$m) & \multicolumn{5}{c}{$F\pm\Delta~F$(10$^{-15}$erg\,cm$^{-2}$\,s$^{-1}$)}\\
\hline\\[-5pt]
                 &                             &       &    knot\,1   &  knot\,2  & knot\,3 & knot\,4        & knot\,5	 \\
H$_2$ & 1--0 S(9) 	                       & 1.688 & $\cdots$      & $\cdots$      & $\cdots$      & $\cdots$       &0.4$\pm$0.2$^*$ \\
H$_2$ & 1--0 S(8) 	                       & 1.715 & $\cdots$      &$\cdots$       & $\cdots$      & $\cdots$       &0.4$\pm$0.2$^*$ \\
H$_2$ & 1--0 S(7) 	                       & 1.748 & $\cdots$      & 1.6$\pm$0.5   & 2.6$\pm$0.4   & 0.4$\pm$0.2$^*$& 1.6$\pm$0.2	  \\
H$_2$ & 1--0 S(6) 	                       & 1.788 & $\cdots$      & $\cdots$      & 2.1$\pm$0.4   &  $\cdots$      & 1.0$\pm$0.3	  \\
H$_2$ & 1--0 S(3) 		               & 1.958 & $\cdots$      & 11$\pm$3      & 14$\pm$4      & 1.2$\pm$0.4    & 4.0$\pm$1.0	  \\
H$_2$ & 2--1 S(4)           	               & 2.004 & $\cdots$      & $\cdots$      & $\cdots$      & $\cdots$       &0.6$\pm$0.2	 \\
H$_2$ & 1--0 S(2)           	               & 2.034 & 3.6$\pm$0.6   & 5.1$\pm$0.6   & 4.7$\pm$0.6   & $\cdots$       & 3.9$\pm$0.3	 \\
H$_2$ & 2--1 S(3)           	               & 2.073 & $\cdots$      & 2.2$\pm$0.6   & 2.1$\pm$0.6   & 1.0$\pm$0.3    & 1.9$\pm$0.3	 \\
H$_2$ & 1--0 S(1)           	               & 2.122 & 9.6$\pm$0.6   & 10.8$\pm$0.6  & 11.0$\pm$0.6  & 2.8$\pm$0.4    & 12.2$\pm$0.3   \\
H$_2$ & 1--0 S(0)           	               & 2.223 & 2.7$\pm$0.8   & 3.3$\pm$0.6   & 3$\pm$0.6     & $\cdots$    & 3.8$\pm$0.6	\\
H$_2$ & 2--1 S(1)           	               & 2.248 & $\cdots$      & 1.9$\pm$0.6   & 2.0$\pm$0.6   & $\cdots$       & 2.6$\pm$0.6	\\
H$_2$ & 1--0 Q(1)           	               & 2.407 & 15$\pm$3      & 15$\pm$3      & 8$\pm$2       &  6$\pm$2       & 19$\pm$3	\\
H$_2$ & 1--0 Q(2)           	               & 2.413 & 10$\pm$3      & 7$\pm$3$^*$   &  6$\pm$2      & $\cdots$       & 10$\pm$3	 \\
H$_2$ & 1--0 Q(3)           	               & 2.424 & 20$\pm$3      & 13$\pm$3      & 10$\pm$3      & 5$\pm$2$^*$    & 21$\pm$3	 \\
H$_2$ & 1--0 Q(4)           	               & 2.437 & $\cdots$      & 6$\pm$3$^*$   &  $\cdots$     & $\cdots$       & 10$\pm$3	 \\
H$_2$ & 1--0 Q(5)           	               & 2.455 & $\cdots$      & 8$\pm$4$^*$   &  $\cdots$     & $\cdots$       & 11$\pm$4	  \\

\hline\\[-5pt]
\hline
\end{tabular}
\tablefoot{$^*$ S/N between 2 and 3.}
\end{center}
\end{table*}

\begin{table}
\caption{\object{BGPS G014.849-00.992} available photometry. \label{G14sed:tab}}

\begin{tabular}{ccc}

\hline \\[-5pt]

\multicolumn{1}{c}{$\lambda$}  &  \multicolumn{1}{c}{F$\pm \Delta$F}  &  Data \\
\multicolumn{1}{c}{($\mu$m)}   &  \multicolumn{1}{c}{(Jy)}           &  source         \\

\hline\\[-5pt]
3.6 & 0.0014$\pm$0.0003  &  Spitzer/GLIMPSE \\ 
4.5 & 0.0086$\pm$0.0009  &  Spitzer/GLIMPSE \\ 
5.8 & 0.01$\pm$0.0008   &  Spitzer/GLIMPSE \\ 
8.0 & 0.011$\pm$0.0009   &  Spitzer/GLIMPSE \\ 
12 & 0.0032$\pm$0.0003    &  WISE            \\ 
22 & 0.48$\pm$0.05     &  WISE            \\ 
24 & 0.37$\pm$0.08  &  Spitzer/MIPSGAL \\ 
70 & 54$\pm$3 & Herschel/Hi-GAL  \\ 
160& 166$\pm$20 & Herschel/Hi-GAL  \\ 
250& 150$\pm$50     & Herschel/Hi-GAL  \\ 
350& 120$\pm$50       & Herschel/Hi-GAL  \\ 
500& 50$\pm$10        & Herschel/Hi-GAL  \\ 
870& 7.6$\pm$0.3       & APEX/ATLASGAL    \\ 
1100 & 2.9$\pm$0.4    & CSO/BGPS         \\ 
\hline\\[-5pt]
\hline
\end{tabular}

\tablefoot{These data are used in the SED analysis shown in Fig.~\ref{G14SED:fig}.}

\end{table}

\subsection{GLIMPSE G035.0393-00.4735}
\label{appendixN:sec}

\begin{figure*}
 \centering
   \includegraphics[width=9.1 cm]{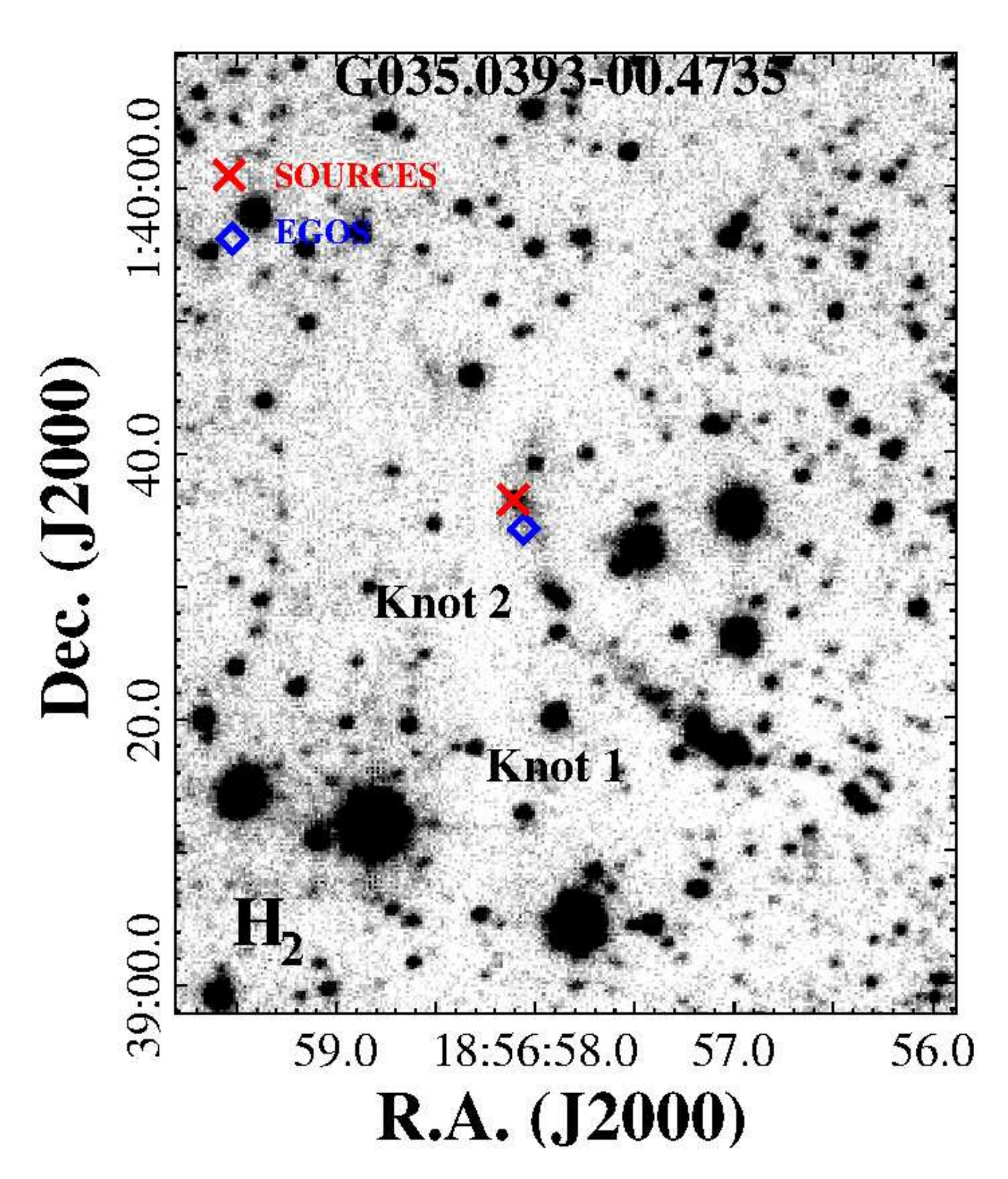} \includegraphics[width=9.1 cm]{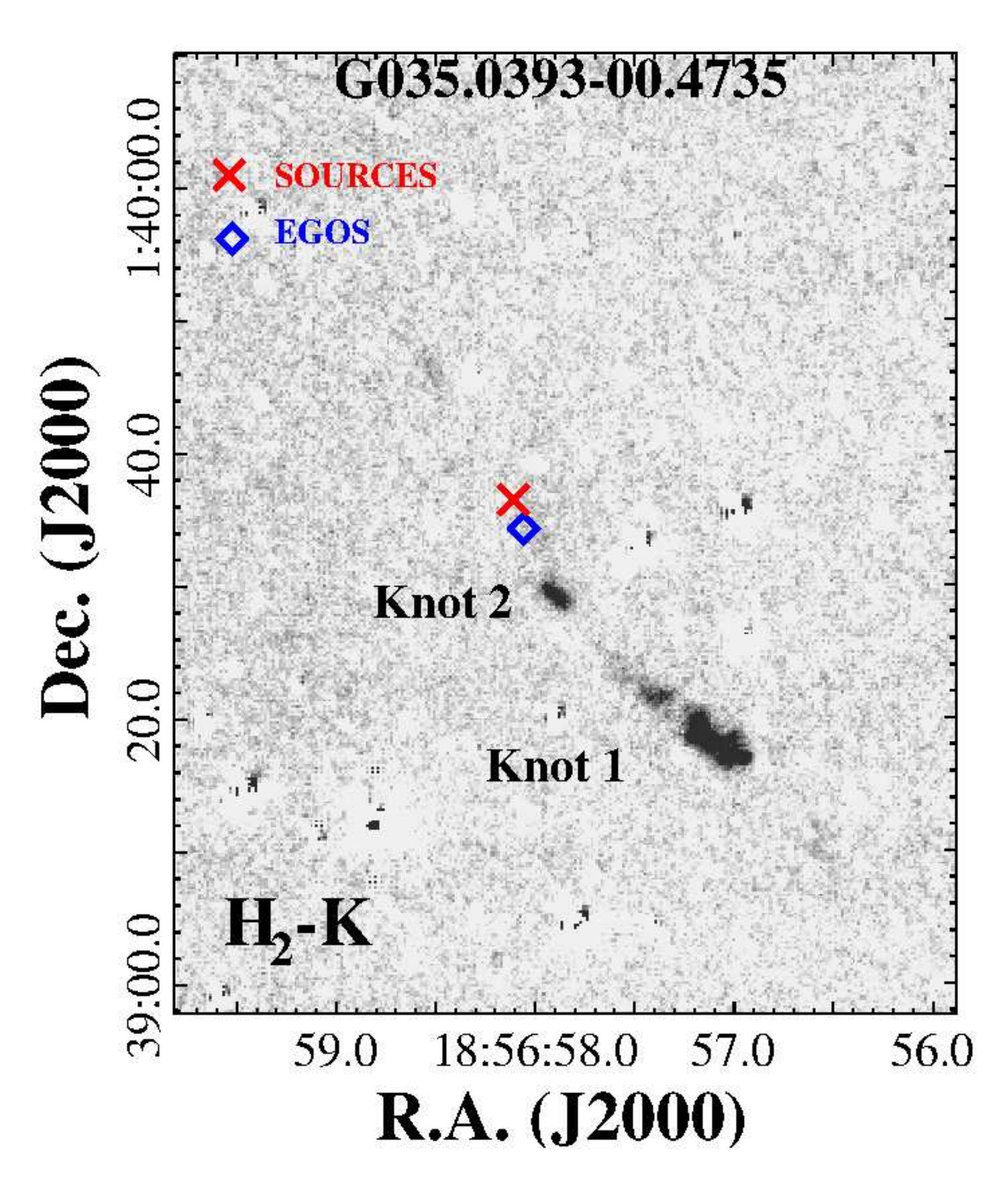}
   \caption{Same as in Figure~\ref{HSL2000ima:fig} but for the \object{GLIMPSE G035.0393-00.4735} flow.
   \label{BS17ima:fig}}
\end{figure*}

\begin{figure}
 \centering
   \includegraphics[width=8 cm]{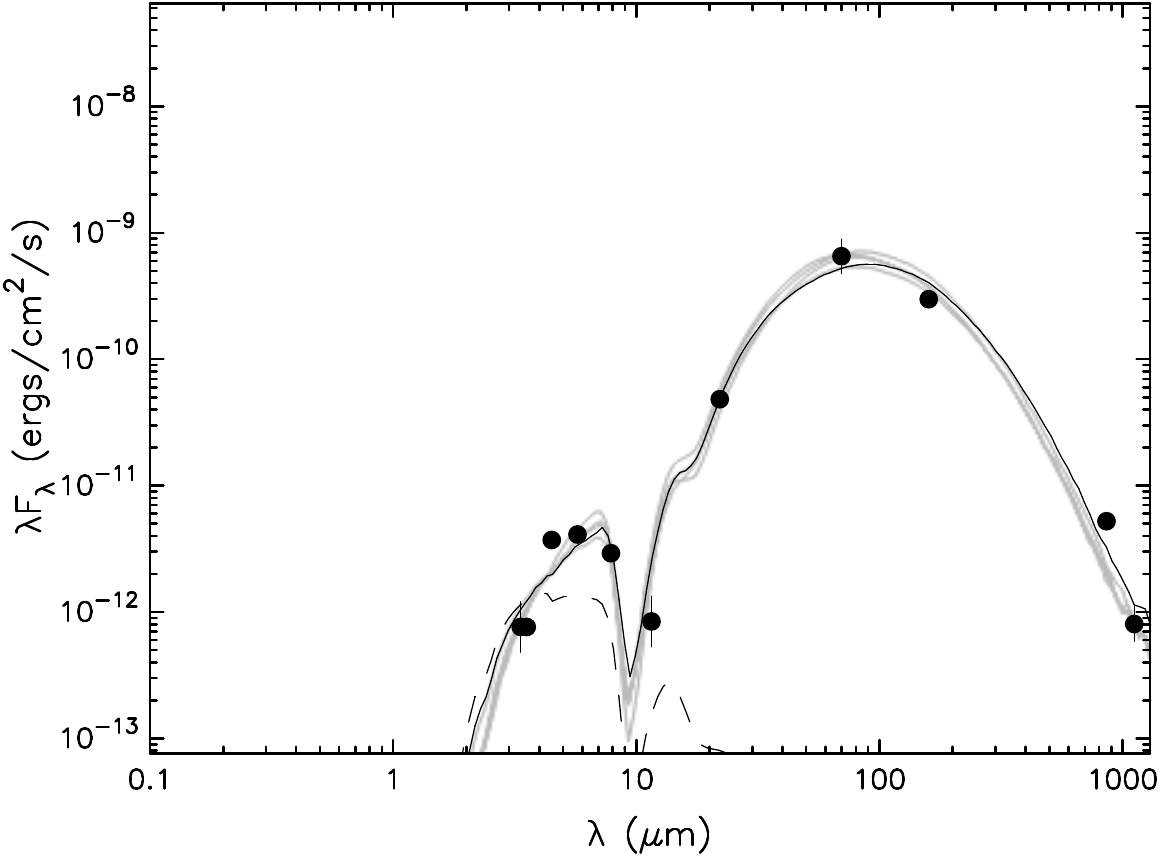} 
   \caption{Spectral energy distribution (SED) of \object{GLIMPSE G035.0393-00.4735} constructed with all photometric data available from the literature,
   namely from 3.6\,$\mu$m to 1.1\,mm, and assuming a distance of 2.5\,kpc to the source~\citep[][]{lim}. 
\label{G35SED:fig}}
\end{figure}

GLIMPSE G035.0393-00.4735 is an IR source, observed from mid-IR wavelengths longward.
It belongs to the IR dark cloud \object{SDC G35.041-0.471}~\citep[][]{peretto}.
EGO emission (EGO G035.04-0.47) was detected close to the source position~\citep[][]{cyganowski}. 
\citet{lee12} detect H$_2$ emission associated with the IR source, named \object{MHO 2429}.
The H$_2$ emission is precessing and shows two lobes (see their Figure~14), a bright one (MHO 2429 A), possibly the blue-shifted one, and
a faint one (MHO 2429 B), possibly the red-shifted one. In our continuum-subtracted H$_2$ image (Fig.~\ref{BS17ima:fig}, right), we detect
two bright knots (1 and 2) in the lobe coincident with MHO 2429 A, and a third knot, with S/N$<$3, at the location of MHO 2429 B.
As for other sources in our sample, the jet emission from one of the two flow lobes is likely hampered by the high visual extinction.
The jet P.A. is $\sim$205$\degr$ or $\sim$25$\degr$, depending on whether or not Knot 2 belongs to the blue- or red-shifted lobe, 
and it has a precession angle of 22$\degr$. 
The spectra of the observed knots show only H$_2$ emission (see Table~\ref{spec_BS17:tab}).
Collected photometry is reported in Table~\ref{G35sed:tab}. By assuming a distance of 3.4\,kpc, 
our SED analysis in Figure~\ref{G35SED:fig} provides us with an $L_{\rm bol}$ value 
of 2--4$\times$10$^2$\,L$_{\sun}$ and a central mass of $\sim$5\,M$_{\sun}$, suggesting that GLIMPSE G035.0393-00.4735 
is an intermediate-mass YSO and not 
an HMYSO.

\begin{table}
\caption{Observed emission lines in the \object{GLIMPSE G035.0393-00.4735} jet. \label{spec_BS17:tab}}
\begin{tabular}{ccccc}
\hline\\[-5pt]
Species & Term &  $\lambda$($\mu$m) & \multicolumn{2}{c}{$F\pm\Delta~F$(10$^{-15}$erg\,cm$^{-2}$\,s$^{-1}$)}\\
\hline\\[-5pt]
                 &                             &       &    knot\,1   &  knot\,2\\
H$_2$ & 1--0 S(7) 	                       & 1.748 & 1.8$\pm$0.6      &  $\cdots$      \\
H$_2$ & 1--0 S(3) 		               & 1.958 & 8$\pm$2      &  $\cdots$        \\
H$_2$ & 1--0 S(2)           	               & 2.034 & 3.8$\pm$0.5   & 1.0$\pm$0.5$^*$     \\
H$_2$ & 2--1 S(3)           	               & 2.073 & $\cdots$      &  $\cdots$   \\
H$_2$ & 1--0 S(1)           	               & 2.122 & 8.3$\pm$0.5   & 3.6$\pm$0.5  \\
H$_2$ & 1--0 S(0)           	               & 2.223 & 2.6$\pm$0.5    & 1.7$\pm$0.5     \\
H$_2$ & 2--1 S(1)           	               & 2.248 & 1.9$\pm$0.5      &  $\cdots$    \\
H$_2$ & 1--0 Q(1)           	               & 2.407 & 11$\pm$3       & 9$\pm$3           \\
H$_2$ & 1--0 Q(2)           	               & 2.413 & 6$\cdots$3$^*$      &  $\cdots$            \\
H$_2$ & 1--0 Q(3)           	               & 2.424 & 9$\pm$3    & 9$\pm$3         \\
\hline\\[-5pt]
\hline
\end{tabular}
\tablefoot{$^*$ S/N between 2 and 3.}
\end{table}

\begin{table}
\caption{GLIMPSE G035.0393-00.4735 available photometry. \label{G35sed:tab}}

\begin{tabular}{ccc}

\hline\\[-5pt]

\multicolumn{1}{c}{$\lambda$}  &  \multicolumn{1}{c}{F$\pm \Delta$F}  &  Data \\
\multicolumn{1}{c}{($\mu$m)}   &  \multicolumn{1}{c}{(Jy)}           &  source          \\

\hline\\[-5pt]
3.4 & 0.0008$\pm$0.0004  &  WISE            \\ 
3.6 & 0.0009$\pm$0.0002  &  Spitzer/GLIMPSE \\ 
4.5 & 0.0056$\pm$0.0005  &  Spitzer/GLIMPSE \\ 
5.8 & 0.008$\pm$0.0007   &  Spitzer/GLIMPSE \\ 
8.0 & 0.008$\pm$0.001   &  Spitzer/GLIMPSE \\ 
12 & 0.003$\pm$0.001    &  WISE            \\ 
22 & 0.36$\pm$0.03     &  WISE            \\ 
70 & 16$\pm$1 & Herschel/Hi-GAL  \\ 
160& 16$\pm$5 & Herschel/Hi-GAL  \\ 
870& 1.5$\pm$0.2       & APEX/ATLASGAL    \\ 
1100 & 0.3$\pm$0.1    & CSO/BGPS         \\ 
\hline\\[-5pt]
\hline

\end{tabular}

\tablefoot{These data are used in the SED analysis shown in Fig.~\ref{G35SED:fig}.}

\end{table}

\subsection{G35.2N}
\label{appendixO:sec}

G35.20-0.74\,N (G35.2N), also known as IRAS 18556+0136, is a massive star forming region at a distance of 2.2\,kpc~\citep[][]{zhang09}.
Radio continuum emission shows a collimated jet directed northwards~\citep[][]{heaton,gibb}.
Coincident with the radio jet emission a collimated jet has been detected at NIR and MIR wavelengths~\citep[][]{fuller,debuizer06,zhang13}.
An SW-NE poorly collimated CO outflow~\citep[collimation factor of 2.4--2.6, see][]{beuther,wu} was first detected by \citet{dent}, with the
blue-shifted lobe located NE of the IRAS position. Additional CO and SiO observations at higher angular 
resolution~\citep[][]{gibb,birks} suggested the presence of multiple precessing outflows. Both papers consider this interpretation more plausible
than the jet precession to explain the different jet/outflow orientations. On the other hand, \citet{lee12} detect 
an hourglass-shape H$_2$ emission with a large opening angle of $\sim$40$\degr$ (MHO 2431) emanating from the central region 
and extending $\sim$2$\arcmin$ SW-NE, i.e. $\sim$1.3\,pc at a distance of 2.2\,kpc. They interpret this as a single outflow with a large opening angle, rather
than as two distinct jets.

Indeed, it is not clear if there is just a single massive source at the centre of the region~\citep[][]{zhang13} or rather a small cluster~\citep[][]{sanchez}.
A 0.2\,pc interstellar disc-like structure was detected in H$^{13}$CO$^+$, H$^{13}$CN and millimetre continuum emission~\citep [][]{dent,gibb,sanchez}. 
At least three separate cores, namely G35.2N (A and B) and \object{[GHL2003] G35MM2} (hereafter G35MM2), are clearly 
detected with ALMA high angular observations~\citep[see Fig.~2 in][]{sanchez}. \citet{sanchez} estimate a mass of $\sim$18\,M$_{\sun}$ for the A/B G35.2N clumps.

Estimates of the observed bolometric luminosity are between 3.1$\times$10$^4$ and 3.3$\times$10$^4$\,L$_{\sun}$~\citep[][]{lumsden13,zhang13}.
However, \citet{zhang13} provide a more complex SED modelling, which takes into account both the foreground extinction and the so called flash-light effect,
i.e. the loss of photons through the outflow cavities. Their $L_{\rm bol}$ estimate ranges between 7$\times$10$^4$ and 2.2$\times$10$^5$\,L$_{\sun}$,
consistent with a single star of 20--30\,M$_{\sun}$. Such values, however, are in contrast with the momentum flux measured from the CO outflow,
that is more than one order of magnitude smaller than expected from such a luminous object~\citep[][]{zhang13}.
\citet{sanchez} argue that a binary system with a total mass of 18\,M$_{\sun}$ would better fit the measured bolometric luminosities and
outflow momentum flux, as well as it would better explain the jet/outflow axes displacement.

Our images indicate the presence of two main precessing jets (see Figure~\ref{G35ima:fig}). 
The first precessing jet (Knots 1, 2, 3, 4, 5) is launched by G35.2N, most likely from the position of clump B. Our NIR continuum images 
show a conical emission (P.A.$\sim$9$\degr$, opening angle of $\sim$45$\degr$), which is most likely the CO outflow cavity of the blue-shifted lobe 
of flow I~\citep[][see also Fig.~\ref{G35ima:fig} and discussion in Sect.~\ref{imaging:sec}]{gibb,birks}, excavated
by the jet. The current jet position angle (close to the source position, YSO emission + Knot 2) is $\sim$-1$\degr$,
i.e. coincident with the MIR and radio jet emission. The jet is then bending and precessing towards NE (knots 3, 4 and 5) with a precession angle of $\sim$28$\degr$,
in agreement with the opening of the outflow cavity. Knot\,1 and the emission towards SSW seem to depict the red-shifted lobe of the jet.
According to our spectroscopic analysis, the H$_2$ knots are clearly tracing the jet axis, and not oblique shocks from an expanding shell/outflow 
as suggested by \citet{lee12}, because of their excitation conditions. 
Most important, [\ion{Fe}{ii}] emission is clearly detected along the whole blue-shifted part of this flow, namely in the jet inside the outflow cavity and in 
Knots 2, 3, 4 and 5. Such emission is produced by strong dissociative shocks and it always traces the axis of the jets~\citep[][]{nisini02,pyo06,shang06,caratti09a},
therefore it cannot be associated with any outflow cavity or expanding flow (see Table~\ref{spec_G35.2N:tab}).
As in other flows, this jet is highly asymmetric, namely one lobe is more extended than the other (in this case the blue-shifted lobe is $\sim$2.5 times more extended
than the red one).

The second flow is composed of knots 6 and 7, in the red-shifted lobe, and knots 8, 9 and 10, in the blue-shifted lobe. 
This flow has a precession angle of 26$\degr$, and, if we consider the two terminal shocks of the flow (Knot\,6 and 10), the flow position angle is $\sim$45$\degr$.
We do not detect emission from any outflow cavity in our images, nor any source emission in the Spitzer GLIMPSE images, indicating that the driving source must be
very embedded and young. G35.2N is not a feasible candidate because the opening angle of the outflow cavity does not match with the jet position.
On the other hand, G35MM2 position lies very close ($\sim$0\farcs7) to the line intersecting the two terminal shocks, therefore it is the best driving source candidate.
However, given the large precession angle of the jet and the lack of H$_2$ emission close to the launching region, we cannot exclude other embedded sources
as possible candidates.

\begin{landscape}
\begin{table*}
\begin{scriptsize}
\caption{Observed emission lines in the \object{G35.2N} jets. \label{spec_G35.2N:tab}}
\begin{center}
\begin{tabular}{ccccccccccccc}
\hline\\[-5pt]
Species & Term &  $\lambda$($\mu$m) & \multicolumn{10}{c}{$F\pm\Delta~F$(10$^{-15}$erg\,cm$^{-2}$\,s$^{-1}$)}\\
\hline\\[-5pt]
             &          &  	 	&    knot\,1   &   knot\,2+YSO$^1$    & knot\,3 &   knot\,4+5   & knot\,6 & knot\,7 & knot\,8-1 & knot\,8-2  & knot\,9 & knot\,10 \\
{[\ion{C}{i}]} & $^1\!D_{2}-^3\!P_{1}$  & 0.983 &  $\cdots$	& $\cdots$     & $\cdots$    & 6.9$\pm$2 & $\cdots$  & $\cdots$ & $\cdots$  & $\cdots$ & $\cdots$ & $\cdots$   \\  
  & +$^1\!D_{2}-^3\!P_{2}$  & 0.985 & & & & & & & & & & \\
H$_2$ & 2--0 S(5)       & 1.084          &     $\cdots$	& $\cdots$     & $\cdots$    & 2.2$\pm$0.8$^*$ & $\cdots$  & $\cdots$ & $\cdots$  & $\cdots$ & $\cdots$ & $\cdots$   \\    
{[\ion{Fe}{ii}]} & $a^4\!D_{7/2}-a^6\!D_{9/2}$ & 1.257 	& $\cdots$	& $\cdots$     & 2.4$\pm$0.8    & 14.8$\pm$0.8    & $\cdots$  & $\cdots$ & $\cdots$  & $\cdots$ & $\cdots$ & $\cdots$   \\
\ion{H}{i} & Pa$\beta$                       & 1.287    &   $\cdots$	& $\cdots$     & $\cdots$    & 2.2$\pm$0.8$^*$ & $\cdots$  & $\cdots$ & $\cdots$  & $\cdots$ & $\cdots$ & $\cdots$   \\        
{[\ion{Fe}{ii}]} & $a^4\!D_{5/2}-a^6\!D_{5/2}$ & 1.293     &     $\cdots$	& $\cdots$     & $\cdots$    & 2.2$\pm$0.8$^*$ & $\cdots$  & $\cdots$ & $\cdots$  & $\cdots$ & $\cdots$ & $\cdots$   \\        
{[\ion{Fe}{ii}]} & $a^4\!D_{7/2}-a^6\!D_{7/2}$ & 1.321 	&    $\cdots$	& $\cdots$     & $\cdots$    & 5.3$\pm$0.8 & $\cdots$  & $\cdots$ & $\cdots$  & $\cdots$ & $\cdots$ & $\cdots$   \\    
{[\ion{Fe}{ii}]} & $a^4\!D_{5/2}-a^4\!F_{9/2}$ & 1.534     &    $\cdots$	& $\cdots$     & $\cdots$    & 2.4$\pm$1.0$^*$ & $\cdots$  & $\cdots$ & $\cdots$  & $\cdots$ & $\cdots$ & $\cdots$   \\        
{[\ion{Fe}{ii}]} & $a^4\!D_{3/2}-a^4\!F_{7/2}$ & 1.600     & $\cdots$	& $\cdots$     & $\cdots$    & 2.7$\pm$0.8 & $\cdots$  & $\cdots$ & $\cdots$  & $\cdots$ & $\cdots$ & $\cdots$   \\        
{[\ion{Fe}{ii}]} & $a^4\!D_{7/2}-a^4\!F_{9/2}$ & 1.644 	& $\cdots$	&  6.9$\pm$0.9    & 17.6$\pm$0.9    & 37.4$\pm$0.8  & $\cdots$  & $\cdots$ & $\cdots$  & $\cdots$ & $\cdots$ & $\cdots$   \\        
{[\ion{Fe}{ii}]} & $a^4\!D_{1/2}-a^4\!F_{5/2}$ & 1.664    & $\cdots$	& $\cdots$     & 2.6$\pm$0.8    & 2.6$\pm$0.8 & $\cdots$  & $\cdots$ & $\cdots$  & $\cdots$ & $\cdots$ & $\cdots$   \\        
{[\ion{Fe}{ii}]} & $a^4\!D_{5/2}-a^4\!F_{7/2}$ & 1.677  & $\cdots$	& $\cdots$     & $\cdots$    & 2.2$\pm$1.0$^*$ & $\cdots$  & $\cdots$ & $\cdots$  & $\cdots$ & $\cdots$ & $\cdots$   \\        
H$_2$ & 1--0 S(9) 	        & 1.688     & $\cdots$	& $\cdots$     & $\cdots$    & 3.1$\pm$1.0 & $\cdots$  & $\cdots$ & $\cdots$  & $\cdots$ & $\cdots$ & 1.7$\pm$0.7$^*$   \\        
H$_2$ & 1--0 S(8) 	        & 1.715      & $\cdots$	& $\cdots$     & $\cdots$    & 3.0$\pm$1.0 & $\cdots$  & $\cdots$ & $\cdots$  & $\cdots$ & $\cdots$ & $\cdots$   \\         
H$_2$ & 1--0 S(7) 	        & 1.748      &  2.1$\pm$0.8$^*$ & $\cdots$     & $\cdots$    & 9$\pm$1 & 6$\pm$1  & 4$\pm$1 & 2.2$\pm$0.7  & $\cdots$ & 3.8$\pm$0.9 & 5.6$\pm$0.7   \\         
H$_2$ & 1--0 S(6) 	        & 1.788         & $\cdots$	& $\cdots$     & $\cdots$    & 6$\pm$1 & $\cdots$  & $\cdots$ & $\cdots$  & $\cdots$ & $\cdots$ & 2.9$\pm$0.7   \\        \
{[\ion{Fe}{ii}]} & $a^4\!D_{7/2}-a^4\!F_{7/2}$ &1.810 & $\cdots$	& $\cdots$     & $\cdots$    & 9$\pm$2 & $\cdots$  & $\cdots$ & $\cdots$  & $\cdots$ & $\cdots$ & $\cdots$   \\        
               &  +$a^4\!P_{5/2}-a^4\!D_{7/2}$  &1.811  & & & & & & & & & & \\
H$_2$ & 1--0 S(5) 	        & 1.836     & $\cdots$	& $\cdots$     & $\cdots$    & 17$\pm$5 & 9$\pm$4$^*$  & $\cdots$ & $\cdots$  & $\cdots$ & 7$\pm$3$^*$ & 15$\pm$5   \\        
H$_2$ & 1--0 S(3) 		 & 1.958     & 11$\pm$4$^*$	& $\cdots$     & $\cdots$    & 45$\pm$4 & 26$\pm$4  & 18$\pm$4 & 9$\pm$4$^*$  & $\cdots$ & 9$\pm$4$^*$ & 22$\pm$5   \\        
H$_2$ & 1--0 S(2)           	& 2.034    & 3.2$\pm$1.0	& $\cdots$     & 2.4$\pm$1.0$^*$    & 12$\pm$2 & 8$\pm$1  & 8.5$\pm$1.5 & 3$\pm$1  & $\cdots$ & 5$\pm$1 & 10$\pm$2   \\        
\ion{He}{i} & $^1P^{0}1-^1S0$           & 2.059     & $\cdots$	& 5.1$\pm$1.7     & $\cdots$    &$\cdots$   & $\cdots$  & $\cdots$ & $\cdots$  & $\cdots$ & $\cdots$ & $\cdots$   \\   
H$_2$ & 2--1 S(3)           	& 2.073    & 2.1$\pm$1.0$^*$	& $\cdots$     & $\cdots$    & 6$\pm$1 & $\cdots$  & $\cdots$ & $\cdots$  & $\cdots$ & $\cdots$ & 4$\pm$1   \\        
\ion{He}{i} & $^3P^{0}1-^3S0$  & 2.113      & $\cdots$	& 3$\pm$1     & $\cdots$    & $\cdots$ & $\cdots$  & $\cdots$ & $\cdots$  & $\cdots$ & $\cdots$ & $\cdots$   \\  
H$_2$ & 1--0 S(1)           	& 2.122    & 12.5$\pm$1.0	& $\cdots$     & 8.4$\pm$0.7    & 25.8$\pm$0.8 & 20$\pm$1  & 15$\pm$1 & 6$\pm$1  & 3$\pm$1 & 11$\pm$1 & 24.1$\pm$0.8   \\        
H$_2$ & 2--1 S(2)           	& 2.154    & $\cdots$	& $\cdots$     & $\cdots$    & 2.2$\pm$0.8$^*$ & $\cdots$  & $\cdots$ & $\cdots$  & $\cdots$ & $\cdots$ & $\cdots$   \\        
\ion{H}{i} & Br$\gamma$  & 2.166      & $\cdots$	& 9.2$\pm$1.5     & $\cdots$    & 2.0$\pm$0.7$^*$ & $\cdots$  & $\cdots$ & $\cdots$  & $\cdots$ & $\cdots$ & $\cdots$   \\        
\ion{Na}{i} &           & 2.187      & $\cdots$	& 2.4$\pm$1.5$^*$     & $\cdots$    & $\cdots$ & $\cdots$  & $\cdots$ & $\cdots$  & $\cdots$ & $\cdots$ & $\cdots$   \\ 
\ion{Na}{i} &           & 2.190      & $\cdots$	& 4.8$\pm$1.5     & $\cdots$    & $\cdots$ & $\cdots$  & $\cdots$ & $\cdots$  & $\cdots$ & $\cdots$ & $\cdots$   \\ 
H$_2$ & 1--0 S(0)           	& 2.223     & 3.5$\pm$1.0	& 4.0$\pm$1.5$^*$     & 3.5$\pm$1.0    & 9$\pm$1 & 6.9$\pm$1.5  & 4$\pm$1 & $\cdots$  & $\cdots$ & 3$\pm$1 & 6$\pm$1   \\        
H$_2$ & 2--1 S(1)           	& 2.248      & $\cdots$	& $\cdots$     & $\cdots$    & 6$\pm$1 & $\cdots$  & $\cdots$ & $\cdots$  & $\cdots$ & $\cdots$ & 3$\pm$1   \\        
CO & v=2--0                & 2.294      & $\cdots$	& 37$\pm$3     & $\cdots$    & $\cdots$ & $\cdots$  & $\cdots$ & $\cdots$  & $\cdots$ & $\cdots$ & $\cdots$   \\ 
CO & v=3--1                & 2.323      & $\cdots$	& 29$\pm$3     & $\cdots$    & $\cdots$ & $\cdots$  & $\cdots$ & $\cdots$  & $\cdots$ & $\cdots$ & $\cdots$   \\ 
CO & v=4--2                & 2.353      & $\cdots$	& 32$\pm$3     & $\cdots$    & $\cdots$ & $\cdots$  & $\cdots$ & $\cdots$  & $\cdots$ & $\cdots$ & $\cdots$   \\ 
CO & v=5--3                & 2.383      & $\cdots$	& 31$\pm$5     & $\cdots$    & $\cdots$ & $\cdots$  & $\cdots$ & $\cdots$  & $\cdots$ & $\cdots$ & $\cdots$   \\ 
H$_2$ & 1--0 Q(1)           	& 2.407     & 15$\pm$4 & 11$\pm$5$^*$     & $\cdots$    & 23$\pm$5 & 23$\pm$4  & 19$\pm$4 & $\cdots$  & $\cdots$ & 13$\pm$4 & 25$\pm$4   \\        
H$_2$ & 1--0 Q(2)           	& 2.413    & $\cdots$	& $\cdots$     & $\cdots$    & 11$\pm$4$^*$ & $\cdots$  & $\cdots$ & $\cdots$  & $\cdots$ & $\cdots$ & 11$\pm$4$^*$   \\        
H$_2$ & 1--0 Q(3)           	& 2.424    & 12$\pm$4	& $\cdots$     & $\cdots$    & 26$\pm$4 & 21$\pm$4  & 15$\pm$4 & $\cdots$  & $\cdots$ & 13$\pm$4 & 26$\pm$4   \\        
H$_2$ & 1--0 Q(4)           	& 2.437    & $\cdots$	& $\cdots$     & $\cdots$    & 15$\pm$5 & $\cdots$  & $\cdots$ & $\cdots$  & $\cdots$ & $\cdots$ & 10$\pm$4$^*$   \\        
H$_2$ & 1--0 Q(5)           	& 2.455    & 12$\pm$4	& $\cdots$     & $\cdots$    & $\cdots$ & $\cdots$  & 10$\pm$4$^*$ & $\cdots$  & $\cdots$ & $\cdots$ & 15$\pm$5   \\        
\hline\\[-5pt]
\hline
\end{tabular}
\tablefoot{$^*$ S/N between 2 and 3.$^{**}$ YSO = \object{NAME G 35.2N}, namely \object{2MASSJ 18581310+0140399}}
\end{center}
\end{scriptsize}
\end{table*}
\end{landscape}
\end{appendix}
\end{document}